\documentclass[12pt,a4paper,oneside]{book}

\usepackage[T1]{fontenc}
\usepackage[utf8]{inputenc}
\usepackage[brazil,english]{babel}
\usepackage{parskip}

\usepackage[top=3cm,left=3cm,right=2cm,bottom=2cm]{geometry}
\usepackage{setspace}
\usepackage{amsmath,amsthm,amsfonts,amssymb,mathtools}
\usepackage{slashed}   
\usepackage{braket}    
\usepackage{bbold}     

\usepackage{graphicx}
\usepackage{subcaption}
\usepackage{float}
\usepackage{booktabs}
\usepackage{longtable}

\usepackage{xcolor}
\usepackage{url}
\usepackage[hidelinks]{hyperref}
\usepackage{pdfpages}
\usepackage[round,comma,authoryear]{natbib}
\usepackage[titletoc]{appendix}

\begin{document}
\selectlanguage{english}

\pagenumbering{gobble}

\begin{titlepage}
\thispagestyle{empty}
\begin{center}
\vspace*{3.0cm}

{\large Jonathan Tejeda Quartuccio\par}

\vfill

{\Large\bfseries Deformed Compact Objects in General Relativity and Modified Gravity\par}

\vfill

{\large São Paulo\par}
\vspace{0.2cm}
{\large 2026\par}

\end{center}
\end{titlepage}

\begin{titlepage}
\begin{center}
{ UNIVERSIDADE CIDADE DE SÃO PAULO\par}\vspace{0.2cm}
{ PROGRAMA DE PÓS GRADUAÇÃO\par}\vspace{0.2cm}
{ DOUTORADO EM ASTROFÍSICA E FÍSICA COMPUTACIONAL\par}\vspace{3.1cm}

\vspace{1.0cm}
{\large Jonathan Tejeda Quartuccio\par}\vspace{3.5cm}

{\Large \bfseries Deformed Compact Objects in General Relativity and Modified Gravity \par}\vspace{5.5cm}

\end{center}

\vspace{1.0cm}

\begin{center}
{\large São Paulo\par}\vspace{0.2cm}
{\large 2026\par}
\end{center}
\end{titlepage}

\begin{titlepage}
\thispagestyle{empty}
\begin{center}
{\large Jonathan Tejeda Quartuccio\par}
\vspace{4.0cm}

{\Large\bfseries Deformed Compact Objects in General Relativity and Modified Gravity\par}
\end{center}

\vspace{3.0cm}

\begin{flushright}
\begin{minipage}{0.58\textwidth}
\onehalfspacing
Texto apresentado ao Programa de Pós-Graduação em Astrofísica e Física Computacional da Universidade Cidade de São Paulo, para a defesa de Doutorado, sob orientação do Prof. Dr. Pedro H. R. S. Moraes.
\end{minipage}
\end{flushright}

\vfill

\begin{center}
{\large São Paulo\par}
\vspace{0.2cm}
{\large 2026\par}
\end{center}
\end{titlepage}

\cleardoublepage
\includepdf[pages=1]{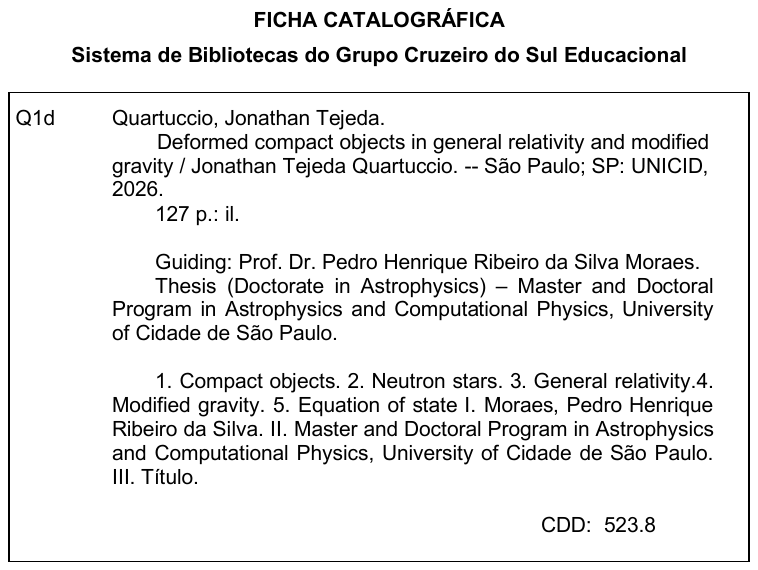}
\cleardoublepage

\begin{titlepage}
\thispagestyle{empty}

\begin{center}
{\bfseries FOLHA DE APROVAÇÃO\par}\vspace{1cm}
\end{center}

\noindent
Tese de Doutorado intitulada \textbf{Deformed Compact Objects in General Relativity and Modified Gravity}, de autoria de \textbf{Jonathan Tejeda Quartuccio}, apresentada ao Programa de Pós-Graduação em Astrofísica e Física Computacional da Universidade Cidade de São Paulo, como parte dos requisitos para obtenção do título de Doutor.

\vspace{0.8cm}

\noindent\textbf{Banca Examinadora:}

\vspace{0.9cm}

\begin{center}
\rule{0.78\textwidth}{0.4pt}\\[-0.05cm]
\textbf{Prof. Dr. Pedro Henrique Ribeiro da Silva Moraes}\\
Orientador -- Universidade Cidade de São Paulo

\vspace{0.9cm}

\rule{0.78\textwidth}{0.4pt}\\[-0.05cm]
\textbf{Prof. Dr. João Pacheco Bicudo Cabral de Melo}\\
Universidade Cidade de São Paulo

\vspace{0.9cm}

\rule{0.78\textwidth}{0.4pt}\\[-0.05cm]
\textbf{Prof. Dr. Kazuo Tsushima}\\
Universidade Cidade de São Paulo

\vspace{0.9cm}

\rule{0.78\textwidth}{0.4pt}\\[-0.05cm]
\textbf{Prof. Dr. José Domingo Arbañil Vela}\\
Universidad Privada del Norte

\vspace{0.9cm}

\rule{0.78\textwidth}{0.4pt}\\[-0.05cm]
\textbf{Prof. Dr. Ronaldo Vieira Lobato}\\
Centro Brasileiro de Pesquisas Físicas
\end{center}

\vfill
\begin{center}
São Paulo, \_\_\_ de \_\_\_\_\_\_\_\_\_\_\_\_\_ de 2026.
\end{center}

\end{titlepage}

\cleardoublepage
\thispagestyle{empty}

\vspace*{\fill}

\begin{flushright}
\textit{Ao meu filho, Heitor,\\
razão de esperança e continuidade.\\[0.3cm]
Aos meus pais,\\
que já não estão mais entre nós,\\
mas permanecem vivos em minha memória,\\
em minha história e em cada passo desta caminhada.}
\end{flushright}

\vspace*{3cm}

\cleardoublepage

\cleardoublepage
\pagenumbering{roman}

\chapter*{Agradecimentos}
\addcontentsline{toc}{chapter}{Agradecimentos}

Agradeço à minha esposa, Stephanie, pelo amor, pela paciência e pelo apoio constante ao longo desta caminhada. Sua presença foi essencial nos momentos de dificuldade, incerteza e cansaço, assim como nas conquistas que marcaram esta trajetória. Esta etapa também carrega muito do seu apoio, da sua compreensão e da sua parceria.

À minha família, agradeço pelo carinho, pelo incentivo e por todo o suporte oferecido ao longo dos anos.

Ao meu orientador, Prof. Dr. Pedro H. R. S. Moraes, agradeço por ter me apoiado desde o início desta jornada, sempre respeitando minhas escolhas de estudo e incentivando minha autonomia científica. Sua confiança e a liberdade que me concedeu para trabalhar com os temas que mais me cativavam foram fundamentais para que esta tese se desenvolvesse de forma tão significativa para mim.

Agradeço especialmente ao Prof. Dr. Fridolin Weber por ter me recebido na San Diego State University durante meu período de doutorado-sanduíche, proporcionando-me uma experiência de aprendizado extremamente enriquecedora. Sua acolhida, orientação e as discussões realizadas nesse período contribuíram de maneira significativa para minha formação acadêmica e científica.

Estendo meus agradecimentos também aos professores, colegas e amigos que, de diferentes formas, contribuíram para minha formação acadêmica, científica e humana. As conversas, discussões, sugestões e momentos compartilhados foram parte importante deste percurso.

Por fim, agradeço à Coordenação de Aperfeiçoamento de Pessoal de Nível Superior (CAPES), pelo apoio financeiro concedido durante o desenvolvimento deste trabalho.

\cleardoublepage

\thispagestyle{empty}
\vspace*{0.55\textheight}

\begin{flushright}
\begin{minipage}{0.58\textwidth}
\itshape
``A mente é amiga daquele que a dominou;\\
mas, para aquele que não a dominou,\\
a própria mente age como inimiga.''

\vspace{0.4cm}
\normalfont--- \textit{Bhagavad Gita}, 6.6
\end{minipage}
\end{flushright}

\cleardoublepage

\tableofcontents
\listoffigures
\listoftables
\cleardoublepage


\chapter*{Resumo}
\addcontentsline{toc}{chapter}{Resumo}

Estrelas de nêutrons e outros objetos compactos constituem laboratórios naturais únicos para o estudo da matéria em densidades supranucleares e da gravidade no regime de campo forte. Nesta tese, investigamos o equilíbrio hidrostático de estrelas compactas em diferentes cenários geométricos e gravitacionais, combinando estrutura estelar relativística, microfísica da matéria densa e modelagem numérica. Inicialmente, revisamos os principais ingredientes microfísicos relevantes para objetos compactos e discutimos equações de estado representativas para matéria hadrônica e matéria de quarks. Em seguida, deduzimos a equação padrão de Tolman-Oppenheimer-Volkoff (TOV) na Relatividade Geral, introduzimos um esquema efetivo de deformação com um parâmetro que conduz ao formalismo TOV deformado ($\mathcal{D}$-TOV) e estendemos o tratamento do equilíbrio hidrostático para a gravidade $f(R,T)$, considerando configurações esféricas e deformadas.

A partir das equações de estrutura obtidas, calculamos sequências de equilíbrio para estrelas de nêutrons e estrelas estranhas descritas pela equação de estado GM1, pelo Modelo de Sacola do MIT e por uma equação de estado politrópica. Na Relatividade Geral, mostra-se que o parâmetro de deformação $\mathcal{D}$ exerce impacto significativo sobre as propriedades globais estelares, com configurações oblatas $(\mathcal{D}<1)$ sustentando, em geral, maiores massas e raios do que o caso esférico, enquanto configurações prolatas $(\mathcal{D}>1)$ levam a estrelas menos massivas e mais compactas. No modelo otimizado de $f(R,T)$, as correções dependentes do traço produzem configurações de equilíbrio ligeiramente mais massivas e mais extensas do que suas correspondentes na Relatividade Geral. No modelo linear $f(R,T)=R+2\lambda T$, a ação combinada entre deformação e acoplamento matéria-geometria modifica adicionalmente a relação massa-raio, com valores positivos de $\lambda$ aumentando sistematicamente a massa máxima suportada para um valor fixo de $\mathcal{D}$.

De modo geral, os resultados indicam que tanto a deformação efetiva quanto o acoplamento matéria-geometria dependente do traço podem afetar de maneira significativa a estrutura de equilíbrio de estrelas compactas.

\vspace{0.5cm}

\noindent\textbf{Palavras-chave:} Objetos compactos; Estrelas de nêutrons; Relatividade Geral; Gravidade modificada; Equação de estado.

\cleardoublepage

\chapter*{Abstract}
\addcontentsline{toc}{chapter}{Abstract}

Neutron stars and related compact objects provide unique laboratories for probing matter at supranuclear densities and gravity in the strong-field regime. In this thesis, we investigate the hydrostatic equilibrium of compact stars in different geometric and gravitational scenarios, combining relativistic stellar structure, dense-matter microphysics, and numerical modeling. We first review the main microphysical ingredients relevant to compact stars and discuss representative equations of state for hadronic and quark matter. We then derive the standard Tolman-Oppenheimer-Volkoff (TOV) equation in General Relativity, introduce an effective one-parameter deformation scheme leading to the deformed TOV ($\mathcal{D}$-TOV) formalism, and extend the hydrostatic equilibrium framework to $f(R,T)$ gravity, considering both spherical and deformed configurations.

Using the resulting structure equations, we compute equilibrium sequences for neutron stars and strange stars described by the GM1 equation of state, the MIT Bag Model, and a polytropic equation of state. In General Relativity, the deformation parameter $\mathcal{D}$ is shown to have a significant impact on the global stellar properties, with oblate configurations $(\mathcal{D}<1)$ generally supporting larger masses and radii than the spherical case, while prolate configurations $(\mathcal{D}>1)$ lead to less massive and more compact stars. In the optimized $f(R,T)$ model, the trace-dependent corrections produce equilibrium configurations that are slightly more massive and more extended than their General Relativity counterparts. In the linear model $f(R,T)=R+2\lambda T$, the combined action of deformation and matter-geometry coupling further modifies the mass-radius relation, with positive values of $\lambda$ systematically increasing the maximum supported mass for fixed $\mathcal{D}$.

Overall, the results indicate that both effective deformation and trace-dependent matter-geometry coupling can significantly affect the equilibrium structure of compact stars.

\vspace{0.5cm}

\noindent\textbf{Keywords:} Compact objects; Neutron stars; General relativity; Modified gravity; Equation of state.

\chapter*{Publications related to this thesis}
\addcontentsline{toc}{chapter}{Publications related to this thesis}

Some of the results presented in this thesis have been published in peer-reviewed journals during the doctoral work. These publications reflect the development of the main research lines explored here, namely the study of compact stars in modified gravity and the investigation of deformed compact configurations. The published works directly related to this thesis are listed below:

\begin{itemize}
    \item Quartuccio, J. T., Moraes, P. H. R. S., \& Arbanil, J. D. V., ``Deformed Compact Objects,'' \textit{International Journal of Theoretical Physics} \textbf{64}, 23 (2025).
    
    \item Quartuccio, J. T., Moraes, P. H. R. S., Zeminiani, G. N., \& Lapola, M. M., ``The equilibrium configurations of neutron stars in the optimized \(f(R,T)\) gravity,'' \textit{Astrophysics and Space Science} \textbf{370}, 37 (2025).

    \item Quartuccio, J. T., \& Moraes, P. H. R. S., ``Deformed compact objects in modified gravity,'' \textit{The European Physical Journal Plus} \textbf{141}, 447 (2026).
\end{itemize}

These articles are closely connected to the central themes of the present thesis and provide part of the theoretical and numerical foundation for the analyses developed in the following chapters.

\cleardoublepage

\pagenumbering{arabic}
\selectlanguage{english}

\chapter{Introduction}

Compact objects represent the terminal outcomes of stellar evolution. For stars with
relatively low initial masses (typically $\lesssim 8\,M_\odot$), the end state is usually a white dwarf,
formed after the star expels its outer layers and leaves behind a degenerate core.
For more massive progenitors (roughly $\sim 8$ to $\sim 20\,M_\odot$), core collapse followed by a supernova can produce a neutron star (NS). For sufficiently massive stars (often $\gtrsim 20\,M_\odot$), core collapse is more likely to result in a black hole, either after a supernova with substantial fallback
or, in some cases, via direct collapse \citep{carroll2017introduction}.

In this work, we focus on neutron stars (NSs) and possible variations such as quark stars (commonly referred to as \emph{strange stars}). Neutron stars can be regarded as natural laboratories for investigating nuclear and particle physics, as well as gravitation, under conditions that are far more extreme than those attainable in terrestrial experiments. This is evident from the expected densities in NS interiors, which may exceed the nuclear saturation density by up to an order of magnitude \citep{liweber}. Consequently, NSs provide key systems for probing the behavior of matter in ultra-high density regimes.

The physics of matter under extreme conditions, such as those expected in NS interiors, remains uncertain. Microphysics determines how dense matter responds to the enormous compression at supranuclear densities. Understanding this microphysics is therefore essential for establishing the relation between pressure and energy density, $p(\varepsilon)$, through the equation of state (EoS). In turn, the EoS largely determines macroscopic stellar properties, including the maximum mass a compact object can support against gravitational collapse and the corresponding radii along equilibrium sequences.

From a theoretical perspective, NSs lie at the interface between microphysics and macrophysics. On the one hand, the EoS encodes the relevant degrees of freedom and interactions of strongly interacting matter. On the other hand, once an EoS is specified, the stellar-structure equations connect it to observables such as the mass-radius relation, the compactness, and (depending on the context) properties related to stellar oscillations.

Realistic NS modeling is challenged by several factors. First, the true EoS of
ultra-dense nuclear matter remains uncertain \citep{oertel2017equations,ozel2016masses}. Second, a complete description of NSs must account for additional physical ingredients, such as rotation, intense magnetic fields, temperature effects, superfluidity, and superconductivity. These effects can significantly influence the structure and evolution of compact stars and make their theoretical treatment substantially more complex \citep{stergioulas2003rotating,oertel2017equations,haskell2019superfluidity}.

For a given EoS, solving the stellar-structure equations yields a theoretical mass-radius relation $M(R)$, which can be directly confronted with observational data. Among the most important constraints are mass measurements of NSs in binary systems, particularly those obtained through pulsar timing \citep{demorest2010two, antoniadis2013massive}. A canonical reference value for neutron-star masses is $\sim 1.4\,M_\odot$, while inferred radii are typically of order $\sim 10$--$15$ km \citep{ozel2016masses, steiner2013neutron}. The discovery of pulsars with masses close to $2\,M_\odot$ has provided a particularly stringent test for candidate equations of state, since any viable EoS within General Relativity (GR) must be able to support such masses without inducing collapse to a black hole \citep{demorest2010two, antoniadis2013massive, cromartie2020relativistic}.

\begin{figure}[h]
    \centering
    \includegraphics[width=0.9\linewidth]{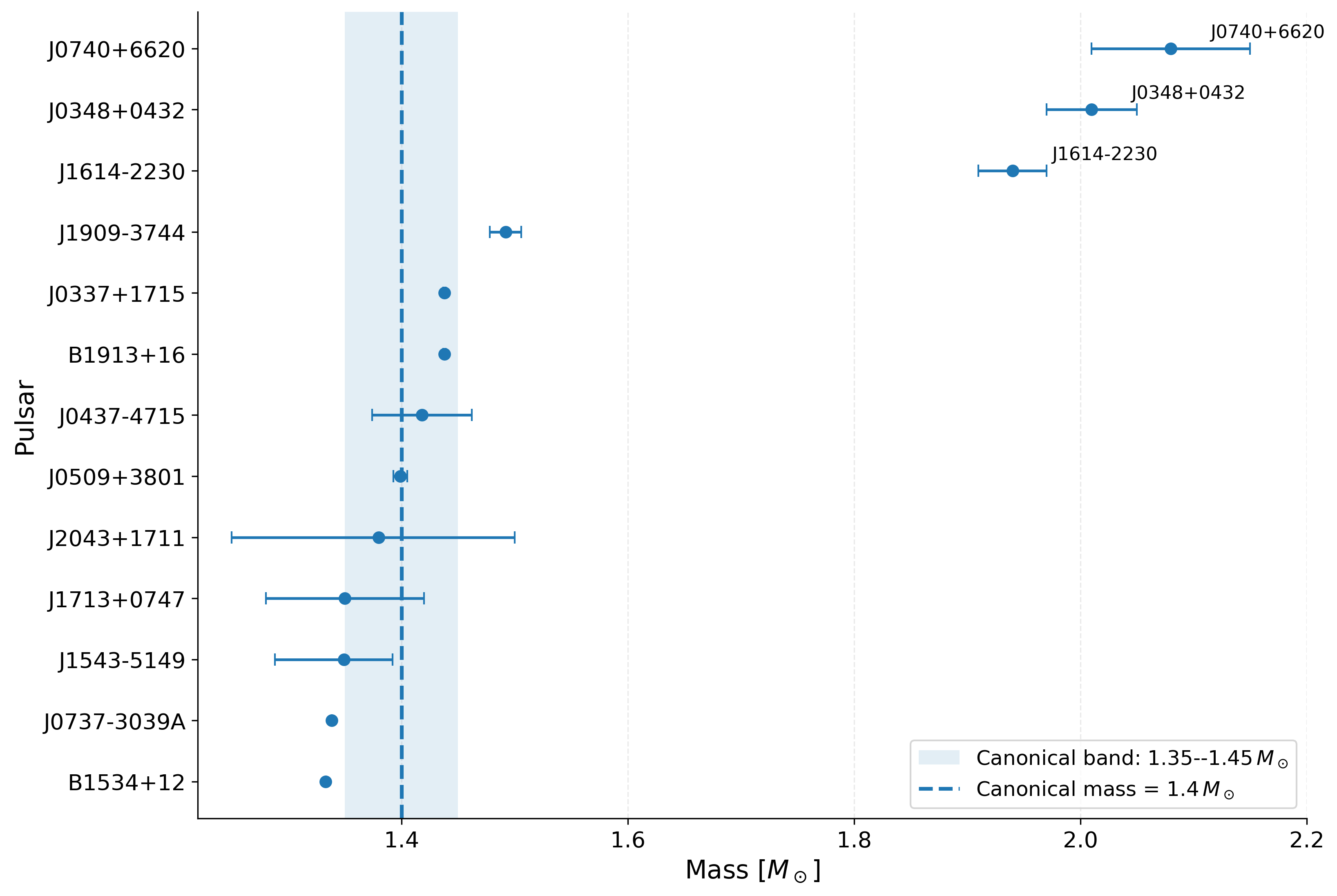}
    \caption{Observed masses of selected pulsars, highlighting the canonical mass scale around \(1.4\,M_\odot\) and the existence of massive systems with masses close to or above \(2\,M_\odot\). The shaded band indicates the canonical interval \(1.35\)--\(1.45\,M_\odot\), while the dashed line marks \(1.4\,M_\odot\). Error bars, when shown, represent the reported uncertainties in the corresponding mass measurements. Data are based on representative pulsar mass measurements compiled in the literature \citep{ozel2016masses, freire2024pulsar}, including high-mass systems such as PSR~J1614--2230, PSR~J0348+0432, and PSR~J0740+6620 \citep{demorest2010two, antoniadis2013massive, cromartie2020relativistic, fonseca2021refined}.}
    \label{fig:pulsar_masses}
\end{figure}

More recently, joint constraints on mass and radius have been obtained through X-ray pulse-profile modeling with NICER, in some cases combined with XMM-Newton observations. A remarkable example is the massive pulsar PSR~J0740+6620, for which the inferred mass is \(M = 2.08 \pm 0.07\,M_\odot\), with radius estimates of order \(12\)--\(14\) km obtained from independent NICER/XMM-Newton analyses \citep{miller2021radius, riley2021nicer}. NICER measures both energy-resolved and phase-resolved X-ray pulse profiles from rotating millisecond pulsars. In the strong-gravity regime, relativistic effects such as light bending, Doppler boosting, and time delays reshape the observed waveform. By fitting theoretical pulse-profile models, which depend on the stellar compactness \(M/R\) and on the geometry of the emitting hot spots, one can infer posterior distributions for \(M\) and \(R\). XMM-Newton observations provide complementary spectral information and help constrain background and calibration systematics, thereby improving the robustness of the inferred parameters \citep{gendreau2016neutron}. Representative pulsar mass measurements, including canonical and massive systems, are displayed in Fig.~\ref{fig:pulsar_masses}.

Complementary to these observational constraints, a classic theoretical benchmark is the upper bound derived by Rhoades and Ruffini \citep{rhoades1974maximum}. Assuming that GR is valid, that the EoS is reliably known up to a fiducial density, and that matter at higher densities satisfies causality, they obtained an upper limit of order \(\sim 3.2\,M_\odot\) for the maximum mass of a nonrotating NS. Although this result is model-independent only within those assumptions, it remains historically important because it shows that the maximum-mass problem is constrained not only by microphysics, but also by very general principles such as relativistic hydrostatic equilibrium and causality. In this sense, the Rhoades-Ruffini limit provides a useful theoretical reference when discussing whether particularly massive compact objects may still be interpreted as NSs.

In parallel, gravitational-wave detections have provided additional information about the properties of compact objects. Gravitational waves are dynamical perturbations of spacetime emitted by accelerated systems with a time-varying mass quadrupole moment, such
as compact binaries. Predicted by GR, the first direct detection of a gravitational-wave signal was made in 2015 and reported in 2016 \citep{abbott2016observation}. The component masses are inferred by matching the observed signal to the relativistic waveform templates.
In particular, the phase evolution of the inspiral depends sensitively on the chirp mass, defined as
\begin{equation}
\mathcal{M}
=
\frac{(m_1m_2)^{3/5}}{(m_1+m_2)^{1/5}},
\label{eq:chirp_mass}
\end{equation}
where \(m_1\) and \(m_2\) are the component masses of the binary. Equivalently, in terms of the total mass \(M=m_1+m_2\) and the symmetric mass ratio \(\eta=m_1m_2/M^2\), the chirp mass can be written as
\begin{equation}
\mathcal{M}=\eta^{3/5}M.
\end{equation}
Because \(\mathcal{M}\) controls the leading-order frequency evolution of the inspiral, it is usually one of the most accurately measured mass parameters in a compact-binary coalescence. Combined with information on the binary mass ratio, it allows one to estimate the individual component masses of the system.

\begin{figure}[H]
    \centering
    \includegraphics[width=0.82\linewidth]{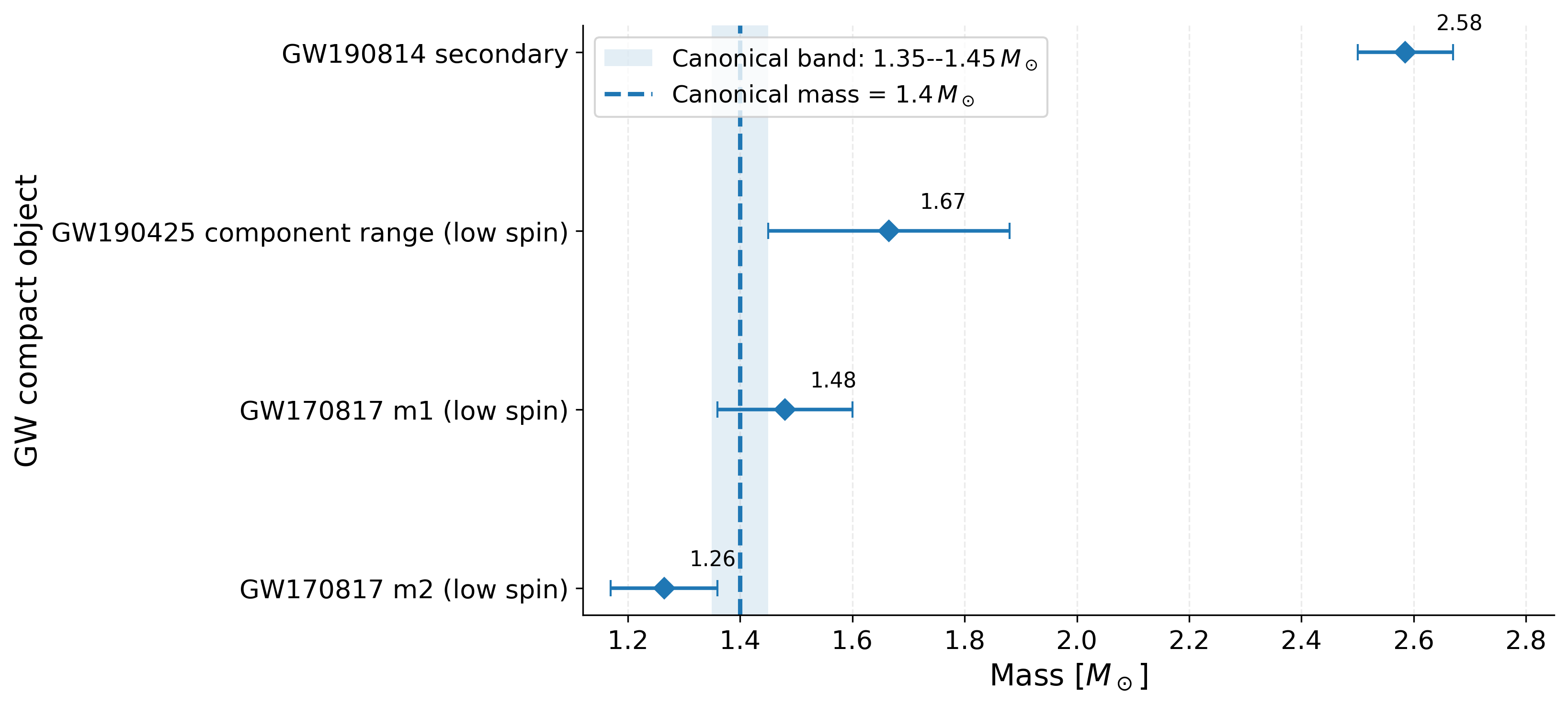}
    \caption{Representative masses of compact objects inferred from selected gravitational-wave events, compared with the canonical neutron-star mass scale. The figure highlights the secondary component of GW190814, whose inferred mass lies in the range commonly associated with the lower mass gap. The shaded band indicates the canonical interval \(1.35\)--\(1.45\,M_\odot\), while the dashed line marks \(1.4\,M_\odot\). Error bars indicate the corresponding 90\% credible intervals reported for each event. Data are taken from GW170817, GW190425, and GW190814 analyses \citep{abbott2017gw170817, abbott2020gw190425, abbott2020gw190814}
    .}
    \label{fig:gw_masses}
\end{figure}

Of particular interest for NS physics is the event GW190814. The signal is consistent with the merger of a black hole with mass in the range \(\sim 22.2\text{--}24.3\,M_\odot\) and a compact companion with estimated mass \(\sim 2.50\text{--}2.67\,M_\odot\) \citep{abbott2020gw190814}. The nature of this secondary object remains uncertain: its mass is low for a typical black hole, yet high for a conventional NS interpretation, placing it in what is commonly referred to as the lower mass gap. Some works have argued that such a massive NS could be supported by sufficiently stiff equations of state \citep{huang2020possibility}, whereas others have explored a low-mass black-hole interpretation \citep{vattis2020could}. If the secondary object is indeed an NS, GW190814 further motivates the investigation of the physical mechanisms that determine the maximum mass that an NS can support before collapsing into a black hole. Although its inferred mass does not exceed the classic Rhoades-Ruffini upper bound, it remains unusually high from the standpoint of standard NS models. A comparison with representative masses inferred from selected gravitational-wave events is shown in Fig.~\ref{fig:gw_masses}, highlighting the unusually high mass of the secondary component in GW190814.

\section{Effective deformation scheme}

Compact objects, such as NSs and white dwarfs, are supported against gravity by the
pressure of dense matter. In white dwarfs, this pressure is provided mainly by degenerate
electrons, whereas in NSs it is associated with degenerate baryonic matter together with
strong nuclear interactions at supranuclear densities. Their equilibrium structure can be
described, in the relativistic regime, by the hydrostatic equilibrium condition known as
the Tolman--Oppenheimer--Volkoff (TOV) equation
\citep{tolman1939static,oppenheimer1939massive}. The standard TOV framework assumes
static, isotropic configurations with perfect spherical symmetry. However, this
approximation may not be sufficient for a more general description of compact stars,
since such objects may exhibit rapid rotation \citep{wen2011properties, stergioulas2003rotating}, strong magnetic fields \citep{rizaldy2018magnetized}, anisotropic
stresses \citep{ruderman1972pulsars}, or other mechanisms capable of breaking spherical symmetry.

Motivated by the need to account, at least effectively, for deviations from spherical symmetry, we adopt as a first approximation a one-parameter deformation scheme. In this approach, the global shape of the object is encoded in a single dimensionless parameter defined as \citep{zubairi2017}
\begin{equation}
\mathcal{D} \equiv \frac{z}{r},
\end{equation}
where \(z\) and \(r\) denote effective polar and equatorial length scales, respectively. It is important to emphasize that this prescription does not represent a fully axisymmetric treatment of the stellar geometry. Instead, the angular sector of the metric is still described in terms of the spherical areal radius \(r\), while deviations from spherical symmetry are effectively incorporated through the parameter \(\mathcal{D}\). Therefore, \(z\) should be understood as an effective polar scale used to characterize the global deformation of the object, rather than as an independent coordinate.

\begin{figure}[H]
    \centering
    \includegraphics[scale=0.7]{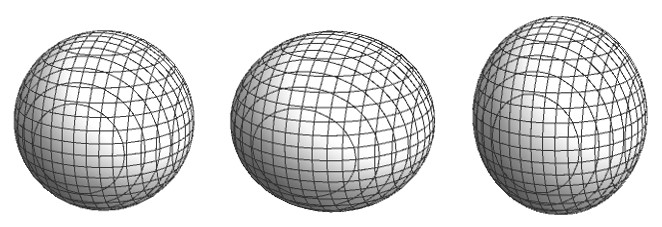}
    \caption{Schematic representation of the deformation parameter: spherical configuration for \(\mathcal{D}=1\) (left panel), oblate configuration for \(\mathcal{D}=0.8\) (middle panel), and prolate configuration for \(\mathcal{D}=1.2\) (right panel) \citep{quartuccio2025deformed}.}
    \label{fig:deformation_shapes}
\end{figure}

Within this framework, the spherical limit is recovered for \(\mathcal{D}=1\), while \(\mathcal{D}<1\) and \(\mathcal{D}>1\) correspond to the oblate and prolate configurations, respectively \citep{zubairi2017,quartuccio2025deformed}. Since this approach is based on an effective deformation of an otherwise spherically symmetric background, it is mainly intended to describe moderate deviations from spherical symmetry. This behavior is illustrated in Fig.~\ref{fig:deformation_shapes}.

This effective description is not intended to replace fully self-consistent two-dimensional treatments of rotation or magnetization. Rather, it provides a simple and tractable way to parametrize departures from spherical symmetry and to investigate how such departures affect the equilibrium structure of compact stars. As will be discussed in a later chapter, the introduction of the deformation parameter into the stellar-structure equations modifies the mass-radius relation. In particular, for sufficiently compact configurations, the resulting shifts in the equilibrium sequences may contribute, at least in principle, to accommodating heavier NSs within a given microphysical description.

\section{Modified gravity}

Beyond geometric effects such as deformation, another important line of research on compact objects considers extensions of GR. These so-called modified gravity theories explore mechanisms capable of affecting stellar structure and, in particular, may help accommodate larger maximum masses for a given EoS \citep{astashenok2013further,olmo2020stellar}. A variety of modified gravity frameworks have been proposed in the literature, including $f(R)$ theories, whose early developments can be traced back to nonlinear curvature Lagrangians and to the Starobinsky $R+R^2$ model \citep{buchdahl1970non, starobinsky1980new, capozziello2009f}, $f(R,T)$ models \citep{harko2011f}, $f(R,\mathcal{L})$ constructions \citep{harko2010f}, more general extensions such as $f(R,\mathcal{L},T)$ \citep{haghani2021generalizing}, and alternative proposals based on a deeper matter-geometry coupling, such as Entangled Relativity \citep{minazzoli2018rethinking}.

From the perspective of compact-star physics, these theories are particularly relevant because they introduce corrections to the gravitational sector that may alter the hydrostatic equilibrium equations and, consequently, modify the predicted mass-radius relation, the maximum supported mass, and other global stellar properties. In this sense, modified gravity provides a complementary route to the microphysical description of dense matter, since changes in stellar observables may arise not only from the EoS itself, but also from the underlying gravitational theory.


In this thesis, we investigate stellar structure in two main settings: (i) General Relativity, and (ii) the subclass of $f(R,T)$ theories of the form $f(R,T)=R+f(T)$. Within this subclass, we analyze two specific models for $f(T)$: a linear model and an optimized nonlinear model. We consider perfectly spherical configurations and configurations with broken spherical symmetry described through an effective deformation scheme. This organization allows for a systematic investigation of the separate and combined effects of effective deformation and modified gravity on the equilibrium properties of compact stars.

\section{Objectives and contributions}

The general objective of this thesis is to investigate the hydrostatic equilibrium of compact objects under different geometric and gravitational assumptions, with special emphasis on the effects of deformation and modified gravity on the global properties of compact stars.

More specifically, this thesis aims:
\begin{itemize}
    \item to formulate and analyze the stellar-structure equations in four distinct situations: (i) spherically symmetric configurations in General Relativity, governed by the standard Tolman--Oppenheimer--Volkoff equations; (ii) non-spherical configurations in General Relativity, described through the effective deformation scheme ($\mathcal{D}$-TOV); (iii) spherically symmetric configurations in $f(R,T)$ gravity of the form $f(R,T)=R+f(T)$; and (iv) non-spherical configurations in the linear model $f(R,T)=R+2\lambda T$, described through the same effective deformation scheme;

    \item to solve these structure equations numerically for different equations of state, obtaining equilibrium sequences, mass--radius relations, central-density relations, and internal stellar profiles;
    
    \item to analyze systematically the isolated effects of the deformation parameter \(\mathcal{D}\) and of the \(f(R,T)\) model parameters, as well as their combined impact on the equilibrium structure of compact stars;
    
    \item to compare the theoretical predictions with observationally motivated constraints on masses and radii.
\end{itemize}

The main contribution of this thesis lies in providing a unified investigation of compact-star equilibrium across spherical and deformed configurations, both in GR and in \(f(R,T)\) gravity. By organizing the analysis in this way, the present work makes it possible to identify separately, and also jointly, the roles played by effective deformation and modified gravity in shaping the macroscopic properties of compact stars.

\subsection{Thesis outline}

Chapter 2 introduces the main microphysical ingredients and the equations of state employed throughout this thesis. Chapter 3 presents the hydrostatic equilibrium formalism in GR for both spherical and deformed configurations. Chapter 4 develops the modified-gravity framework, with emphasis on the \(f(R,T)\) formalism and related extensions considered in this work. Chapter 5 presents the solutions of the structure equations, discussing the resulting mass--radius relations, central-density sequences, and internal profiles for the different equations of state and gravitational scenarios analyzed. Finally, Chapter 6 is devoted to a discussion of the main results and to future perspectives. Complete algebraic derivations of the equilibrium equations are collected in the Appendices.

\chapter{Microphysics and Equations of State}

The stability of a star is governed by hydrostatic equilibrium \citep{carroll2017introduction}, which results from the balance between two competing effects. On the one hand, gravity tends to compress the stellar material inward, driving it toward collapse. On the other hand, an outward pressure gradient counteracts this gravitational pull. When the pressure gradient exactly balances gravity, the star can maintain a stable configuration.

In main-sequence stars, the required pressure support is ultimately sustained by the thermal energy generated by nuclear-fusion reactions in the core. In compact remnants such as white dwarfs and NSs, however, nuclear burning has ceased and thermal pressure alone is insufficient to prevent collapse. Instead, hydrostatic equilibrium is maintained primarily by degeneracy pressure \citep{shapiro2024black}. Dense (cold) matter becomes degenerate when the particle Fermi energies greatly exceed the thermal energy scale, a situation naturally realized in compact objects. In NSs, the baryon number density in the core can reach several times the nuclear saturation density, $n_0 \approx 0.16~\mathrm{fm^{-3}}$ (corresponding to a saturation mass density $\rho_0 \approx (2\text{--}3)\times 10^{14}~\mathrm{g~cm^{-3}}$) \citep{lattimer2000nuclear}. At such extreme densities, the interparticle spacing becomes comparable to the particles' de Broglie wavelengths and their wave functions overlap significantly. Because baryons are fermions, they obey the Pauli exclusion principle, which forbids identical fermions from occupying the same quantum state. This quantum constraint gives rise to an effective pressure support, the degeneracy pressure, that helps counteract gravitational collapse \citep{shapiro2024black,haensel2007neutron}.

In white dwarfs this pressure is provided primarily by degenerate electrons, whereas in NSs it is associated mainly with neutrons. However, at supranuclear densities, the pressure is not determined by degeneracy alone: strong nuclear interactions among baryons also contribute substantially to the total pressure \citep{sagert2006compact}.

Neutron-star matter spans a wide range of densities: it is subnuclear in the crust and increases toward the center, reaching several times the nuclear saturation density $n_0$ in the core, where additional degrees of freedom (e.g., hyperons or deconfined quark matter) may become relevant \citep{chamel2008physics}. These possibilities will be discussed later.

During core collapse and NS formation, matter undergoes rapid neutronization through electron capture on protons (and on heavy nuclei). In its simplest form, this weak-interaction process can be written as
\begin{equation}
    p + e^- \rightarrow n + \nu_e,
\end{equation}
where $p$ denotes the proton, $e^-$ the electron, $n$ the neutron, and $\nu_e$ the electron neutrino. This reaction reduces the electron fraction and drives the composition toward neutron-rich matter. In the early stages of collapse the produced neutrinos can stream out, but as the density rises the neutrino mean free path decreases and neutrinos become trapped in the nascent proto-neutron star, diffusing out on longer timescales \citep{yamada2024physical}.

As the neutron fraction increases, adding more neutrons forces them to occupy progressively higher-momentum states because neutrons are fermions. Consequently, the characteristic Fermi energies become much larger than the thermal energy scale, motivating the common approximation of treating an evolved NS as a cold object, in the sense that \[ k_B T \ll E_F, \] or equivalently \(T \ll T_F\), where \(T_F \equiv E_F/k_B\) is the Fermi temperature. This approximation is not appropriate immediately after birth, when the remnant is a hot proto-neutron star and finite-temperature effects are important \citep{kumar2025constraints}.

\section{Composition of compact objects}
\subsection{Beta equilibrium}

 The increasing neutron excess, together with the occupation of progressively higher Fermi-momentum states, raises the symmetry energy of the system. Under these conditions, weak interactions drive the system toward beta equilibrium by allowing neutrons and protons to interconvert. Schematically, one may write the beta process as
\begin{equation}
    n \leftrightarrow p + e^- + \bar{\nu}_e.
\end{equation}

In a mature (cold) NS, neutrinos are not trapped and can escape. Consequently, their chemical potentials can be taken as $\mu_\nu \approx 0$. In neutrino-free matter, beta equilibrium implies the familiar condition
\begin{equation}
    \mu_n = \mu_p + \mu_e,
\end{equation}
and, once muons are present, the additional equilibrium condition
\begin{equation}
    \mu_e = \mu_\mu.
\end{equation}

These relations are supplemented by two global constraints: (i) baryon-number conservation,
\begin{equation}
n_B=\sum_i b_i n_i,
\end{equation}
and (ii) charge neutrality,
\begin{equation}
\sum_i q_i n_i = 0,
\end{equation}
where $b_i$ and $q_i$ denote the baryon number and electric charge of each species $i$ \citep{lattimer2014neutron}.

\begin{figure}[H]
    \centering
    \includegraphics[scale=0.5]{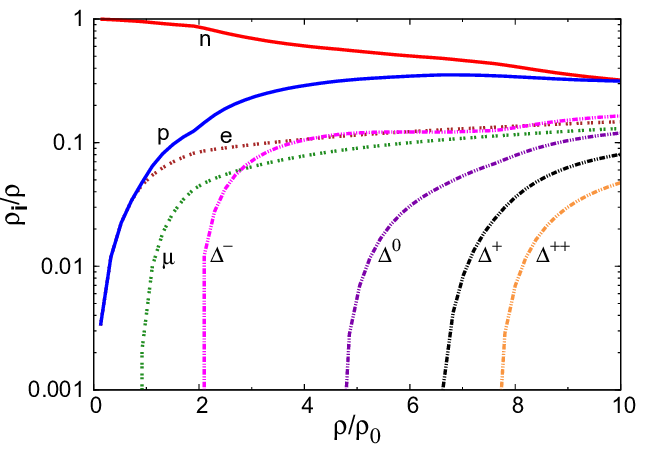}
    \caption{Relative particle abundances in neutron-star matter as a function of the baryon density normalized to the nuclear saturation density, \(\rho/\rho_0\). Here, \(\rho\) denotes the total baryon density, \(\rho_i\) the density of the particle species \(i\), and \(\rho_0\) the nuclear saturation density. Reproduced from \cite{weber2007pulsars}.}
    \label{fig:ns_composition}
\end{figure}

When the electron chemical potential grows to exceed the muon rest-mass energy ($m_\mu c^2 \approx 105.66~\mathrm{MeV}$), it becomes energetically favorable for part of the negative charge to be carried by muons in addition to electrons \citep{haensel2007neutron}. At still higher densities, new baryonic degrees of freedom may become competitive, including excited states such as $\Delta$ isobars, once the relevant chemical-potential combinations satisfy the corresponding threshold conditions \citep{motta2020delta}. In general, the onset of additional particle species tends to soften the EoS, since the system can redistribute its composition in a way that typically reduces the pressure increase at a given energy density. In many model realizations, the $\Delta^-$ is often favored at relatively low densities because its negative charge can help reduce the electron chemical potential \citep{motta2020delta}.

\subsection{Structure and composition}

A NS can be broadly divided into distinct regions. The outermost layers consist of a thin atmosphere and envelope, whose composition depends on the star's evolutionary history (e.g., fallback and accretion) and is often modeled as light elements (H/He) or, in some cases, heavier elements. Beneath this, the outer crust is well described as a Coulomb lattice of increasingly neutron-rich nuclei immersed in a highly degenerate electron gas. Deeper still, in the inner crust, nuclei coexist with relativistic degenerate electrons and a gas of ``dripped'' neutrons, which are expected to form a neutron superfluid \citep{chamel2008physics}.

At higher densities the star transitions to the core. In the outer core, the standard minimal composition includes neutrons, a smaller fraction of protons, and leptons (electrons and muons) in beta equilibrium. The inner core remains uncertain: depending on the true high-density EoS, it may contain additional degrees of freedom such as hyperons, $\Delta$ isobars, meson condensates, or deconfined quark matter, possibly in color-superconducting phases \citep{rajagopal2001condensed,lattimer2000nuclear,chamel2008physics}. Nucleon pairing is also expected in the core, with neutrons forming a superfluid and protons becoming superconducting over part of the density range.

While the microphysics of the crust is comparatively better constrained and has been modeled in substantial detail, the dominant uncertainty in NS modeling lies in the composition and interactions at supranuclear densities in the core. Consequently, the fundamental high-density EoS of NS matter remains unknown.

\begin{figure}[h]
    \centering
    \includegraphics[scale=0.5]{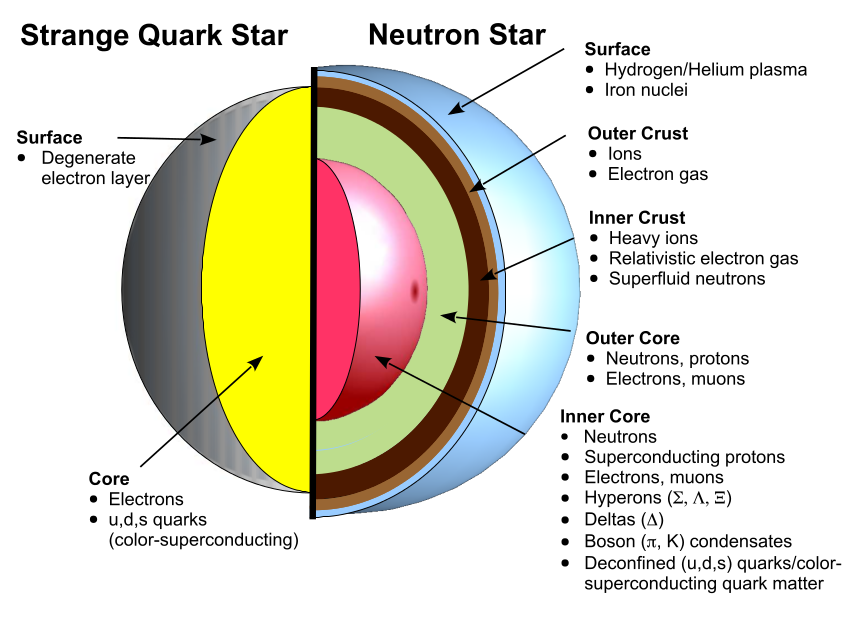}
    \caption{Schematic structures of a strange quark star (left) and a neutron star (right). Adapted from \citet{weber2014properties}; see also the Astrobites summary by \citet{lam2014peeling}.}
\label{fig:ns_qs_schematic}
    \label{fig:ns_qs_schematic}
\end{figure}

In the outer crust, the pressure is dominated by a strongly degenerate electron gas, and the equilibrium nuclear composition at each depth is obtained by minimizing the Gibbs free energy per baryon at fixed pressure under the constraint of local charge neutrality \citep{haensel2007neutron,baym1971ground}. As the density increases, the electron chemical potential rises and favors electron-capture reactions, driving nuclei toward progressively more neutron-rich isotopes \citep{chamel2008physics}. At the neutron-drip density, $\rho_{\rm ND}\approx 4\times 10^{11}~\mathrm{g~cm^{-3}}$, it becomes energetically favorable for some neutrons to become unbound, leading to the appearance of a neutron gas in the interstices of the lattice. This marks the transition from the outer to the inner crust \citep{baym1971ground}. Deeper in the inner crust, the competition between Coulomb and nuclear energies may give rise to non-spherical nuclear structures collectively known as ``nuclear pasta,'' particularly near the crust-core interface \citep{ravenhall1983structure}. Finally, the dripped neutrons are expected to form Cooper pairs and become superfluid at sufficiently low temperatures \citep{haensel2007neutron}.

Because NS interiors are composed of strongly interacting, highly degenerate fermionic
matter, their theoretical description requires quantum many-body methods. In practice,
the EoS depends sensitively on how the underlying interactions and many-body correlations
are modeled, ranging from microscopic approaches to effective field-theory and
phenomenological mean-field descriptions commonly used in NS applications
\citep{glendenning2012compact}.

\begin{figure}[h]
    \centering
    \includegraphics[width=0.85\textwidth]{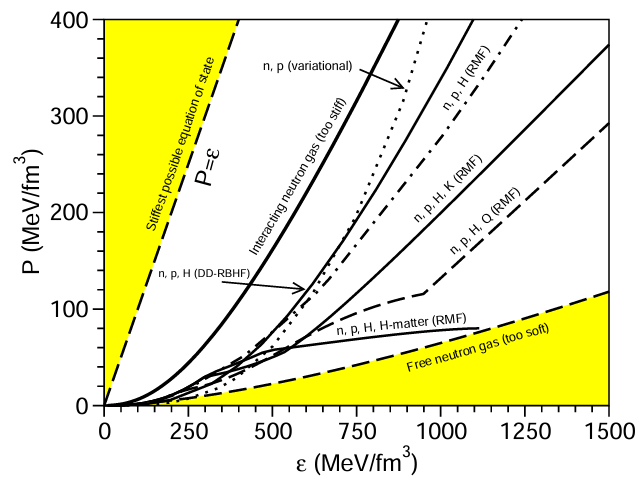}
    \caption{Pressure as a function of energy density for representative equations of state.
    The line $P=\varepsilon$ corresponds to the causal-limit slope (in units $c=1$).
    Reproduced from \citet{weber2009neutron}.}
    \label{fig:weber_eos_landscape}
\end{figure}

As a general overview, Fig.~\ref{fig:weber_eos_landscape} illustrates representative
equations of state in the $p(\varepsilon)$ plane and highlights how different microphysical
assumptions lead to markedly different stiffness at supranuclear densities. To understand
the origin of these differences, it is useful to briefly review how such equations of state
are constructed within common nuclear many-body frameworks.

One widely used starting point is the relativistic Hartree (RH), or mean-field approximation,
in which each particle moves independently in average fields generated by the other particles.
At the level of the energy functional, this corresponds to retaining only the direct (Hartree)
contributions while neglecting exchange (Fock) terms. In relativistic nuclear models, this is
the standard implementation of relativistic mean-field (RMF) theory \citep{serot1992relativistic}.

For fermions, however, the many-body state must be antisymmetric under the exchange of two identical particles, i.e.,
\begin{equation}
    \Psi(\dots, \mathbf{r}_i,\dots,\mathbf{r}_j,\dots) = -\Psi(\dots,\mathbf{r}_j,\dots,\mathbf{r}_i,\dots).
\end{equation}
A convenient way to enforce this property is to approximate the many-body wave function by a Slater determinant built from single-particle orbitals,
\begin{equation}
    \Psi(\mathbf{r}_1,\dots,\mathbf{r}_N) = \frac{1}{\sqrt{N!}}\det \left[\psi_i(\mathbf{r}_j) \right],
\end{equation}
which automatically implements the Pauli principle. The Hartree-Fock approximation then follows from a variational procedure within this class of antisymmetrized states, leading to single-particle equations that include exchange (Fock) contributions \citep{fetter2012quantum}.

In the relativistic context, the explicit inclusion of Fock terms defines relativistic Hartree-Fock (RHF) approaches, including density-dependent variants, which provide a more complete mean-field treatment and can affect both the stiffness of the EoS and the particle composition at supranuclear densities \citep{serot1992relativistic}.

The relativistic Brueckner-Hartree-Fock (RBHF) approach goes beyond RHF for dense nuclear matter by treating the effective interaction between nucleons in the nuclear medium. Rather than employing the bare nucleon-nucleon interaction directly, RBHF constructs an in-medium interaction by resumming repeated two-body scatterings while enforcing Pauli blocking of occupied intermediate states. This procedure defines the reaction matrix (the \(G\)-matrix). With a self-consistent single-particle spectrum, the \(G\)-matrix accounts for important short-range and tensor correlations that are absent in Hartree and Hartree-Fock descriptions, and it underlies Brueckner-type predictions for the NS EoS at supranuclear densities \citep{day1967elements}.

A widely used relativistic description of dense nuclear matter is provided by the RMF framework, rooted in the seminal work of Walecka \citep{walecka1974nuclear} and subsequent developments. In RMF theory, the effective interaction among baryons is modeled as being mediated by mesonic mean fields. At a minimal level that already captures the main qualitative features of the EoS, one introduces an isoscalar-scalar field $\sigma$ (responsible for an effective attraction), an isoscalar-vector field $\omega_\mu$ (providing a repulsion that becomes increasingly important at high density), and an isovector-vector field $\vec{\rho}_\mu$ (controlling the isospin dependence of the interaction and therefore closely connected to the symmetry energy, the proton fraction, and related thresholds such as the onset of fast neutrino-emission channels). Leptons are typically included as degenerate Fermi gases to ensure charge neutrality and beta equilibrium \citep{hui1982relativistic,serot1992relativistic}.

A convenient starting point is the RMF Lagrangian density,
\begin{equation}
\begin{split}
    \mathcal{L} = \sum_B \bar{\psi}_B \Big[i \gamma^\mu \partial_\mu - m_B + g_{\sigma B}\sigma 
    - g_{\omega B}\gamma^\mu \omega_\mu 
    - g_{\rho B}\gamma^\mu \vec{\tau}\cdot \vec{\rho}_\mu \Big]\psi_B 
    + \mathcal{L}_\sigma + \mathcal{L}_\omega + \mathcal{L}_\rho \\
    + \sum_l\bar{\psi}_l\left(i\gamma^\mu \partial_\mu - m_l\right)\psi_{l},
\end{split}
\end{equation}
in which baryons couple to the $\sigma$, $\omega_\mu$, and $\vec{\rho}_\mu$ fields through coupling constants $g_{\sigma B}$, $g_{\omega B}$, and $g_{\rho B}$. The mesonic sector contains the kinetic and mass terms for the mediator fields and, in many realistic parametrizations, additional nonlinear self-interactions (e.g., a potential $U(\sigma)$) and/or density-dependent couplings introduced to reproduce nuclear saturation properties and finite-nucleus systematics. Varying the action yields a Dirac equation for baryons moving in the background mean fields, with an effective mass $m^*_B = m_B - g_{\sigma B}\sigma$ and vector mean-field potentials associated with $\omega_\mu$ and $\vec{\rho}_\mu$. The meson fields satisfy Klein-Gordon-type (for $\sigma$) and Proca-type (for vector mesons) equations with baryonic source currents. In uniform infinite matter, relevant for constructing NS equations of state, the mean fields reduce to constant expectation values (typically only time-like components for the vectors), turning the field equations into coupled algebraic self-consistency conditions \citep{ring1996relativistic}.

Building an EoS then amounts to solving these mean-field self-consistency relations together with the conditions appropriate for NS matter. For a prescribed total baryon density, one determines the particle composition by imposing beta equilibrium and charge neutrality, which fix the Fermi momenta of all species. The mean fields and the composition are obtained iteratively until convergence. With the converged solution, the energy density $\varepsilon$ and pressure $p$ are computed from the energy-momentum tensor (equivalently, from thermodynamically consistent relations), including both kinetic contributions from degenerate baryons and leptons and field contributions from the mesonic sector. Repeating this procedure over a range of baryon densities yields $p(n)$ and $\varepsilon(n)$. Eliminating $n$ provides the final barotropic relation $p(\varepsilon)$, which is the required input for stellar-structure calculations \citep{haensel2007neutron}.

\subsection{Strange quark matter}

Bodmer \citep{bodmer1971collapsed}, Witten \citep{witten1984cosmic}, and Terezawa \citep{terezawa1989336} proposed that strange quark matter could be the true ground state of the strong interaction, rather than $^{56}\mathrm{Fe}$. In this picture, deconfined quark matter composed of three flavors ($u,d,s$) can have a lower energy per baryon than two-flavor ($u,d$) quark matter because the presence of the strange quark opens an additional Fermi sea and can reduce the overall Fermi energy at fixed baryon density. If the energy per baryon of three-flavor quark matter at zero pressure satisfies
\begin{equation}
    \frac{E}{A} < 930~\mathrm{MeV},
\end{equation}
then strange matter would be more stable than ordinary nuclear matter, implying that self-bound strange quark stars could exist as stable compact-object configurations \citep{bodmer1971collapsed, witten1984cosmic, jaffe1984strange}.

\begin{figure}[h]
    \centering
    \includegraphics[scale=0.75]{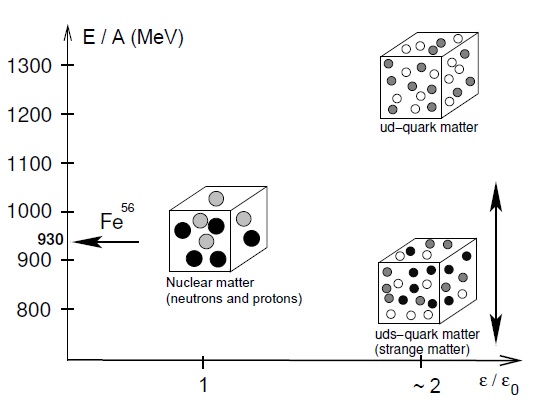}
    \caption{Energy per baryon $E/A$ for different forms of matter \citep[see][]{weber2005strange}.}
    \label{fig:e_over_a_weber2005}
\end{figure}

For strange quark matter, typical model estimates yield an energy per baryon of order $E/A \simeq 829~\mathrm{MeV}$, i.e., about $100~\mathrm{MeV}$ lower than the corresponding two-flavor ($u,d$) case (see Fig.~\ref{fig:e_over_a_weber2005}).

Many studies have suggested that, under sufficiently high pressures, nucleons in the cores of NSs may dissociate, leading to the appearance of deconfined quark matter \citep{ivanenko1965hypothesis,itoh1970hydrostatic,fritzsch197316th}. As deconfinement sets in, weak interactions tend to drive the composition toward three-flavor quark matter by converting some of the $u$ and $d$ quarks into $s$ quarks, consistent with beta equilibrium.

Heavier quark flavors ($c$, $b$, and $t$) are not expected to be populated in NS interiors because of their large current masses and the corresponding high threshold chemical potentials. For example, producing charm quarks would require densities of order $\gtrsim 10^{17}~\mathrm{g~cm^{-3}}$, i.e., far above typical NS central densities \citep{weberbrasil}.

One of the most widely used descriptions of deconfined quark matter is the MIT Bag Model \citep{jaffe1984strange}. In this model, quarks are treated as a (nearly) free Fermi gas inside a ``bag,'' while confinement is modeled phenomenologically by a uniform vacuum energy density $B$ (the bag constant). Mechanical equilibrium can be interpreted as the balance between the kinetic pressure of the quark gas and the external bag pressure associated with $B$. In the massless and non-interacting limit, the EoS takes the simple form
\begin{equation}
p = \frac{1}{3}\left(\varepsilon - 4B\right), \label{eq:mit_bag_massless}
\end{equation}
where $p$ is the total (isotropic) pressure of the quark fluid, $\varepsilon$ is the corresponding total energy density, and $B$ is the vacuum energy density. Despite its simplicity, the MIT Bag Model provides an effective phenomenological description of quark matter and is widely used in studies of quark stars and hybrid stars \citep{jaffe1984strange,orsaria2014quark}.

In the MIT Bag Model, the bag constant acts as a uniform vacuum contribution: it subtracts from the pressure and adds to the energy density. From Eq.~\eqref{eq:mit_bag_massless}, a larger $B$ lowers $p(\varepsilon)$, corresponding to a softer EoS; equilibrium configurations then tend to have smaller radii and support lower maximum masses. A smaller $B$ has the opposite effect, yielding a stiffer EoS.

\section{Neutron-star cooling}

The thermal evolution of NSs is largely controlled by neutrino emission from the stellar interior during the first $\sim 10^{3}$--$10^{5}$ years, after which photon emission from the surface becomes the dominant cooling channel. In the neutrino-cooling stage, the main energy-loss mechanisms are weak-interaction processes operating in dense, degenerate matter, such as modified Urca reactions and nucleon-nucleon bremsstrahlung. If the core composition allows it, most notably if the proton fraction exceeds the kinematic threshold required by momentum conservation, the direct Urca process can operate, leading to much faster cooling. Neutron superfluidity and proton superconductivity can strongly modify these emissivities, suppressing standard neutrino processes at low temperatures and introducing an additional channel associated with Cooper-pair breaking and formation (PBF) \citep{yakovlev2004neutron}.

\subsection{Direct Urca}

The nucleonic direct Urca process is the simplest beta cycle involving neutrino emission,
\begin{equation}
    n\rightarrow p + e^- + \bar{\nu}_e, \qquad p + e^- \rightarrow n + \nu_e.
\end{equation}
It is an efficient (fast) cooling mechanism because it involves fewer degenerate fermions and therefore has a comparatively large available phase space. However, it can only proceed if momentum conservation can be satisfied with all participating particles near their respective Fermi surfaces, which leads to a kinematic condition of the form
\begin{equation}
    k_{F,n} \le k_{F,p} + k_{F,e},
\end{equation}
and translates into a minimum proton fraction in charge-neutral beta-equilibrated matter \citep{lattimer1991direct}.

\subsection{Modified Urca}

If the direct Urca triangle inequality is not satisfied, neutrino emission can still proceed through the modified Urca process, in which an additional ``spectator'' nucleon is required to share momentum. Representative neutron-branch reactions are
\begin{equation}
    n + n \rightarrow n + p + e^- + \bar{\nu}_e, \qquad
   n + p + e^- \rightarrow n + n + \nu_e,
\end{equation}
with analogous proton-branch reactions. Because the process involves more degenerate particles, the available phase space is strongly reduced and the emissivity is much lower than in direct Urca, making modified Urca the standard cooling mechanism in many ``minimal cooling'' scenarios \citep{friman1979neutrino}.

\subsection{Nucleon--nucleon bremsstrahlung}

Another important neutrino source is nucleon-nucleon bremsstrahlung, in which a neutrino-antineutrino pair is emitted during a strong-interaction collision,
\begin{equation}
    N + N \rightarrow N + N + \nu + \bar{\nu}.
\end{equation}
In degenerate matter, these reactions provide an additional neutrino luminosity channel (often comparable to or below modified Urca, depending on density and composition), with neutrinos escaping freely in mature, cold neutron stars and efficiently carrying energy away from the core \citep{yakovlev2004neutron}.

\subsection{Effects of superfluidity and superconductivity}

In a highly degenerate Fermi system, the presence of an attractive component of the interaction in any pairing channel renders the Fermi surface unstable to the formation of Cooper pairs. As a result, fermions with opposite momenta near the Fermi surface preferentially form correlated pairs with (approximately) zero total momentum, and below a critical temperature the system undergoes a BCS-like transition to a superfluid/superconducting state characterized by an energy gap $\Delta$ in the quasiparticle spectrum. In NS matter, this attraction originates from the nuclear force and leads to density-dependent pairing patterns, such as ${}^{1}S_{0}$ neutron pairing in the inner crust, ${}^{1}S_{0}$ proton pairing in parts of the core, and ${}^{3}P_{2}$ neutron pairing at higher core densities \citep{haskell2019superfluidity}.

These pairing gaps have a direct impact on neutrino cooling. For temperatures below the critical temperature $T_c$, processes that require single-particle excitations near the Fermi surface, such as modified Urca and nucleon-nucleon bremsstrahlung, are exponentially suppressed because creating quasiparticles costs an energy of order $\Delta$. Near the onset of pairing, $T \approx T_c$, an additional neutrino-emission channel becomes efficient through \textit{pair breaking and formation} (PBF): thermal formation and breakup of Cooper pairs can radiate $\nu\bar{\nu}$ pairs, producing a transient enhancement of the emissivity before the low-$T$ suppression dominates at $T \ll T_c$ \citep{haskell2019superfluidity}.

\section{Equations of state adopted in this thesis}

In the stellar-structure calculations developed in the following chapters, the microphysical input is the barotropic relation \(p(\varepsilon)\), which closes the system of hydrostatic equilibrium equations. In this thesis we employ three representative equations of state that span different levels of microphysical modeling: a hadronic EoS based on the GM1 parametrization, a quark-matter EoS described by the MIT Bag Model, and a polytropic EoS as a controlled baseline.

The GM1 EoS is a relativistic mean-field hadronic model that provides a standard description of cold, charge-neutral, beta-equilibrated matter composed of nucleons and leptons. Its use is motivated by its simplicity, its widespread adoption in compact-star studies, and its suitability as a reference hadronic model for isolating the effects of deformation and modified gravity on the stellar structure. Although equations of state constrained by multi-messenger observations, including gravitational-wave data, are essential for realistic astrophysical modeling, the main purpose of the present work is not to perform an EoS inference analysis. Rather, we aim to investigate how different gravitational and geometrical assumptions modify the equilibrium configurations for representative and commonly used microphysical inputs.

The polytropic EoS is used as a flexible parametrization to explore generic trends and to perform consistency checks. Although it does not encode the detailed composition of dense matter, it captures the global stiffness of the stellar fluid and provides a transparent way to analyze how the structure equations respond to controlled changes in the pressure-density relation. In this sense, the polytropic model serves as a useful baseline for testing the numerical implementation and for comparing the qualitative behavior of the standard, deformed, and modified-gravity stellar models. For quark-matter and self-bound strange-quark-star configurations, we adopt the MIT Bag Model discussed in Sec.~2.1.3.

\chapter{Hydrostatic equilibrium in General Relativity}

Neutron stars inhabit the strong-field regime of gravity, where Newtonian theory of gravity is no longer adequate. In such environments, gravity must be described relativistically as the manifestation of spacetime curvature sourced by matter and energy, as encoded in Einstein's theory \citep{einstein1915feldgleichungen}.
The purpose of this chapter is twofold: first, we briefly outline how Einstein's field equations follow from an action principle; second, we derive the relativistic condition for hydrostatic balance in static, spherically (and non spherically) symmetric configurations, culminating in the Tolman--Oppenheimer--Volkoff equation that governs compact-star structure in GR.

\section{Einstein field equations from an action principle}
\label{subsec:einstein_from_action}

General Relativity can be formulated as a dynamical theory of the metric field $g_{\mu\nu}$ derived from an action principle. The basic idea is that the physical spacetime geometry is obtained by demanding that the total action is stationary under arbitrary metric variations that preserve the chosen boundary conditions (Hamilton's principle) \citep{carroll2017introduction}.

\subsection{Einstein--Hilbert action and the definition of $T_{\mu\nu}$}

We consider the Einstein--Hilbert action supplemented by a matter action,
\begin{equation}
S \;=\; S_{\rm EH} + S_m
\;=\;
\frac{1}{16\pi}\int d^4x\,\sqrt{-g}\,(R-2\Lambda)
\;+\;
\int d^4x\,\sqrt{-g}\,\mathcal{L}_m,
\label{eq:EH_action_thesis}
\end{equation}
where $R$ is the Ricci scalar, $\Lambda$ is the cosmological constant, and $\mathcal{L}_m$ is the matter
Lagrangian density. Throughout this thesis we employ geometrized units, $G=c=1$, unless stated otherwise.

The energy--momentum tensor is defined as the metric functional derivative of the matter action,
\begin{equation}
T_{\mu\nu}\;\equiv\; -\frac{2}{\sqrt{-g}}\,
\frac{\delta(\sqrt{-g}\,\mathcal{L}_m)}{\delta g^{\mu\nu}}.
\label{eq:Tmunu_def_thesis}
\end{equation}
This definition ensures that $T_{\mu\nu}$ correctly encodes the local energy density, momentum density, and
stresses measured by observers, and it provides the source term appearing in the gravitational field equations
\citep{poisson2004relativist}.

Two standard identities are central to the derivation. First, the variation of the metric determinant is
\begin{equation}
\delta\sqrt{-g} \;=\; -\frac{1}{2}\sqrt{-g}\,g_{\mu\nu}\,\delta g^{\mu\nu}.
\label{eq:delta_sqrtg}
\end{equation}
Second, the variation of the Ricci scalar can be expressed using the Palatini identity,
\begin{equation}
\delta R_{\mu\nu} \;=\; \nabla_\alpha\delta\Gamma^\alpha_{\mu\nu}-\nabla_\nu\delta\Gamma^\alpha_{\mu\alpha},
\label{eq:palatini}
\end{equation}
which implies
\begin{equation}
\delta R \;=\; R_{\mu\nu}\,\delta g^{\mu\nu} \;+\; \nabla_\alpha V^\alpha,
\label{eq:delta_R}
\end{equation}
where $V^\alpha$ is a vector built from $\delta\Gamma^\alpha_{\mu\nu}$ and $\delta g^{\mu\nu}$
\citep{carroll2017introduction}. The second term in Eq.~\eqref{eq:delta_R} is a total divergence
and therefore contributes only a boundary term to the action.

Varying Eq.~\eqref{eq:EH_action_thesis} with respect to $g^{\mu\nu}$, using
Eqs.~\eqref{eq:delta_sqrtg}--\eqref{eq:delta_R}, and discarding the boundary term, yields
\begin{equation}
\delta S_{\rm EH}
=
\frac{1}{16\pi}\int d^4x\,\sqrt{-g}\,
\Bigl(R_{\mu\nu}-\frac{1}{2}R\,g_{\mu\nu}+\Lambda g_{\mu\nu}\Bigr)\,\delta g^{\mu\nu}.
\label{eq:delta_SEH}
\end{equation}
The matter part varies as
\begin{equation}
\delta S_m
= -
\frac{1}{2}\int d^4x\,\sqrt{-g}\,T_{\mu\nu}\,\delta g^{\mu\nu},
\label{eq:delta_Sm}
\end{equation}
directly from the definition \eqref{eq:Tmunu_def_thesis}.
Imposing stationarity of the total action, $\delta S=0$, for arbitrary $\delta g^{\mu\nu}$ then gives
\begin{equation}
R_{\mu\nu}-\frac{1}{2}R\,g_{\mu\nu}+\Lambda g_{\mu\nu}
=
8\pi\,T_{\mu\nu}.
\label{eq:Einstein_eq_thesis}
\end{equation}
Equation~\eqref{eq:Einstein_eq_thesis} is Einstein's field equation: it states that the spacetime curvature
(through the Einstein tensor $G_{\mu\nu}\equiv R_{\mu\nu}-\tfrac12 R g_{\mu\nu}$) is sourced by the local
matter-energy content described by $T_{\mu\nu}$ (a detailed deduction can be found in Appendix~A).

\subsection{Bianchi identity, local conservation, and the bridge to the TOV equation}

A central geometric identity is the contracted Bianchi identity,
\begin{equation}
\nabla^\mu G_{\mu\nu}=0.
\label{eq:bianchi_thesis}
\end{equation}
Taking the covariant divergence of Eq.~\eqref{eq:Einstein_eq_thesis} and using Eq.~\eqref{eq:bianchi_thesis},
one obtains the standard local conservation law
\begin{equation}
\nabla^\mu T_{\mu\nu}=0,
\label{eq:conservation_thesis}
\end{equation}
(valid when $\Lambda$ is constant). This relation is not an additional assumption: it follows from the differential structure of the field equations and ensures the internal consistency of GR
\citep{carroll2017introduction}.

For static stellar configurations, Eq.~\eqref{eq:conservation_thesis} provides the relativistic Euler equation (hydrostatic balance). In particular, the $\nu=r$ component of $\nabla^\mu T_{\mu\nu}=0$ relates the radial pressure gradient to the gravitational potential in a static metric. Combined with the metric potentials determined by Eq.~\eqref{eq:Einstein_eq_thesis} for a spherically symmetric spacetime, this leads directly to the Tolman--Oppenheimer--Volkoff (TOV) equation derived in the next section. 

For the compact-star applications considered in this thesis, we set the cosmological constant to zero, $\Lambda=0$. While a nonzero $\Lambda$ can be consistently included in the field equations, its associated curvature scale is set by the Hubble radius and its contribution to the structure of NSs is negligible compared to the matter-induced curvature in the stellar interior \citep{zubairi2014solutions}. Therefore, throughout the stellar-structure derivations (including the TOV equation) we work with the $\Lambda=0$ limit, keeping $\Lambda$ only in the general action-level presentation for completeness.

\subsection{From local conservation to the relativistic Euler equation}

A crucial step in the derivation of the Tolman-Oppenheimer-Volkoff equation is the local conservation of the energy-momentum tensor, Eq.~\eqref{eq:conservation_thesis}.  Thus, for any tensor quantity, the covariant derivative incorporates the effects of spacetime curvature through the Christoffel symbols.

Let us consider a static and spherically symmetric spacetime with line element
\begin{equation}
ds^2=e^{2\Phi(r)}dt^2-e^{2\Lambda(r)}dr^2+r^2\left(d\theta^2-\sin^2\theta\,d\varphi^2\right),
\end{equation}
where \(\Phi(r)\) and \(\Lambda(r)\) are metric functions depending only on the radial coordinate. Here, \(\Lambda(r)\) denotes the radial metric potential and should not be confused with the cosmological constant. We further assume that the stellar matter is described by a perfect fluid. In mixed form, the energy-momentum tensor may be written as
\begin{equation}
T^\mu_{\phantom{\mu}\nu}=(\varepsilon+p)u^\mu u_\nu-p\,\delta^\mu_{\phantom{\mu}\nu},
\end{equation}
where \(\varepsilon\) is the energy density, \(p\) is the pressure, and \(u^\mu\) is the four-velocity of the fluid, satisfying
\begin{equation}
u^\mu u_\mu=1.
\end{equation}

Since the star is static, the fluid is at rest in the chosen coordinates. Therefore,
\begin{equation}
u^\mu=\left(e^{-\Phi},0,0,0\right),
\qquad
u_\mu=\left(e^\Phi,0,0,0\right).
\end{equation}
It then follows that the nonvanishing mixed components of the energy-momentum tensor are
\begin{equation}
T^t_{\phantom{t}t}=\varepsilon,
\qquad
T^r_{\phantom{r}r}=T^\theta_{\phantom{\theta}\theta}=T^\varphi_{\phantom{\varphi}\varphi}=-p.
\end{equation}

To obtain the equation of hydrostatic equilibrium, we evaluate the radial component of the conservation law,
\begin{equation}
\nabla_\mu T^\mu_{\phantom{\mu}r}=0.
\end{equation}
Using the definition of the covariant derivative of a mixed tensor,
\begin{equation}
\nabla_\mu T^\mu_{\phantom{\mu}\nu}
=
\partial_\mu T^\mu_{\phantom{\mu}\nu}
+
\Gamma^\mu_{\mu\lambda}T^\lambda_{\phantom{\lambda}\nu}
-
\Gamma^\lambda_{\mu\nu}T^\mu_{\phantom{\mu}\lambda},
\end{equation}
and taking \(\nu=r\), we obtain
\begin{equation}
\partial_\mu T^\mu_{\phantom{\mu}r}
+
\Gamma^\mu_{\mu\lambda}T^\lambda_{\phantom{\lambda}r}
-
\Gamma^\lambda_{\mu r}T^\mu_{\phantom{\mu}\lambda}
=0.
\end{equation}

Because the tensor is diagonal and the configuration is static, the only nonvanishing contribution from the first term is
\begin{equation}
\partial_\mu T^\mu_{\phantom{\mu}r}
=
\partial_r T^r_{\phantom{r}r}
=
-\frac{dp}{dr}.
\end{equation}
For the second term, only \(\lambda=r\) contributes, so that
\begin{equation}
\Gamma^\mu_{\mu\lambda}T^\lambda_{\phantom{\lambda}r}
=
\Gamma^\mu_{\mu r}\,T^r_{\phantom{r}r}
=
-\Gamma^\mu_{\mu r}\,p.
\end{equation}
For the last term, since \(T^\mu_{\phantom{\mu}\lambda}\) is diagonal, one finds
\begin{equation}
-\Gamma^\lambda_{\mu r}T^\mu_{\phantom{\mu}\lambda}
=
-\Gamma^t_{tr}T^t_{\phantom{t}t}
-\Gamma^r_{rr}T^r_{\phantom{r}r}
-\Gamma^\theta_{\theta r}T^\theta_{\phantom{\theta}\theta}
-\Gamma^\varphi_{\varphi r}T^\varphi_{\phantom{\varphi}\varphi}.
\end{equation}
Substituting the diagonal components of \(T^\mu_{\phantom{\mu}\nu}\), this becomes
\begin{equation}
-\Gamma^\lambda_{\mu r}T^\mu_{\phantom{\mu}\lambda}
=
-\Gamma^t_{tr}\varepsilon
+\Gamma^r_{rr}p
+\Gamma^\theta_{\theta r}p
+\Gamma^\varphi_{\varphi r}p.
\end{equation}

Combining all terms, the contributions proportional to \(p\) cancel except for the one involving \(\Gamma^t_{tr}\), yielding
\begin{equation}
\frac{dp}{dr}+\Gamma^t_{tr}(\varepsilon+p)=0.
\end{equation}
For the metric above, the relevant Christoffel symbol is
\begin{equation}
\Gamma^t_{tr}=\Phi'(r).
\end{equation}
Therefore, the conservation law reduces to
\begin{equation}
\frac{dp}{dr}=-(\varepsilon+p)\Phi'(r).
\label{Euler_static}
\end{equation}

Equation \eqref{Euler_static} is the relativistic Euler equation for a static perfect fluid in a spherically symmetric spacetime. It provides the direct link between local energy-momentum conservation and the Tolman-Oppenheimer-Volkoff equation. Once \(\Phi'(r)\) is obtained from the gravitational field equations, Eq.~\eqref{Euler_static} immediately leads to the equation governing hydrostatic equilibrium inside the star.

\section{The Tolman-Oppenheimer-Volkoff equation}
At each point inside a compact star, the inward pull of gravity is balanced by an outward pressure gradient \citep{weberbrasil}. In compact objects, this pressure support is provided primarily by degenerate matter and is encoded in the EoS. This balance defines the condition of hydrostatic equilibrium. For physically acceptable stellar configurations, the pressure profile decreases monotonically with the radial coordinate measured from the center, i.e., $dp(r)/dr<0$ for $0<r<R$. The stellar radius $R$ is then defined as the location where the pressure vanishes,
\begin{equation}
p(R)=0.
\end{equation}

To derive the relativistic hydrostatic equilibrium equation in GR, we start from the static, spherically symmetric line element introduced in Eq.~(3.11),
\begin{equation}
	ds^2 = e^{2\Phi(r)}dt^2 - e^{2\Lambda(r)}dr^2 - r^2 \left(d\theta^2 + \sin^2 \theta\, d\varphi^2\right),
    \label{eq:spherical_metric}
\end{equation}
which describes a spherical mass distribution in curvature coordinates. As discussed previously, \(\Phi(r)\) and \(\Lambda(r)\) are metric potentials depending only on the radial coordinate and encoding the gravitational field generated by the stellar matter. Once the matter content is specified and an EoS is chosen, these functions are determined by Einstein's field equations.

Since the metric \eqref{eq:spherical_metric} is diagonal, its nonvanishing components and their inverses follow immediately. For readability, we present a guided derivation in the main text, while the complete algebraic expressions (including the explicit metric components, Christoffel symbols, and curvature tensors associated with \eqref{eq:spherical_metric}) are collected in Appendix~B.

From the metric \eqref{eq:spherical_metric}, the Levi-Civita connection is defined by
\begin{equation}
\Gamma^\alpha_{\mu\nu}= \frac{1}{2}g^{\alpha \lambda}\left(\partial_\mu g_{\nu \lambda}
+ \partial_\nu g_{\mu \lambda}-\partial_\lambda g_{\mu \nu} \right).
\end{equation}

From the Levi-Civita connection one constructs the curvature tensors and the Einstein tensor \(G_{\mu\nu}\). In what follows, it is convenient to work with the mixed components \(G^\mu_{\ \nu}\), since the energy-momentum tensor of a perfect fluid is diagonal in mixed form. Therefore, we only record the independent mixed components of \(G^\mu_{\ \nu}\) that are required to derive the stellar-structure equations.

The $tt$ component can be written as
\begin{equation}
	G^{t}{}_{t}(r)
	= \frac{1}{r^{2}} - e^{-2\Lambda(r)}\left(\frac{1}{r^{2}} - 2\,\frac{\Lambda'(r)}{r}\right),
	\label{eq:Gtt_mixed}
\end{equation}
while the $rr$ component reads
\begin{equation}
	G^{r}{}_{r}(r)
	= \frac{1}{r^{2}} - e^{-2\Lambda(r)}\left(2\,\frac{\Phi'(r)}{r} + \frac{1}{r^{2}}\right).
	\label{eq:Grr_mixed}
\end{equation}
Here a prime denotes differentiation with respect to the radius $r$.

Assuming an isotropic perfect-fluid energy-momentum tensor in mixed form,
\begin{equation}
T^{\mu}{}_{\nu}(r)=\mathrm{diag}\!\left(\varepsilon(r),\,-p(r),\,-p(r),\,-p(r)\right),
\label{eq:T_mixed_perfectfluid}
\end{equation}
Einstein's field equations $G^{\mu}{}_{\nu}=8\pi T^{\mu}{}_{\nu}$ (in units $G=c=1$) yield, from the $tt$ and $rr$ components,
\begin{equation}
\frac{1}{r^{2}} - e^{-2\Lambda(r)}\left(\frac{1}{r^{2}}-2\,\frac{\Lambda'(r)}{r}\right)
= 8\pi \varepsilon(r),
\label{eq:einstein_tt}
\end{equation}
\begin{equation}
\frac{1}{r^{2}} - e^{-2\Lambda(r)}\left(2\,\frac{\Phi'(r)}{r}+\frac{1}{r^{2}}\right)
= -8\pi p(r).
\label{eq:einstein_rr}
\end{equation}

To obtain the hydrostatic equilibrium equation, we combine the $tt$ and $rr$ components of Einstein's equations, Eqs.~\eqref{eq:einstein_tt} and \eqref{eq:einstein_rr}. It is convenient to introduce the standard mass function $m(r)$ by writing the radial metric coefficient as 
\begin{equation} 
e^{-2\Lambda(r)} \equiv 1 - \frac{2m(r)}{r}.
\label{eq:def_mass_function}
\end{equation} 
This definition is not an \emph{ansatz}: it follows from integrating the $tt$ field equation and it encodes the gravitational mass enclosed within the radius $r$, so that the combination $1-2m(r)/r$ is directly tied to the geometry through
\[
g_{rr}=-e^{2\Lambda(r)}
=
-\left(1-\frac{2m(r)}{r}\right)^{-1}.
\]
Eliminating $\Lambda'(r)$ between Eqs.~\eqref{eq:einstein_tt} and \eqref{eq:einstein_rr}, one obtains, after straightforward algebra, the gradient of the gravitational potential, 
\begin{equation}
\Phi'(r) = \frac{m(r) + 4\pi r^3 p(r)}{r^2\left(1 - \frac{2m(r)}{r}\right)}.
\label{eq:Phi_prime} 
\end{equation}

The second ingredient follows from local energy-momentum conservation, $\nabla_\mu T^{\mu}{}_{\nu}=0$. For a static perfect fluid, the $\nu=r$ component yields the relativistic Euler (hydrostatic balance) equation,
\begin{equation} 
\frac{dp(r)}{dr} = -\left[\varepsilon(r) + p(r)\right]\Phi'(r).
\label{eq:euler_static} 
\end{equation} 

Substituting Eq.~\eqref{eq:Phi_prime} into Eq.~\eqref{eq:euler_static} leads to the Tolman-Oppenheimer-Volkoff (TOV) equation \citep{tolman1939static, oppenheimer1939massive},
\begin{equation} 
\frac{dp(r)}{dr} = -\left[\varepsilon(r) + p(r)\right]\, \frac{m(r) + 4\pi r^3 p(r)}{r^2\left(1 - \frac{2m(r)}{r}\right)}.
\label{eq:TOV} 
\end{equation}

Equation~\eqref{eq:TOV}, together with an EoS $p=p(\varepsilon)$ and the mass-continuity relation 
\begin{equation} 
\frac{dm(r)}{dr}=4\pi r^2\varepsilon(r), \label{eq:mass_continuity} 
\end{equation} 
determines the internal structure and global properties of static, spherically symmetric compact stars in GR. Throughout this work we adopt geometrized units, $G=c=1$, so that mass, length, and time share the same dimension. 

At the center of the star one specifies the central pressure $p(0)\equiv p_c$ (and the corresponding central energy density $\varepsilon(0)\equiv \varepsilon_c$) through the equation of state $p=p(\varepsilon)$. Equation~\eqref{eq:TOV} provides the fundamental condition for hydrostatic equilibrium of static stellar configurations in GR. In the Newtonian (weak-field, low-pressure) limit, one assumes $p(r)\ll \varepsilon(r)$, $4\pi r^{3}p(r)\ll m(r)$, and $2m(r)/r \ll 1$. Moreover, the total energy density reduces to the rest-mass density, $\varepsilon(r)\simeq \rho(r)c^{2}$ (and $\varepsilon\simeq \rho$ in units $c=1$). Under these assumptions, Eq.~\eqref{eq:TOV} reduces to the familiar Newtonian hydrostatic equilibrium equation, \begin{equation} 
\frac{dp(r)}{dr} = -\rho(r)\,\frac{m(r)}{r^{2}}, \label{eq:newtonian_hse} 
\end{equation} 
(or equivalently $dp/dr=-Gm(r)\rho(r)/r^{2}$ in SI units). 

The relativistic corrections in Eq.~\eqref{eq:TOV} increase the magnitude of the pressure gradient compared to the Newtonian treatment and are ultimately responsible for the existence of a maximum mass in NS sequences: as the central density increases, equilibrium solutions reach a turning point beyond which configurations become unstable and collapse gravitationally. Finally, since all factors on the right-hand side of Eq.~\eqref{eq:TOV} are positive for physically reasonable matter ($\varepsilon(r)>0$, $p(r)>0$) and regular stellar interiors ($r>2m(r)$), the pressure necessarily decreases outward, $dp(r)/dr<0$. If the EoS is microscopically stable, $dp/d\varepsilon \ge 0$, this also implies that $\varepsilon(r)$ decreases monotonically with radius.

The standard TOV framework describes a \emph{static} configuration: all thermodynamic and geometric quantities depend only on $r$ and are time independent. In practice, this corresponds to modeling the star as non-rotating and in hydrostatic equilibrium, neglecting oscillations and secular evolution. This provides an excellent first approximation for many isolated NSs and for slowly rotating systems. However, it becomes insufficient when one aims to include effects such as rapid rotation (which breaks spherical symmetry and renders the problem intrinsically axisymmetric) \citep{wen2011properties}, ultra-strong magnetic fields \citep{rizaldy2018magnetized}, or other sources of deformation and stress that require a suitable generalization of the equilibrium formalism.

\subsection{Rotation as a motivation for deformed compact configurations}

The structure equations of rotating compact stars are considerably more involved than those describing static configurations. This additional complexity arises because rotation deforms the stellar shape, producing an oblate configuration characterized by flattening at the poles and an enlarged equatorial region. As a consequence, the spacetime is no longer described by metric functions depending only on the radial coordinate, but must also depend on the polar angle \(\theta\). In addition, rotation provides centrifugal support against gravitational collapse, so that a rotating star can sustain a larger mass than its nonrotating counterpart. Since the stellar mass distribution and the equilibrium geometry are modified by rotation, the corresponding metric functions depend explicitly on the rotational frequency \citep{weberbrasil}.

Another important relativistic effect is the dragging of local inertial frames. In rotating stars, this phenomenon introduces a non-diagonal metric component \(g_{t\varphi}\), which reflects the coupling between stellar rotation and spacetime geometry. In this case, the determination of the stellar structure becomes a self-consistent problem, since the extent to which local inertial frames are dragged depends on quantities that are themselves part of the unknown solution, such as the stellar mass and rotational frequency \citep{weberbrasil}.

A convenient form of the metric for a uniformly rotating compact star is given by
\begin{equation}
g_{tt} = e^{2\nu} - e^{2\psi}\omega^2,
\qquad
g_{t\varphi} = e^{2\psi}\omega,
\qquad
g_{rr} = -e^{2\lambda},
\qquad
g_{\theta\theta} = -e^{2\mu},
\qquad
g_{\varphi\varphi} = -e^{2\psi},
\end{equation}
which leads to the line element
\begin{equation}
ds^2 = e^{2\nu}dt^2 - e^{2\psi}(d\varphi - \omega dt)^2 - e^{2\mu}d\theta^2 - e^{2\lambda}dr^2.
\label{rot_metric}
\end{equation}
The metric functions \(\nu\), \(\psi\), \(\mu\), and \(\lambda\), as well as the angular velocity \(\omega\) of local inertial frames, depend on the coordinates \(r\) and \(\theta\), and implicitly on the stellar angular velocity \(\Omega\).

It is useful to define the relative angular velocity
\begin{equation}
\bar{\omega}(r,\theta,\Omega) \equiv \Omega - \omega(r,\theta,\Omega),
\end{equation}
which corresponds to the angular velocity of the stellar fluid measured with respect to a locally inertial observer. This quantity is particularly relevant because the centrifugal support experienced by a fluid element is determined by its rotation relative to the local inertial frame rather than by \(\Omega\) alone \citep{weberbrasil}.

In GR, there is no simple universal stability criterion for rapidly rotating stellar configurations. Nevertheless, an absolute upper bound on stable rotation is provided by the Kepler frequency, \(\Omega_K\), which corresponds to the maximum angular velocity that a star can sustain before mass shedding begins at the equator \citep{weberbrasil}.

In Newtonian gravity, this limit is obtained by equating the centrifugal and gravitational forces at the equator:
\begin{equation*}
\begin{split}
F_c &= F_G,\\
m\Omega^2R &= \frac{GMm}{R^2},
\end{split}
\end{equation*}
which yields
\begin{equation}
\Omega_K = \sqrt{\frac{GM}{R^3}}.
\end{equation}
In units \(G=1\), this expression becomes
\begin{equation}
\Omega_K = \sqrt{\frac{M}{R^3}}.
\end{equation}

The relativistic generalization is more involved. To derive it, one applies the extremal principle to the circular orbit of a pointlike particle moving along the equator of the star. For such a particle, one has \(dr=d\theta=0\), and the line element \eqref{rot_metric} reduces to
\begin{equation}
ds^2 = \left[e^{2\nu} - e^{2\psi}(\Omega-\omega)^2\right]dt^2.
\end{equation}
For timelike motion and signature \((+ - - -)\), the corresponding proper-time functional can be written as
\begin{equation}
J = \int dt\, \sqrt{e^{2\nu} - e^{2\psi}(\Omega-\omega)^2}.
\label{functionalJ}
\end{equation}

Now let
\begin{equation}
V = e^{\psi-\nu}(\Omega-\omega),
\label{defV}
\end{equation}
so that
\begin{equation}
\Omega-\omega = V e^{\nu-\psi}.
\end{equation}
Using this definition, the functional becomes
\begin{equation}
J = \int dt\, e^\nu \sqrt{1-V^2}.
\end{equation}

We now write the integrand of \eqref{functionalJ} as
\begin{equation}
L = \left[e^{2\nu} - e^{2\psi}(\Omega-\omega)^2\right]^{1/2}.
\end{equation}
Assuming \(\Omega=\mathrm{const.}\), the extremum condition implies \(\partial L/\partial r = 0\), which gives
\begin{equation*}
\frac{\partial L}{\partial r}
=
\frac{1}{2L}
\left[
2e^{2\nu}\nu'
-2e^{2\psi}(\Omega-\omega)^2\psi'
+2e^{2\psi}(\Omega-\omega)\omega'
\right]
=0.
\end{equation*}
Therefore,
\begin{equation}
e^{2\nu}\nu'
-
e^{2\psi}(\Omega-\omega)^2\psi'
+
e^{2\psi}(\Omega-\omega)\omega'
=0.
\end{equation}
Substituting \(\Omega-\omega = Ve^{\nu-\psi}\), one obtains
\begin{equation}
\psi' e^{2\nu}V^2 - \omega' e^{\nu+\psi}V - \nu' e^{2\nu} = 0.
\label{quadV}
\end{equation}
Equation \eqref{quadV} is quadratic in the orbital velocity \(V\).

The Kepler frequency is obtained by solving \eqref{quadV} together with Eq.~\eqref{defV}. Writing \(x\equiv \Omega_K-\omega\), one finds
\begin{equation}
\psi' x^2 - \omega' x - \nu' e^{2(\nu-\psi)}=0,
\end{equation}
whose solution is
\begin{equation}
x
=
\frac{\omega' \pm \sqrt{\omega'^2 + 4\psi'\nu'e^{2(\nu-\psi)}}}{2\psi'}.
\end{equation}
Hence,
\begin{equation}
\Omega_K
=
\omega
+
\frac{\omega' \pm \sqrt{\omega'^2 + 4\psi'\nu'e^{2(\nu-\psi)}}}{2\psi'}.
\end{equation}
Taking the physically relevant positive branch, this may be rewritten as
\begin{equation}
\Omega_K
=
\omega
+
\frac{\omega'}{2\psi'}
+
e^{\nu-\psi}
\sqrt{
\frac{\nu'}{\psi'}
+
\left(
\frac{\omega'}{2\psi'}e^{\psi-\nu}
\right)^2
}.
\label{OmegaK_rel}
\end{equation}
This expression must be evaluated at the stellar equator, \(r=R_{\rm eq}\) and \(\theta=\pi/2\).

Equation \eqref{OmegaK_rel} shows explicitly that the limiting rotational frequency depends on the metric functions and on the frame-dragging angular velocity. Therefore, a fully relativistic treatment of rapidly rotating compact stars requires solving a significantly more complicated system than that associated with static and spherically symmetric stars.

For the purposes of the present work, this observation provides the physical motivation for introducing a simpler effective description of deformed stellar configurations. Instead of explicitly solving the full rotational problem, we adopt a phenomenological deformation parameter \(\mathcal{D}\), which captures deviations from strict spherical symmetry in a much simpler way. Although the \(\mathcal{D}\)-TOV formalism does not explicitly impose rotation as the unique origin of deformation, rotation offers a natural and physically well-motivated interpretation for such deviations. In this sense, the \(\mathcal{D}\)-TOV framework may be understood as an effective approach to the study of deformed compact objects, motivated in part by the complexity of the fully rotational problem.

\section{Deformed compact objects}

Spherical symmetry is the standard assumption in compact-star modeling, and the background metric underlying the TOV equation describes an exactly spherical configuration. In realistic astrophysical environments, however, compact objects are not expected to be perfectly spherical. Deviations from sphericity may arise from several effects, most notably rotation \citep{wen2011properties} and ultra-strong magnetic fields \citep{rizaldy2018magnetized}, as well as from internal stresses that can effectively break isotropy \citep{ruderman1972pulsars}. Consequently, deformed compact-star models have been proposed in the literature to capture departures from exact spherical symmetry in a controlled manner \citep{zubairi2017,quartuccio2025deformed}. Within these approaches, equilibrium configurations may become either oblate or prolate, depending on the physical mechanism responsible for the deformation.

Solutions describing non-spherical compact stars generally involve a substantially higher degree of mathematical and numerical complexity than their spherical counterparts, owing to the structure of the underlying field equations. This motivates the use of simplified or effective descriptions capable of capturing global departures from spherical symmetry without solving the full multidimensional problem.

At supranuclear densities, such as those expected in the interiors of NSs, it is physically plausible that anisotropic stresses may arise in the stellar interior, leading to different radial and tangential pressures, \(p_r(r)\) and \(p_t(r)\) \citep{ruderman1972pulsars}. It is important, however, to distinguish pressure anisotropy from geometrical deformation. In the usual spherically symmetric treatment, such anisotropy does not by itself imply a geometrically deformed star. Rather, it means only that the pressure along the radial direction differs from that in the angular directions, while the two tangential directions remain equivalent. In this sense, anisotropic matter can still be consistently described within a spherically symmetric stellar configuration.

A different situation occurs when the anisotropy is associated with a preferred spatial direction, as may happen in the presence of rapid rotation, strong magnetic fields, or other axisymmetric effects. In such cases, the stresses along the polar and equatorial directions need not be equivalent, and the anisotropy may then be interpreted as a signature of genuine stellar deformation. Even in the simpler spherical case, when \(p_r(r)\neq p_t(r)\), the hydrostatic equilibrium equation is modified by an additional term proportional to the anisotropy \(\Delta(r)=p_t(r)-p_r(r)\) \citep{bowers1974anisotropic}. This contribution can provide extra support against gravity when \(p_t(r)>p_r(r)\), thereby altering the internal structure and potentially allowing more massive equilibrium configurations. For this reason, pressure anisotropy is often regarded as an important physical ingredient in the modeling of compact stars, both as an effective correction within spherical symmetry and as a possible manifestation of mechanisms capable of inducing stellar deformation.

A complementary illustration of the anisotropy-deformation connection appears in studies of strongly magnetized white dwarfs. In that context, the magnetic field introduces a preferred direction and splits the stress into components parallel and perpendicular to the field, naturally leading to non-spherical (axisymmetric) equilibrium configurations. Although the microphysics and typical densities differ from the NS regime, magnetized white dwarfs provide an example in which a well-defined physical agent (the magnetic field) generates anisotropic stresses that can be encoded effectively as a global spheroidal deformation (see, e.g., \citealt{terrero2019modeling}).

In the present work, rather than modeling pressure anisotropy explicitly (i.e., $p_r\neq p_t$), we adopt a one-parameter effective deformation scheme that preserves a tractable one-dimensional description while capturing global departures from spherical symmetry.

A comparatively simple effective framework to model non-spherical compact objects was introduced by
\citet{zubairi2017}. In that formalism, the stellar deformation is encoded through a dimensionless
parameter $\mathcal{D}$, defined as the ratio between the polar radius $z$ and the equatorial radius $r$, 
\begin{equation}
\mathcal{D} \equiv \frac{z}{r}.
\end{equation}
If $z>r$ (i.e., $\mathcal{D}>1$), the configuration is elongated along the polar direction and is therefore
prolate. Conversely, if $z<r$ (i.e., $\mathcal{D}<1$), the star is flattened at the poles and extended along
the equator, corresponding to an oblate configuration.

\subsection{Deformed hydrostatic equilibrium equation (\(\mathcal{D}\)-TOV)}

Following \citet{zubairi2017}, we describe deviations from spherical configurations through an effective one-parameter deformation of the radial metric sector. This construction should be understood as a phenomenological prescription rather than as a fully axisymmetric solution of Einstein's equations. In particular, the angular sector is still written in terms of the spherical areal radius \(r\), while the effects of the global deformation are incorporated through the parameter \(\mathcal{D}\). We adopt the parametrized line element
\begin{equation}
ds^2 = e^{2\Phi(r)}dt^2
      - \left(1-\frac{2m(r)}{r}\right)^{-\mathcal{D}}dr^2
      - r^2\left(d\theta^2+\sin^2\theta\,d\varphi^2\right),
\label{eq:metric_D}
\end{equation}
where \(\mathcal{D}\) is a constant deformation parameter and \(m(r)\) is the enclosed gravitational mass function. For convenience, we define
\begin{equation}
a(r)\equiv 1-\frac{2m(r)}{r},
\qquad
g_{rr}=-a(r)^{-\mathcal{D}},
\qquad
g^{rr}=-a(r)^{\mathcal{D}}.
\label{eq:a_def}
\end{equation}

In this effective description, \(\mathcal{D}=1\) recovers the standard spherical case. Values \(\mathcal{D}<1\) and \(\mathcal{D}>1\) are interpreted as mimicking oblate and prolate global configurations, respectively, through the relation between effective polar and equatorial length scales discussed previously. However, since the metric retains a spherical angular sector, \(\mathcal{D}\) should not be regarded as an arbitrary measure of a fully axisymmetric deformation. Instead, the model is intended to describe moderate departures from spherical symmetry and to investigate how such departures affect the stellar equilibrium equations.

It is also important to emphasize that, for \(\mathcal{D}\neq1\), the metric should not be interpreted as providing an exact exterior vacuum solution. In the exactly spherically symmetric case, Birkhoff's theorem implies that the exterior vacuum geometry is Schwarzschild. A strongly nonspherical configuration, on the other hand, would require a consistent exterior spacetime and matching conditions, for example within a slow-rotation Hartle--Thorne treatment or a fully axisymmetric framework. In the present work, we restrict ourselves to the effective interior description and define the stellar surface by the condition \(p(R)=0\). A complete treatment of the exterior matching for genuinely deformed configurations lies beyond the scope of this thesis.

Applying Einstein's equations to the deformed metric \eqref{eq:metric_D}, one obtains a compact expression
for the $rr$ component of the Einstein tensor (see Appendix~C for the intermediate algebra). For a static,
isotropic perfect fluid,
\begin{equation}
T^{\mu}{}_{\nu}(r)=\mathrm{diag}\!\left(\varepsilon(r),\,-p(r),\,-p(r),\,-p(r)\right),
\end{equation}
the field equation in the radial sector yields an explicit relation for the gravitational potential gradient,
\begin{equation}
\Phi'(r)=\frac{4\pi r^2 p(r)+\tfrac{1}{2}\left[1-a(r)^{\mathcal{D}}\right]}{r\,a(r)^{\mathcal{D}}},
\label{eq:Phi_prime_D}
\end{equation}
where $a(r)\equiv 1-2m(r)/r$.

At this stage, the hydrostatic equilibrium equation follows from local energy-momentum conservation.
For a static perfect fluid, the relativistic Euler equation reads
\begin{equation}
\frac{dp(r)}{dr}= -\left[\varepsilon(r)+p(r)\right]\Phi'(r).
\label{eq:euler_D}
\end{equation}
Substituting Eq.~\eqref{eq:Phi_prime_D} into Eq.~\eqref{eq:euler_D} leads to the deformed
Tolman-Oppenheimer-Volkoff equation ($\mathcal{D}$--TOV),
\begin{equation}
\frac{dp(r)}{dr}
= -\left[\varepsilon(r)+p(r)\right]\,
\frac{8\pi r^2 p(r)-\left(1-\dfrac{2m(r)}{r}\right)^{\mathcal{D}}+1}
{2r\left(1-\dfrac{2m(r)}{r}\right)^{\mathcal{D}}}.
\label{eq:DTOV}
\end{equation}
By construction, the standard spherical TOV equation is recovered in the limit $\mathcal{D}\to 1$.

Finally, within the same effective spheroidal prescription of \citet{zubairi2017}, the enclosed mass is modeled by introducing a deformed volume element. It is important to emphasize that this mass-continuity relation is not derived directly from the \(tt\) component of the field equations for the metric \eqref{eq:metric_D}. Rather, it should be understood as an effective prescription that rescales the standard spherical mass integral in order to account, at least phenomenologically, for the global deformation of the configuration.

In the spherical case, the mass element is proportional to \(4\pi r^2 dr\). In the deformed description, this radial factor is replaced by the spheroidal combination \(4\pi r z\,dr\). Using \(z=\mathcal{D}r\), one obtains
\begin{equation}
m(r)
=
\int_0^r 4\pi \tilde{r}^{\,2}\mathcal{D}\,
\varepsilon(\tilde{r})\,d\tilde{r},
\qquad
\Rightarrow
\qquad
\frac{dm(r)}{dr}
=
4\pi \mathcal{D} r^2\varepsilon(r),
\label{eq:mass_continuity_D}
\end{equation}
where \(m(r)\) should be interpreted as an effective enclosed mass function associated with the deformed configuration. For \(\mathcal{D}=1\), the standard spherical mass-continuity equation is recovered. Therefore, the role of \(\mathcal{D}\) in Eq.~\eqref{eq:mass_continuity_D} is to encode an effective rescaling of the mass distribution, consistently with the phenomenological character of the deformation scheme.

\subsection{Physical interpretation of the deformation parameter $\mathcal{D}$}
\label{sec:physical_interpretation_D}

The \(\mathcal{D}\)-TOV formalism provides an effective description of departures from
exact spherical symmetry while preserving a one-dimensional hydrostatic structure. In
this approach, the deformation is not obtained from a fully axisymmetric solution of
Einstein's equations. Instead, it is introduced phenomenologically through a single
dimensionless parameter, \(\mathcal{D}\), which modifies the radial gravitational sector
and rescales the effective mass distribution associated with the stellar configuration.

At the metric level, the deformation enters through the factor
\begin{equation}
a(r)^{\mathcal{D}},
\qquad
a(r)\equiv 1-\frac{2m(r)}{r},
\end{equation}
which modifies the effective gravitational response in the deformed balance condition.
For physically relevant stellar interiors, one typically has \(0<a(r)<1\). This follows
from the fact that the enclosed mass is positive, \(m(r)>0\), so that \(a(r)<1\), while
regular stellar configurations must also satisfy \(2m(r)/r<1\) in order to avoid the
formation of a trapped surface or horizon inside the star. Thus, \(a(r)\) encodes the
local compactness of the configuration: the smaller \(a(r)\) is, the stronger the
relativistic gravitational field associated with the ratio \(m(r)/r\). Near the stellar
center, where \(m(r)\sim r^3\), one has \(2m(r)/r\sim r^2\) and therefore
\(a(r)\to1\). Moving outward, \(a(r)\) reflects the cumulative contribution of the
enclosed mass to the local geometry.

Since \(0<a(r)<1\), the exponent \(\mathcal{D}\) changes the radial metric sector in a
physically transparent way. In the oblate branch \((\mathcal{D}<1)\), one finds
\begin{equation*}
a(r)^{\mathcal{D}}>a(r),
\end{equation*}
whereas in the prolate branch \((\mathcal{D}>1)\), one has
\begin{equation*}
a(r)^{\mathcal{D}}<a(r).
\end{equation*}

Inspecting the $\mathcal{D}$-TOV equation, Eq.~\eqref{eq:DTOV}, one sees that increasing $a(r)^{\mathcal{D}}$ (as occurs for $\mathcal{D}<1$) tends to reduce the magnitude of the pressure gradient, making $p(r)$ decrease more slowly with radius when compared to the spherical TOV case. This argument is qualitative, since the final mass-radius sequence is determined by the coupled integration of the structure equations for a given EoS. Nevertheless, it provides useful intuition for why the oblate branch of the model often shifts equilibrium sequences toward less rapidly decreasing pressure profiles and, consequently, toward configurations that may support larger gravitational masses for the same microphysics.

Geometrically, the parameter $\mathcal{D}$ is motivated by the ratio between an effective polar length scale \(z\) and an effective equatorial length scale \(r\), Eq.~(3.57). Thus, $\mathcal{D}=1$ recovers the spherical limit, $\mathcal{D}<1$ corresponds to oblate configurations ($z<r$), and $\mathcal{D}>1$ corresponds to prolate configurations ($z>r$). However, this identification should not be interpreted as a complete geometrical reconstruction of an axisymmetric star. Since the angular sector of the metric remains written in terms of the spherical areal radius \(r\), the parameter $\mathcal{D}$ should be understood as an effective global measure of deformation rather than as an exact eccentricity, quadrupole moment, or coordinate transformation.

In this sense, $\mathcal{D}$ plays the role of a phenomenological parameter that summarizes the net effect of nonspherical features on the radial hydrostatic balance. Physically, such nonspherical features may arise from rotation, strong magnetic fields, anisotropic stresses, elastic stresses in the crust, or other internal mechanisms capable of selecting a preferred direction inside the star. The present formalism does not attempt to model these mechanisms individually. Instead, it asks a more limited question: how would the equilibrium structure be modified if the global departure from spherical symmetry could be encoded effectively in the radial sector through a single parameter?

This interpretation is also important when comparing different values of $\mathcal{D}$. The parameter should not be varied arbitrarily as if each value represented an exact relativistic stellar solution with a well-defined exterior spacetime. Rather, it should be regarded as a controlled deformation parameter, useful for exploring qualitative trends around the spherical limit. For large deviations from spherical symmetry, a fully axisymmetric treatment, together with appropriate exterior matching conditions, would be required.

Therefore, the $\mathcal{D}$-formalism should be interpreted as a phenomenological generalization of the standard TOV framework. It preserves the simplicity of a one-dimensional treatment while incorporating, in an effective manner, the possibility that the star is not exactly spherical. Within this scope, $\mathcal{D}$ provides a useful diagnostic tool for investigating how global deformation may affect pressure gradients, mass-radius sequences, and the maximum mass supported by a given EoS.

\chapter{Modified Gravity}

General Relativity, proposed by Albert Einstein in 1915 \citep{einstein1915feldgleichungen}, provides the standard geometric description of gravity and has been successfully tested in a wide range of regimes, from Solar-System scales to strong-field compact-binary systems. Nevertheless, open problems in cosmology and astrophysics, such as the nature of dark matter, the physical origin of the accelerated expansion of the Universe, and the cosmological constant problem, have motivated the investigation of possible extensions of GR.

In cosmology, the accelerated expansion of the Universe is commonly described within the standard \(\Lambda\)CDM model through the cosmological constant \(\Lambda\), which acts as an effective dark-energy component. Although this framework is highly successful phenomenologically, the physical interpretation of \(\Lambda\) remains an open problem. In particular, if \(\Lambda\) is associated with vacuum energy, its observed value differs enormously from naive quantum-field-theory estimates, leading to the well-known cosmological constant problem \citep{adler1995vacuum, ryden2017introduction}. This difficulty has motivated the study of alternative scenarios in which cosmic acceleration may arise from modifications of the gravitational sector itself, rather than from an additional dark-energy component.

The underlying idea of modified gravity is that, since gravity in GR is encoded in spacetime geometry, departures from the Einstein-Hilbert action may generate effective contributions to the field equations. At cosmological scales, such contributions can mimic a dark-energy sector, while in strong-field regimes they may modify the equilibrium structure of compact objects. In this sense, modified gravity theories provide a useful theoretical framework for exploring whether deviations from GR could become relevant in regimes of high curvature, high density, or strong matter-geometry coupling.

Modified gravity theories have also been applied in astrophysics, particularly in studies of NS structure \citep{olmo2020stellar}. As discussed in the introduction of this thesis, the GW190814 event raised renewed interest in the question of the maximum mass that NSs can support. In this context, modified gravity has been explored as a possible mechanism capable of producing more massive compact-star configurations for a given microphysical EoS \citep{astashenok2013further}.

A wide variety of modified gravity models has been proposed in the literature. Among them are \(f(R)\) theories, whose early developments can be traced back to nonlinear curvature Lagrangians and to the Starobinsky model \citep{buchdahl1970non, starobinsky1980new, capozziello2009f}; \(f(R,T)\) models \citep{harko2011f}; \(f(R,\mathcal{L})\) frameworks \citep{harko2010f}; and more general extensions such as \(f(R,\mathcal{L},T)\) \citep{haghani2021generalizing}. Here \(f\) denotes a generic function of its argument(s), \(R\) is the Ricci scalar, \(T\) is the trace of the energy-momentum tensor, and \(\mathcal{L}\) is the matter Lagrangian density. In these approaches, the Einstein-Hilbert action is generalized so that the resulting field equations contain additional terms with respect to GR.

In strong-field regimes, such as NS interiors, these additional terms may modify the hydrostatic equilibrium equations and, in some scenarios, increase the maximum supported mass \citep{moraes2016stellar,quartuccio2025equilibrium,quartuccio2026deformed}. At the same time, consistency with Solar-System tests typically requires deviations from GR to be negligible in the weak-field regime.

In this chapter we derive the hydrostatic equilibrium equations within the class of theories \(f(R,T)=R+f(T)\) for static, isotropic, spherically symmetric stars, and we then extend the formalism to configurations with broken spherical symmetry within the effective deformation scheme introduced in the previous chapter.

\section{The $f(R,T)$ gravity}

$f(R,T)$ gravity is a modified extension of GR in which the gravitational action is allowed to depend not only on the Ricci scalar $R$, which encodes spacetime curvature, but also on $T$, the trace of the energy-momentum tensor. The $f(R,T)$ framework was originally proposed as a phenomenological avenue to address fundamental problems in cosmology, as discussed above, without invoking new particles or additional exotic energy components. In this approach, departures from the standard $\Lambda$CDM dynamics may emerge from the additional terms associated with the explicit $T$-dependence, which become relevant whenever matter sources are present.

\subsection{Action and field equations in $f(R,T)$ gravity}

We start from the action of $f(R,T)$ gravity \citep{harko2011f},
\begin{equation}
S=\frac{1}{16\pi }\int d^4x\,\sqrt{-g}\,f(R,T)+S_m,
\label{eq:action_fRT}
\end{equation}
where $R$ is the Ricci scalar, $T\equiv g^{\mu\nu}T_{\mu\nu}$ is the trace of the energy--momentum tensor,
and the matter action is
\begin{equation}
S_m=\int d^4x\,\sqrt{-g}\,\mathcal{L}_m.
\label{eq:matter_action}
\end{equation}
The energy--momentum tensor is defined by
\begin{equation}
T_{\mu\nu}\equiv -\frac{2}{\sqrt{-g}}\frac{\delta(\sqrt{-g}\mathcal{L}_m)}{\delta g^{\mu\nu}}.
\label{eq:Tmunu_def}
\end{equation}

Varying the action \eqref{eq:action_fRT} with respect to \(g^{\mu\nu}\) yields the field equations in the form
\begin{equation}
f_R R_{\mu\nu}
-\frac{1}{2} f\,g_{\mu\nu}
+\left(g_{\mu\nu}\Box-\nabla_\mu\nabla_\nu\right)f_R
=
8\pi \,T_{\mu\nu}
-
f_T\left(T_{\mu\nu}+\Theta_{\mu\nu}\right),
\label{eq:fRT_field_general}
\end{equation}
where \(f_R\equiv\partial f/\partial R\), \(f_T\equiv\partial f/\partial T\),
\(\Box\equiv\nabla^\alpha\nabla_\alpha\) is the covariant d'Alembertian operator, and
\begin{equation}
\Theta_{\mu\nu}
\equiv
g^{\alpha\beta}
\frac{\delta T_{\alpha\beta}}{\delta g^{\mu\nu}}.
\label{eq:Theta_def}
\end{equation}
The intermediate variational steps are presented in Appendix~D.

\subsection{$f(R,T)=R+f(T)$ and perfect fluids}

In this thesis we focus on the subclass
\begin{equation}
f(R,T)=R+f(T),
\label{eq:fT_model}
\end{equation}
for which $f_R=1$ and the operator term in Eq.~\eqref{eq:fRT_field_general} vanishes. The field equations become
\begin{equation}
G_{\mu\nu}
=
8\pi \,T_{\mu\nu}
-f_T\left(T_{\mu\nu}+\Theta_{\mu\nu}\right)
+\frac12 f(T)\,g_{\mu\nu}.
\label{eq:fT_field_basic}
\end{equation}

For a perfect fluid, we adopt the standard choice $\mathcal{L}_m=-p$, which implies
\begin{equation}
\Theta_{\mu\nu}=-2T_{\mu\nu}-p\,g_{\mu\nu}.
\label{eq:Theta_Lm_minus_p}
\end{equation}
Substituting Eq.~\eqref{eq:Theta_Lm_minus_p} into Eq.~\eqref{eq:fT_field_basic} yields
\begin{equation}
G_{\mu\nu}
=
8\pi \,T_{\mu\nu}
+f_T\left(T_{\mu\nu}+p\,g_{\mu\nu}\right)
+\frac12 f(T)\,g_{\mu\nu}.
\label{eq:fT_field_final}
\end{equation}

Taking the covariant divergence of Eq.~\eqref{eq:fT_field_basic} and using $\nabla^\mu G_{\mu\nu}=0$,
one obtains a modified conservation law. For the class $f(R,T)=R+f(T)$ it can be written as
\begin{equation}
\nabla^\mu T_{\mu\nu}
=
\frac{f_T}{8\pi -f_T}
\left[
\left(T_{\mu\nu}+\Theta_{\mu\nu}\right)\nabla^\mu \ln f_T
+\nabla^\mu\Theta_{\mu\nu}
-\frac12 g_{\mu\nu}\nabla^\mu T
\right].
\label{eq:noncons_general}
\end{equation}

Equation~\eqref{eq:noncons_general} shows that, in general, the energy-momentum tensor is not covariantly conserved in $f(R,T)$ gravity. This is a direct consequence of the explicit dependence of the gravitational action on the trace $T$, which induces an effective matter-geometry coupling. Physically, this nonconservation can be interpreted as an exchange between the matter sector and the gravitational sector, or equivalently as the presence of an extra force acting on the fluid elements. As a result, the hydrostatic equilibrium condition is modified with respect to GR, since the pressure gradient is no longer determined only by the standard balance between gravity and pressure.

In the stellar applications developed in this thesis, we restrict the analysis to static, isotropic perfect fluids. In this case, once a matter Lagrangian is specified, the tensor $\Theta_{\mu\nu}$ is fixed and Eq.~\eqref{eq:noncons_general} leads to the modified TOV equation used below. For more general matter sources, such as anisotropic fluids, magnetized matter, viscous fluids, or elastic stresses in the crust, the form of $\Theta_{\mu\nu}$ and its divergence would be different. Consequently, the nonconservation law would generate additional terms in the equilibrium equation, potentially affecting the pressure profile, stability conditions, mass-radius relation, and matching with the exterior spacetime. Therefore, the perfect-fluid assumption adopted here should be understood as a controlled simplification that allows us to isolate the effects of the $f(R,T)$ coupling on compact-star equilibrium.

\subsection{Static spherical configuration and the modified TOV equation}

We now specialize to static, spherically symmetric configurations and adopt the line element
\begin{equation}
ds^2=e^{\upsilon(r)}dt^2-e^{\omega(r)}dr^2-r^2\left(d\theta^2+\sin^2\theta\,d\varphi^2\right).
\label{eq:spherical_metric_fT}
\end{equation}
Here, \(\upsilon(r)\) and \(\omega(r)\) are metric potentials depending only on the radial coordinate, associated respectively with the temporal and radial components of the metric. In this signature, the perfect-fluid trace reads $T=\varepsilon-3p$. For a static configuration, the $\nu=r$
component of the modified conservation law can be cast in the form
\begin{equation}
p'(r)+\big[\varepsilon(r)+p(r)\big]\frac{\upsilon'(r)}{2}
=
\frac{f_T}{8\pi +f_T}\left[\frac{1}{2}p'(r)-\frac{1}{2}\varepsilon'(r)\right],
\label{eq:noncons_spherical}
\end{equation}
where a prime denotes $d/dr$.

To determine $\upsilon'(r)$, we use the $rr$ component of Eq.~\eqref{eq:fT_field_final}. For the metric
\eqref{eq:spherical_metric_fT} one has
\begin{equation}
G_{rr}=\frac{1-e^{\omega(r)}+r\,\upsilon'(r)}{r^2}.
\label{eq:Grr_spherical}
\end{equation}
Moreover, $T_{rr}=p\,e^{\omega}$ and $g_{rr}=-e^{\omega}$, so that $T_{rr}+p\,g_{rr}=0$ and the $f_T$ term
cancels in the $rr$ component. Therefore,
\begin{equation}
\frac{1-e^{\omega}+r\,\upsilon'}{r^2}
=
8\pi \,p\,e^{\omega}-\frac{1}{2}f(T)\,e^{\omega}.
\end{equation}
Solving for $\upsilon'(r)$ gives
\begin{equation}
\upsilon'(r)=8\pi \,p(r)\,r\,e^{\omega(r)}-\frac{1}{2}f(T)\,r\,e^{\omega(r)}+\frac{e^{\omega(r)}-1}{r}.
\label{eq:upsilon_prime_correct}
\end{equation}

Substituting Eq.~\eqref{eq:upsilon_prime_correct} into Eq.~\eqref{eq:noncons_spherical} and using
$\varepsilon'(r)=(d\varepsilon/dp)\,p'(r)$, one obtains
\begin{equation}
p'(r)=-(\varepsilon(r)+p(r))\,
\frac{\left[\,4\pi \,p\,r\,e^{\omega}-\frac14 f(T)\,r\,e^{\omega}+\frac{e^{\omega}-1}{2r}\right]}
{1-\dfrac{f_T}{2(8\pi +f_T)}\left(1-\dfrac{d\varepsilon}{dp}\right)}.
\label{eq:pprime_intermediate}
\end{equation}

Finally, introducing the standard mass function through
\begin{equation}
e^{\omega(r)}=\left(1-\frac{2m(r)}{r}\right)^{-1},
\label{eq:ew_mass_def}
\end{equation}
and using the identity
\begin{equation}
\frac{e^{\omega}-1}{2r}=\frac{m(r)}{r^2}\,e^{\omega},
\end{equation}
Eq.~\eqref{eq:pprime_intermediate} can be written in the compact form

\begin{equation}
\frac{dp}{dr}
=
-(\varepsilon(r)+p(r))\,
\frac{\left[4\pi p(r)r-\dfrac{f(T)\,r}{4}+\dfrac{m}{r^2}\right]}
{\left(1-\dfrac{2m}{r}\right)\left[1-\dfrac{f_T}{2(8\pi+f_T)}\left(1-\dfrac{d\varepsilon}{dp}\right)\right]}.
\label{eq:TOV_fT_final_G1}
\end{equation}

In the GR limit $f(T)\to 0$ (and thus $f_T\to 0$), Eq.~\eqref{eq:TOV_fT_final_G1} reduces to the standard
Tolman-Oppenheimer-Volkoff equation.

\subsection{Mass equation}

For the static spherical metric \eqref{eq:spherical_metric_fT}, the $tt$ component of the Einstein tensor is
\begin{equation}
G_{tt}=\frac{e^{\upsilon(r)-\omega(r)}\left(-1+e^{\omega(r)}+r\,\omega'(r)\right)}{r^2}.
\end{equation}
For a perfect fluid, $T_{tt}=\varepsilon(r)e^{\upsilon(r)}$ and $g_{tt}=e^{\upsilon(r)}$. Substituting these expressions
into the field equations \eqref{eq:fT_field_final} and introducing the mass function through
$e^{\omega(r)}=\left(1-\frac{2m(r)}{r}\right)^{-1}$, one obtains the modified mass-continuity relation
\begin{equation}
m'(r)=4\pi r^2\varepsilon(r)+r^2\left[\frac{f_T}{2}\big(\varepsilon(r)+p(r)\big)+\frac{f(T)}{4}\right].
\label{eq:mass_fT}
\end{equation}
The intermediate algebra is deferred to Appendix~D.

In this thesis we restrict to the subclass $f(R,T)=R+f(T)$, which preserves the Einstein-Hilbert
dependence on the Ricci scalar (i.e., the gravitational Lagrangian remains linear in $R$) and introduces
modifications through an explicit dependence on the matter trace $T$. In this sense, the geometric sector
is kept in its simplest form, while the matter sector acquires an additional channel through which the
local properties of the source can affect the field equations. As a result, the effective relation between
matter-energy and spacetime curvature is altered, and hydrostatic equilibrium in compact stars can be
modified even for the same underlying microphysical EoS. In particular, these corrections
may shift equilibrium sequences and, in some scenarios, change the maximum mass supported by NS
configurations.


\subsection{Optimized $f(R,T)$ gravity}

In this work, one of the trace functionals $f(T)$ used to investigate NS equilibrium configurations was originally constructed in a cosmological context. We adopt the functional form obtained by
\citet{fortunato2024search} via Gaussian-process reconstruction applied to observational determinations of
the Hubble parameter $H(z)$. The resulting trace sector is
\begin{equation}
f(T)=\alpha T^2 + A\tanh\!\left[\lambda\left(T+T_0\right)\right]+\beta T+\gamma,
\label{eq:fT_optimized}
\end{equation}
where $\alpha$, $A$, $\lambda$, $T_0$, $\beta$, and $\gamma$ are constant parameters.

The implications of this choice for NS structure are discussed in the next chapter, where we compare equilibrium sequences obtained with this nonlinear trace sector to the corresponding GR baseline.

\subsection{Linear trace coupling: $f(R,T)=R+2\lambda T$}
\label{subsec:linear_fT}

As a useful benchmark, we also consider the linear trace-coupling model
\begin{equation}
f(R,T)=R+2\lambda T,
\label{eq:fT_linear}
\end{equation}
where $\lambda$ is a constant coupling parameter. This choice represents the simplest realization of a
trace-dependent correction and has been widely employed in compact-star applications as a minimal
phenomenological extension of GR in the matter sector \citep{moraes2016stellar,astashenok2013further,olmo2020stellar}.
In contrast to nonlinear $f(T)$ prescriptions, the linear model satisfies
\begin{equation}
f(T)=2\lambda T,
\qquad
f_T=2\lambda=\mathrm{const.},
\qquad
f_{TT}=0,
\end{equation}
so that the modified structure equations simplify considerably.

In particular, substituting $f(T)=2\lambda T$ and $f_T=2\lambda$ into the general spherical equilibrium
equation derived previously yields
\begin{equation}
\frac{dp}{dr}
=
-(\varepsilon+p)\,
\frac{\left[4\pi p\,r-\dfrac{\lambda (\varepsilon -3p)\,r}{2}+\dfrac{m}{r^2}\right]}
{\left(1-\dfrac{2m}{r}\right)\left[1-\dfrac{\lambda}{8\pi+2\lambda}\left(1-\dfrac{d\varepsilon}{dp}\right)\right]},
\label{eq:TOV_linear}
\end{equation}
with the trace evaluated from the perfect-fluid relation (in the present conventions)
\begin{equation}
T=\varepsilon-3p.
\end{equation}
Here, as throughout the stellar-structure analysis, 
\(\varepsilon\), \(p\), and \(m\) are understood as radial functions, 
\(\varepsilon=\varepsilon(r)\), \(p=p(r)\), and \(m=m(r)\). 
For compactness, the explicit radial dependence will not always be written.

Similarly, the modified mass-continuity equation becomes
\begin{equation}
m'(r)=4\pi r^2\varepsilon+ \frac{\lambda \Big[3\varepsilon - p\Big]r^2}{2}.
\label{eq:mass_linear}
\end{equation}

The linear model \eqref{eq:fT_linear} is particularly convenient for comparisons, since it retains the
trace-induced modification while avoiding additional nonlinear scales associated with $f_{TT}(T)$.
Quantitative comparisons between the linear and nonlinear trace sectors are presented in the next chapter.

\section{Deformed compact objects in $f(R,T)$ gravity}

In Chapter~3 we discussed how departures from exact spherical symmetry can be used as an effective tool
to explore heavier NS configurations, since the deformation parameter $\mathcal{D}$ modifies
the hydrostatic balance by changing the effective pressure-gradient response. In the present chapter we
have also shown that modified gravity theories can affect compact-star structure by altering the
matter-geometry relation encoded in the field equations, which may lead to equilibrium sequences with
different maximum masses for the same microphysical input. In much of the existing compact-star
literature in modified gravity, however, the stellar configuration is still assumed to be perfectly
spherical.

From an astrophysical standpoint, exact spherical symmetry is an idealization. Compact
stars may depart from sphericity due to rotation, strong magnetic fields, anisotropic
stresses, or related mechanisms. It is therefore natural to extend the modified-gravity
description by incorporating an effective deformation, while still preserving the
tractability of a one-dimensional hydrostatic treatment.

In this thesis, we implement this step by adopting the \(\mathcal{D}\)-deformed metric
introduced in Chapter~3 and deriving the corresponding hydrostatic equilibrium equation
in the \(f(R,T)=R+f(T)\) framework. In this sense, the goal is not to replace the
standard spherical description, but to investigate how deviations from exact spherical
symmetry may affect compact-star equilibrium when combined with the matter--geometry
coupling characteristic of \(f(R,T)\) gravity.


We begin by rewriting the line element as
\begin{equation}
ds^2 = e^{2\Phi(r)}dt^2 - e^{2\Lambda(r)}dr^2 - r^2\left(d\theta^2+\sin^2\theta\,d\varphi^2\right),
\label{eq:CAO10}
\end{equation}
and implement the effective deformation through
\begin{equation}
e^{2\Lambda(r)}=\left(1-\frac{2m(r)}{r}\right)^{-\mathcal{D}} \equiv \mathcal{A}(r)^{-\mathcal{D}},
\qquad
\mathcal{A}(r)\equiv 1-\frac{2m(r)}{r}.
\label{eq:A_def}
\end{equation}

Within the same steps used in the spherical case, the modified TOV equation in the deformed geometry can be written as \citep{quartuccio2026deformed}
\begin{equation}
\frac{dp(r)}{dr}
=
-\big(\varepsilon(r)+p(r)\big)\,
\frac{\mathcal{A}(r)^{-\mathcal{D}}
\left[
4\pi p(r)\,r-\frac{f(T)}{4}\,r+\frac{1-\mathcal{A}(r)^{\mathcal{D}}}{2r}
\right]}
{1-\dfrac{f_T}{2\left(8\pi+f_T\right)}
\left(1-\dfrac{d\varepsilon}{dp}\right)}.
\label{eq:DTOV_fT}
\end{equation}

By construction, Eq.~\eqref{eq:DTOV_fT} reduces to the spherical $f(R,T)$ equilibrium equation when
$\mathcal{D}=1$, and to the GR $\mathcal{D}$--TOV equation when $f(T)\to 0$ (and thus $f_T\to 0$).

To obtain the mass equation, we start from the $tt$ component of the field equations and obtain a relation
of the schematic form
\begin{equation}
(r-2m)\big(1-\mathcal{A}^{\mathcal{D}}\big)
+2\mathcal{D}\,\mathcal{A}^{\mathcal{D}}\big(rm'(r)-m(r)\big)
=
2(r-2m)\,S(r),
\label{eq:mass_step_compact}
\end{equation}
where, for convenience, we defined
\begin{equation}
S(r)\equiv
4\pi r^2\varepsilon(r)
+r^2\left[\frac{f_T}{2}\big(\varepsilon(r)+p(r)\big)+\frac{f(T)}{4}\right].
\label{eq:S_def}
\end{equation}
Solving Eq.~\eqref{eq:mass_step_compact} for $m'(r)$ and using $r-2m=r\,\mathcal{A}$ yields
\begin{equation}
m'(r)=\frac{m(r)}{r}+
\frac{S(r)}{\mathcal{D}\,\mathcal{A}(r)^{\mathcal{D}-1}}
-\frac{1-\mathcal{A}(r)^{\mathcal{D}}}{2\mathcal{D}\,\mathcal{A}(r)^{\mathcal{D}-1}}.
\label{eq:mass_deformed_intermediate}
\end{equation}
Finally, substituting the definition \eqref{eq:S_def} leads to the explicit deformed mass equation \citep{quartuccio2026deformed}
\begin{equation}
m'(r)=\frac{m(r)}{r}
+\frac{
4\pi r^2\varepsilon(r)
+r^2\left[\dfrac{f_T}{2}\big(\varepsilon(r)+p(r)\big)+\dfrac{f(T)}{4}\right]
-\dfrac{1-\mathcal{A}(r)^{\mathcal{D}}}{2}
}
{\mathcal{D}\,\mathcal{A}(r)^{\mathcal{D}-1}}.
\label{eq:mass_deformed_fT}
\end{equation}
In the spherical limit $\mathcal{D}\to 1$, Eq.~\eqref{eq:mass_deformed_fT} reduces to the mass equation
obtained previously in the $f(R,T)=R+f(T)$ case.

\chapter{Solutions of the Structure Equations}

In this chapter we present the equilibrium solutions obtained for NSs and strange stars
in both GR and \(f(R,T)\) gravity, considering spherical as well as effectively deformed
compact configurations. Our goal is to investigate how the stellar structure is affected
by the gravitational framework, by the deformation parameter introduced in the
\(\mathcal{D}\)-TOV formalism, and by the choice of EoS.

To this end, we analyze mass--radius relations, the behavior of the stellar mass as a
function of the central energy density, and internal pressure and energy-density profiles
for distinct classes of compact stars. This allows a direct comparison between hadronic
and quark-matter descriptions, as well as between microphysical EoS models and more
schematic parametrizations, such as the polytropic baseline.


In order to probe different physical compositions and modeling strategies, we consider three distinct equations of state. The GM1 parametrization is adopted as a realistic hadronic EoS, since it describes dense nuclear matter within the relativistic mean-field formalism, incorporating the interaction among baryons through meson exchange and providing a well-established framework for NS structure calculations \citep{glendenning1991reconciliation}. This choice is especially relevant when the goal is to model conventional NSs composed predominantly of strongly interacting hadronic matter.

To investigate compact stars composed of deconfined quark matter, we employ the MIT Bag Model. In this description, quarks are treated as quasi-free particles confined inside a region whose vacuum energy is represented by the bag constant. This model is particularly useful in the study of strange stars, since it captures in a simple and effective way the possibility that quark matter may form self-bound compact objects \citep{jaffe1984strange}. Although idealized, the MIT Bag Model remains one of the most widely used approaches in the literature for examining the global properties of quark stars.

In addition, we make use of a polytropic EoS in order to study the general behavior of the solutions in a more controlled and phenomenological setting. Unlike GM1 and MIT, the polytropic model is not intended to reproduce a specific microscopic composition. Instead, it provides a simpler parametrization of the matter sector, which is useful for identifying the qualitative role played by the gravitational modifications themselves, without the additional complexity introduced by detailed nuclear or quark microphysics \citep{tooper1964general}.

\section{Deformed TOV solutions in General Relativity}

In order to obtain equilibrium configurations in GR, we numerically solved the $\mathcal{D}$-TOV system for different equations of state describing hadronic and quark matter. In this section, our goal is to analyze how the deformation parameter affects the global properties of compact stars within the GR framework, by considering both NS and strange-star configurations.

The structure equations solved in this section are given by
\begin{equation}
\frac{dm}{dr} = 4\pi \mathcal{D} r^{2}\varepsilon,
\end{equation}
and
\begin{equation}
\frac{dp}{dr}
=
-\frac{(\varepsilon+p)\left[8\pi r^{2}p-\left(1-\frac{2m}{r}\right)^{\mathcal{D}}+1\right]}
{2r\left(1-\frac{2m}{r}\right)^{\mathcal{D}}},
\end{equation}
where $\mathcal{D}$ is the deformation parameter. The standard GR limit is recovered for $\mathcal{D}=1$, while values $\mathcal{D}\neq1$ describe deformed compact configurations. As in the previous chapter, the radial dependence of the stellar quantities is left implicit unless otherwise stated.

The numerical integration starts at a small nonzero radius $r_{0}$, thus avoiding the coordinate singularity at the stellar center. The central conditions are specified by $p(r_{0})=p_{c}$ and by the regular expansion
\begin{equation}
m(r_{0}) \simeq \frac{4\pi}{3}\mathcal{D} r_{0}^{3}\varepsilon_{c},
\end{equation}
where $\varepsilon_{c}$ is obtained from the corresponding EoS at the chosen central pressure. The system is then integrated outward by means of a fourth-order Runge-Kutta scheme until the pressure reaches a prescribed surface value, which defines the stellar radius. From the resulting solutions, we construct sequences of stellar models by varying the central pressure logarithmically, allowing the determination of the mass-radius relation, the mass as a function of normalized central energy density, and representative normalized internal profiles $p(r)/p_{c}$ and $\varepsilon(r)/\varepsilon_{c}$.

\subsection{GM1 equation of state}

For the hadronic matter sector, we adopt the GM1 parametrization in tabulated form. The EoS is introduced through an interpolation procedure, allowing the energy density to be reconstructed as a function of pressure, \(\varepsilon=\varepsilon(p)\), throughout the stellar integration.

In this work, the energy-density range extends up to
\[
\varepsilon \simeq 2.63\times10^3~\mathrm{MeV/fm^3},
\]
or, equivalently,
\[
\varepsilon \simeq 3.48\times10^{-3}~\mathrm{km}^{-2}
\]
in geometrized units. In practice, the central pressure is varied within the interval allowed by the table, so that the resulting central energy densities remain inside the domain covered by the tabulated EoS. This allows a sequence of equilibrium configurations to be obtained consistently from the same microscopic hadronic description.

\begin{figure}[h]
    \centering
    \includegraphics[width=0.8\textwidth]{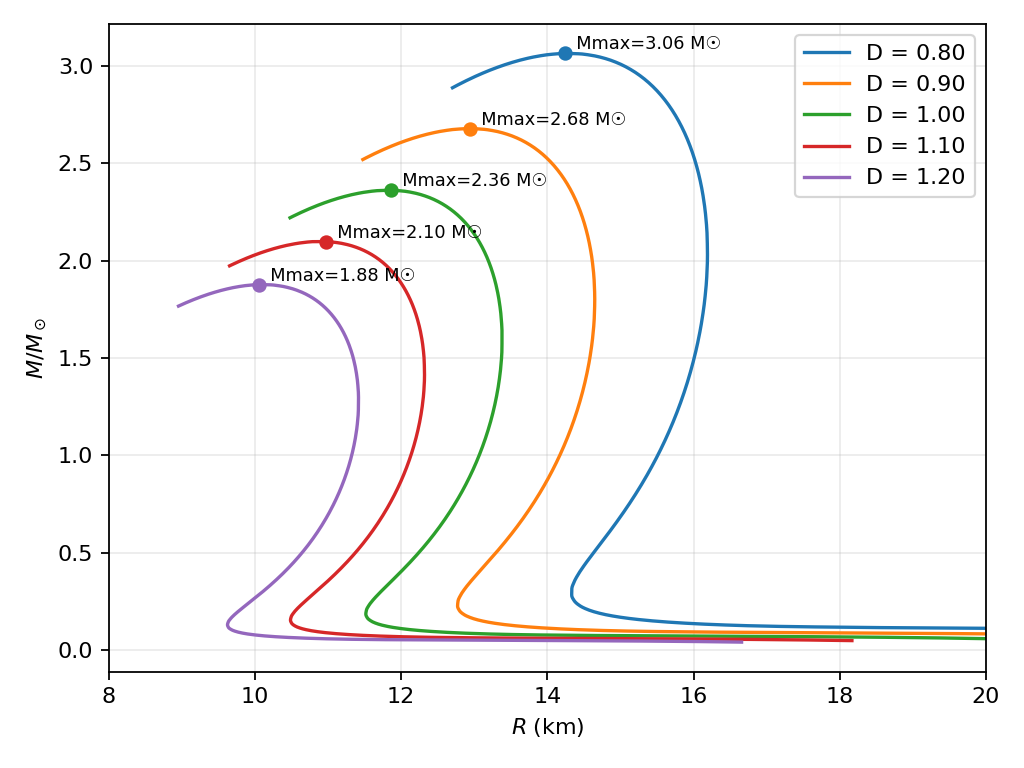}
    \caption{Mass--radius relations obtained from the \(\mathcal{D}\)-TOV equations in General Relativity using the GM1 equation of state, for different values of the deformation parameter \(\mathcal{D}\). The markers indicate the maximum-mass configuration of each sequence.}
    \label{fig:gm1_dtov_mr}
\end{figure}

Figure \ref{fig:gm1_dtov_mr} shows the mass-radius relations obtained from the $\mathcal{D}$-TOV equations in GR using the GM1 EoS for different values of the deformation parameter. The standard spherical result is recovered for $\mathcal{D}=1$, while values different from unity modify both the maximum mass and the typical stellar radius. A clear systematic trend is observed: decreasing $\mathcal{D}$ shifts the mass-radius curve upward and to the right, leading to more massive and larger stellar configurations, whereas increasing $\mathcal{D}$ shifts the sequence downward and to the left, thus producing lighter and more compact stars.

In particular, the maximum mass changes from approximately $3.06\,M_{\odot}$ for $\mathcal{D}=0.8$ to about $1.88\,M_{\odot}$ for $\mathcal{D}=1.2$, with the intermediate cases displaying the expected monotonic behavior. This result indicates that the deformation parameter has a direct impact on the hydrostatic balance between pressure support and gravitational compression. For smaller values of $\mathcal{D}$, the equilibrium configurations are able to sustain larger masses before reaching the turning point of the sequence. On the other hand, larger values of $\mathcal{D}$ reduce the maximum supported mass and favor more compact configurations. Therefore, even within the same hadronic EoS, the introduction of the deformation parameter significantly alters the global properties of the star.

\begin{figure}[h]
    \centering
    \includegraphics[width=0.8\textwidth]{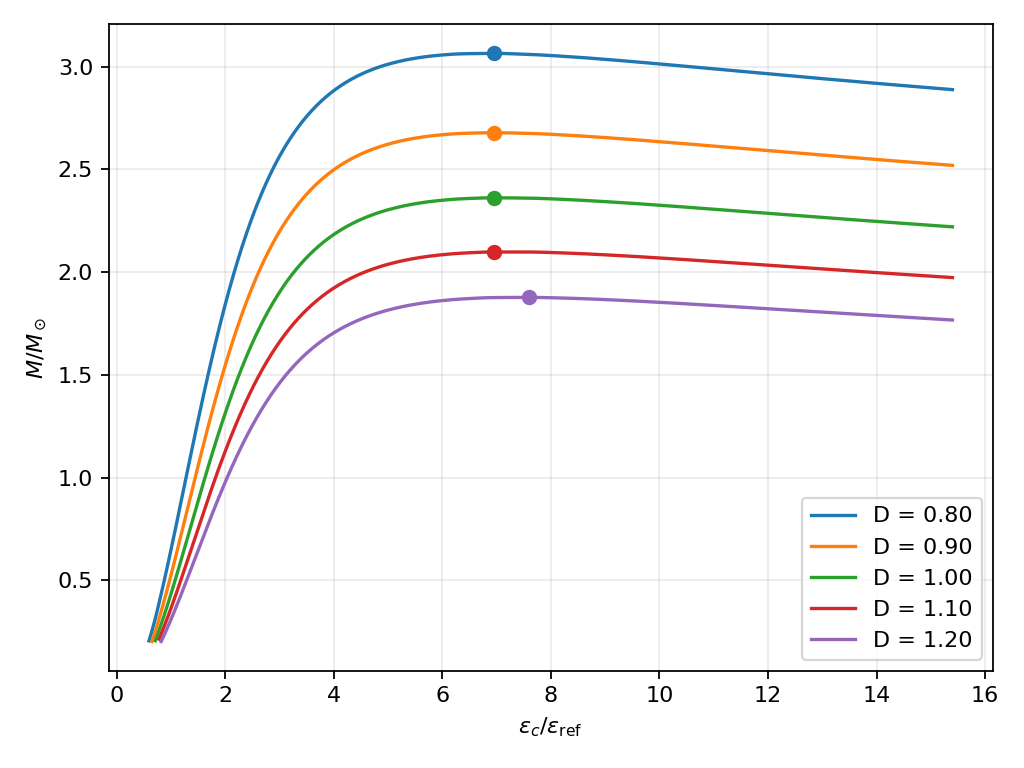}
    \caption{Stellar mass as a function of the normalized central energy density for the GM1 equation of state in the framework of the \(\mathcal{D}\)-TOV formalism. The markers indicate the maximum-mass configuration of each sequence.}
    \label{fig:gm1_dtov_mrho}
\end{figure}

Figure \ref{fig:gm1_dtov_mrho} shows the stellar mass as a function of the normalized central energy density. For all cases, the same qualitative behavior is observed: the mass initially increases with the central energy density, reaches a maximum value, and then undergoes a mild decrease as the central density continues to grow. This pattern is characteristic of equilibrium sequences of relativistic compact stars and reflects the competition between pressure support and gravitational compression.

Another relevant aspect of Fig. \ref{fig:gm1_dtov_mrho} is that the maximum-mass point occurs at relatively similar normalized central densities, although the corresponding masses differ substantially from one sequence to another. This indicates that the introduction of $\mathcal{D}$ does not merely rescale the stellar mass, but changes the way in which the equilibrium structure responds to increasing central density. Therefore, the combined analysis of Figs. \ref{fig:gm1_dtov_mr} and \ref{fig:gm1_dtov_mrho} shows that the deformation parameter affects both the global observables and the internal conditions under which the stellar sequence reaches its limiting configuration.

\begin{figure}[h]
    \centering
    \includegraphics[width=0.8\textwidth]{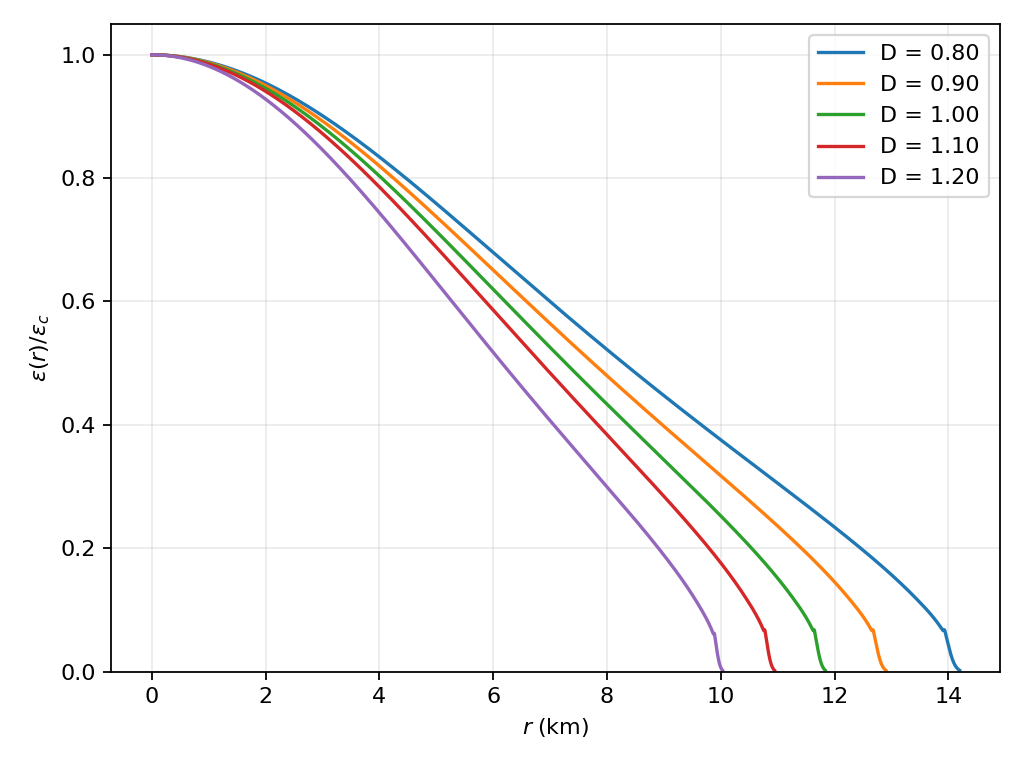}
    \caption{Normalized energy-density profiles for the maximum-mass configurations obtained with the GM1 equation of state in the framework of the \(\mathcal{D}\)-TOV formalism, for different values of the deformation parameter \(\mathcal{D}\).}
    \label{fig:gm1_dtov_rho_profile}
\end{figure}

Figure \ref{fig:gm1_dtov_rho_profile} shows the normalized energy-density profiles corresponding to the maximum-mass configurations obtained. Since the density is normalized by its central value, all curves start from the same point at the stellar center, which makes it possible to compare directly how the internal matter distribution changes along the radius for each configuration.

A clear trend can be identified in the figure. For smaller values of $\mathcal{D}$, the density profile extends to larger radii and decreases more gradually from the center to the surface. In contrast, larger values of $\mathcal{D}$ produce profiles that are more radially compact and exhibit a steeper decay of the energy density. This indicates that the deformation parameter affects not only the global mass and radius of the equilibrium sequence, but also the way in which matter is distributed inside the star.

Therefore, Fig. \ref{fig:gm1_dtov_rho_profile} reinforces the interpretation obtained from the mass-radius analysis: configurations with smaller $\mathcal{D}$ are associated with more extended stars, whereas larger values of $\mathcal{D}$ lead to more compact structures. In this sense, the deformation parameter modifies the macroscopic stellar structure by changing how the same hadronic EoS is realized in the internal density profile of the equilibrium configuration.

\begin{figure}[h]
    \centering
    \includegraphics[width=0.8\textwidth]{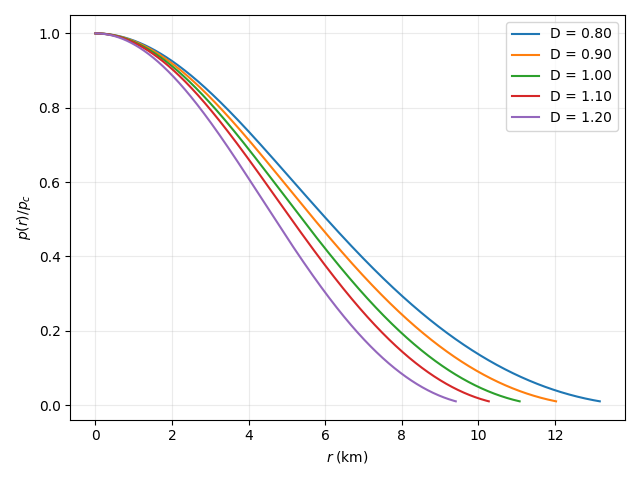}
    \caption{Normalized pressure profiles for the maximum-mass configurations obtained with the GM1 equation of state in the framework of the \(\mathcal{D}\)-TOV formalism, for different values of the deformation parameter \(\mathcal{D}\).}
    \label{fig:gm1_dtov_p_profile}
\end{figure}

Figure \ref{fig:gm1_dtov_p_profile} shows the normalized pressure profiles for the maximum-mass configurations. The behavior displayed in this Figure is consistent with the qualitative interpretation discussed in Sec.~\ref{sec:physical_interpretation_D}. As argued there, the deformation parameter $\mathcal{D}$ modifies the effective hydrostatic balance through the factor $a(r)^{\mathcal{D}}$, which changes the radial decay of the pressure. In this sense, the numerical profiles shown here provide a concrete illustration of that general picture: smaller values of $\mathcal{D}$ are associated with more slowly decreasing pressure profiles, whereas larger values of $\mathcal{D}$ lead to a steeper pressure decay and more compact configurations.


A global interpretation of the mass-radius curves is: at the low-mass end of the sequence, gravity is relatively weak, so the equilibrium configurations are more extended and therefore characterized by larger radii. As the central density increases, the stellar mass grows and the competition between self-gravity and pressure support becomes more pronounced. In the intermediate-mass regime, this competition may lead to a relatively mild variation of the radius and, depending on the stiffness of the EoS, even to a slight increase in radius over part of the sequence. For the GM1 parametrization, this behavior reflects the ability of dense hadronic matter to provide substantial pressure support as the central density rises.

At higher masses, however, gravitational compression becomes increasingly dominant, and the radius starts to decrease more clearly as the sequence approaches its limiting configuration. In this regime, the increase in central density is no longer compensated by pressure support in the same efficient way, so the star becomes progressively more compact. The physically relevant onset of instability is associated not simply with the decrease of the radius, but with the approach to the maximum-mass point, beyond which the sequence enters the branch usually interpreted as unstable against radial perturbations.

\subsection{MIT Bag Model with massless quarks}

To describe self-bound quark stars, we consider the MIT Bag Model in the massless quark approximation. In this case, the EoS is introduced analytically through the relation
\begin{equation}
\varepsilon = 3p + 4B,
\end{equation}
where $B$ is the bag constant. In the present implementation, we adopt $B = 60\,\mathrm{MeV/fm^3}$ \citep{moraes2016stellar}. This choice lies within the standard range commonly considered in MIT Bag Model applications to strange stars and corresponds to \(B^{1/4}\simeq146~\mathrm{MeV}\). 

Since this EoS is analytic, the stellar sequence is constructed by varying the central pressure over the interval
\[
p_c \in [10^{-4}B,\,25B].
\]
For the value \(B=60~\mathrm{MeV/fm^3}\) adopted here, this corresponds to
\[
p_c \in [0.006,\,1500]~\mathrm{MeV/fm^3}.
\]
Using the massless MIT Bag Model relation \(\varepsilon=3p+4B\), the corresponding
central energy-density interval is
\[
\varepsilon_c \in [240.018,\,4740]~\mathrm{MeV/fm^3},
\]
or, equivalently,
\[
\varepsilon_c \in [3.18\times10^{-4},\,6.27\times10^{-3}]~\mathrm{km}^{-2}
\]
in geometrized units. The numerical integration follows the same general strategy
adopted in the hadronic case, with the stellar structure being obtained through the outward integration of the \(\mathcal{D}\)-TOV equations for different values of the central pressure and deformation parameter.


\begin{figure}[h]
    \centering
    \includegraphics[width=0.8\textwidth]{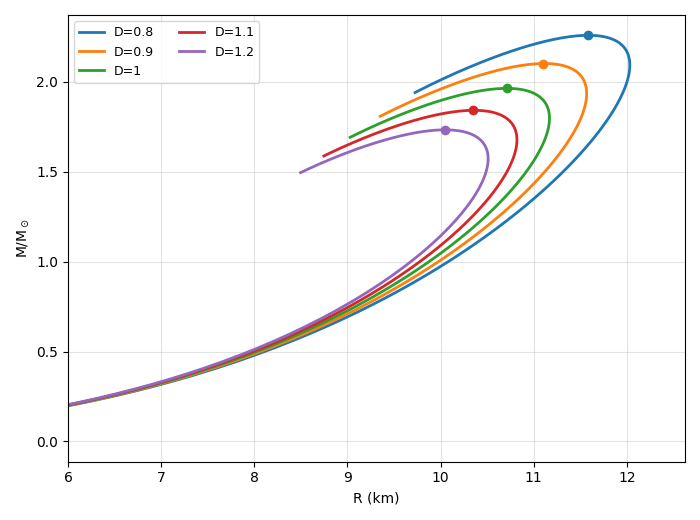}
    \caption{Mass--radius relations obtained from the $\mathcal{D}$-TOV equations in General Relativity using the massless MIT Bag Model, for different values of the deformation parameter $\mathcal{D}$. The markers indicate the maximum-mass configuration of each sequence.}
    \label{fig:mit_dtov_mr}
\end{figure}

Figure \ref{fig:mit_dtov_mr} shows the mass-radius relations obtained from the $\mathcal{D}$-TOV equations in GR using the massless MIT Bag Model for different values of the deformation parameter $\mathcal{D}$. As in the hadronic case, the standard spherical limit is recovered for $\mathcal{D}=1$, while values different from unity modify both the maximum mass and the typical stellar radius.

A systematic trend is again observed. Smaller values of $\mathcal{D}$ shift the mass-radius sequence toward larger masses and larger radii, whereas larger values of $\mathcal{D}$ move the curves toward lower masses and smaller radii. The maximum mass decreases monotonically as $\mathcal{D}$ increases, indicating that the deformation parameter has a direct impact on the range of equilibrium configurations supported by the quark-matter EoS.

An important aspect of Fig. \ref{fig:mit_dtov_mr}, however, is that the overall shape of the sequence differs from that found for the GM1 EoS. In the MIT case, the stellar radius increases over a substantial portion of the stable branch as the mass grows, which is a characteristic feature of self-bound quark stars. In the low-mass regime, this behavior is consistent with the approximate relation $M \propto R^3$, expected for self-bound configurations with nearly uniform density. This reflects the fact that, unlike ordinary hadronic stars, strange stars described by the MIT Bag Model are bound not only by gravity but also by the strong interaction encoded effectively in the bag description. Therefore, the figure illustrates both the effect of the deformation parameter and the distinct structural behavior associated with self-bound quark matter.

\begin{figure}[h]
    \centering
    \includegraphics[width=0.8\textwidth]{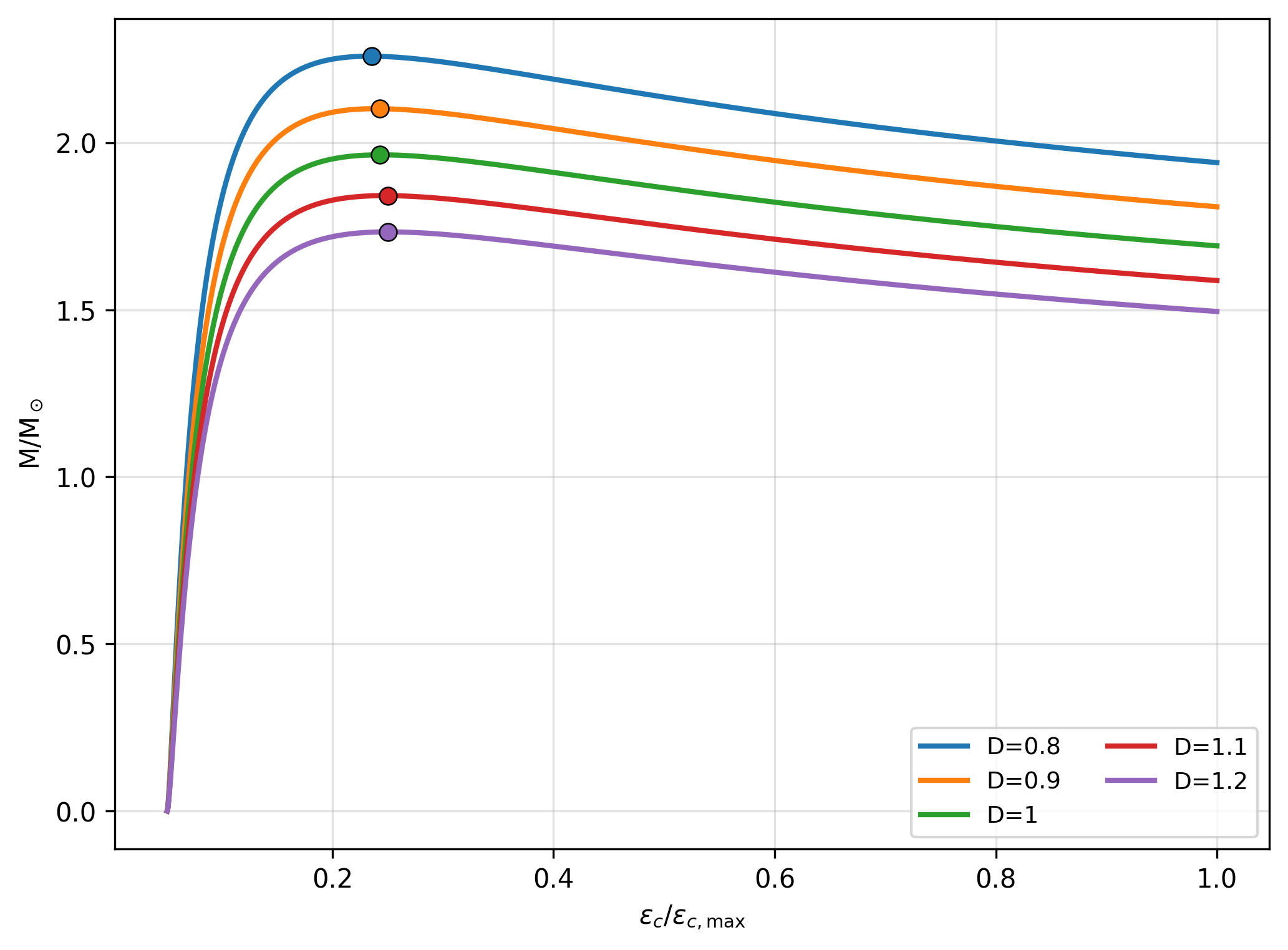}
    \caption{Stellar mass as a function of the normalized central energy density for the massless MIT Bag Model in the framework of the \(\mathcal{D}\)-TOV formalism. The normalization is performed with respect to the maximum central energy density adopted in each sequence.}
    \label{fig:mit_dtov_mrho}
\end{figure}

Figure \ref{fig:mit_dtov_mrho} shows the stellar mass as a function of the normalized central energy density. As in the hadronic case, all sequences display the same qualitative pattern: the mass increases with the central energy density, reaches a maximum value, and then undergoes a gradual decrease as the central density is further increased. This behavior is characteristic of relativistic equilibrium sequences and identifies the limiting configuration associated with the turning point of each branch.

\begin{figure}[h]
    \centering
    \includegraphics[width=0.8\textwidth]{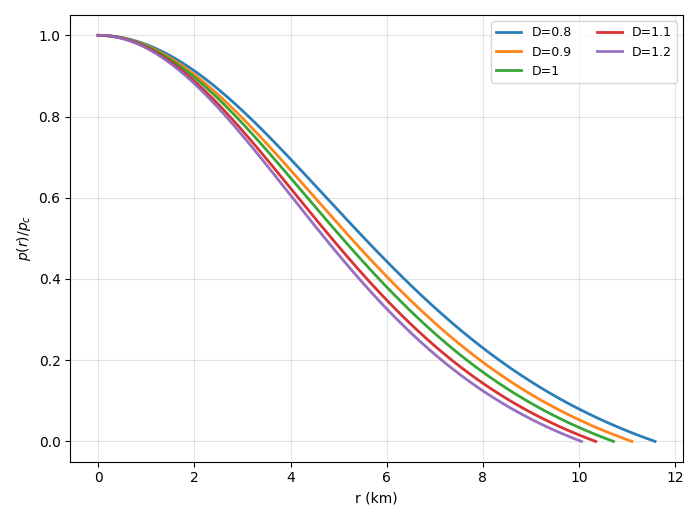}
    \caption{Normalized pressure profiles for the maximum-mass configurations obtained with the massless MIT Bag Model in the framework of the \(\mathcal{D}\)-TOV formalism, for different values of the deformation parameter \(\mathcal{D}\).}
    \label{fig:mit_dtov_p_profile}
\end{figure}

Figure \ref{fig:mit_dtov_p_profile} presents the normalized pressure profiles for the maximum-mass configurations.

Since the pressure is normalized by the central value of each configuration, the figure emphasizes the relative radial decay of the pressure support. It is seen that smaller values of $\mathcal{D}$ allow the pressure to remain significant over a larger radial interval, while larger values of $\mathcal{D}$ produce a more concentrated pressure distribution and smaller radii. 

\begin{figure}[h]
    \centering
    \includegraphics[width=0.8\textwidth]{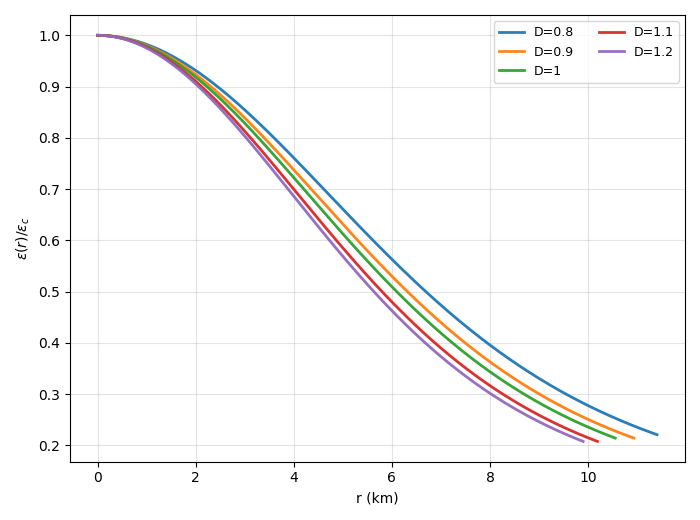}
    \caption{Normalized energy-density profiles for the maximum-mass configurations obtained with the massless MIT Bag Model in the framework of the \(\mathcal{D}\)-TOV formalism, for different values of the deformation parameter \(\mathcal{D}\).}
    \label{fig:mit_dtov_rho_profile}
\end{figure}


Figure \ref{fig:mit_dtov_rho_profile} shows the normalized energy-density profiles for the maximum-mass configurations. A particularly important feature is that the energy density does not vanish at the stellar surface. Instead, the profiles terminate at a finite value of $\varepsilon(r)/\varepsilon_c$, which is a characteristic signature of self-bound quark stars. This behavior follows directly from the MIT Bag Model EoS,
\begin{equation}
\varepsilon = 3p + 4B,
\end{equation}
since at the surface the pressure goes to zero while the energy density remains finite, $\varepsilon_{\rm surf}=4B$. Therefore, unlike ordinary hadronic stars, strange stars described by the MIT Bag Model possess a sharp surface with nonzero density.


\section{Neutron stars in optimized $f(R,T)$ gravity}

In this section we present the equilibrium solutions obtained in the optimized
\(f(R,T)=R+f(T)\) gravity introduced in Chapter~4. In this model, the trace sector is
described by the functional form
\begin{equation}
f(T)=\alpha T^2 + A\tanh[\lambda(T+T_0)] + \beta T + \gamma,
\end{equation}
whose behavior was originally reconstructed in a cosmological context through Gaussian
Processes applied to observational measurements of the Hubble parameter, and subsequently
represented by the analytic form above \citep{fortunato2024search}. 

Since \(f(R,T)\) must have the same physical dimension as the Ricci scalar, the constants
appearing in \(f(T)\) are understood to carry the appropriate dimensions so that each term
in the functional has dimension \(L^{-2}\) in geometrized units. In particular, the
combination \(\lambda(T+T_0)\) is dimensionless, as required by the argument of the
hyperbolic tangent.

In the stellar application considered here, the same functional structure is retained,
but the parameters are treated phenomenologically and adjusted in order to produce
physically acceptable neutron-star equilibrium configurations. This step is necessary
because the original reconstruction was performed in a cosmological setting, while the
interior of compact stars probes a very different curvature and density regime.

A comparison between the cosmological and stellar parameter sets is shown in Table \ref{tab:optimized_frt_parameters}. An important outcome of this analysis is that the parameters $\alpha$, $\lambda$, and $T_0$ can be kept fixed at their cosmological values, whereas $A$, $\beta$, and $\gamma$ must be strongly suppressed in the compact-star regime. This result suggests that, although the functional form itself can be extended from cosmology to stellar structure, part of its parameter content may depend on the physical scale under consideration.

\begin{table}[h]
\centering
\caption{Parameters of the optimized \(f(T)\) functional adopted in this thesis. The
cosmological values correspond to the parameters reconstructed by Fortunato et al. through
Gaussian Processes, while the stellar values correspond to the phenomenological
neutron-star application developed in \cite{quartuccio2025equilibrium}. The parameters
are understood to carry the appropriate dimensions required for each term in \(f(T)\) to
have the same dimension as the Ricci scalar.}
\label{tab:optimized_frt_parameters}
\small
\begin{tabular}{ccc}
\hline
Parameter & Cosmological value & Stellar value \\
\hline
\(\alpha\)     & \(-1.83\times10^{-5}\) & \(-1.83\times10^{-5}\) \\
\(A\)          & \(-1.05\times10^{4}\)  & \(5.0\times10^{-4}\) \\
\(\lambda\)  & \(-2.39\times10^{3}\)  & \(-2.39\times10^{3}\) \\
\(T_0\)        & \(2.58\times10^{3}\)   & \(2.58\times10^{3}\) \\
\(\beta\)      & \(-2.99\)              & \(1.05\times10^{-3}\) \\
\(\gamma\)     & \(-1.61\times10^{4}\)  & \(1.05\times10^{-3}\) \\
\hline
\end{tabular}
\end{table}

The analysis presented in this section, following the work of \cite{quartuccio2025equilibrium}, utilizes a polytropic EoS as a controlled baseline to investigate the stellar effects of the optimized \(f(R,T)\) functional. The polytropic relation is written as
\begin{equation}
p = K \rho_b^\Gamma,
\end{equation}
where \(K\) is the polytropic constant, \(\Gamma\) is the adiabatic index, and \(\rho_b\) denotes the baryonic, or rest-mass, density. This auxiliary density should not be confused with the total energy density \(\varepsilon\), which is the quantity entering the structure equations.

For the relativistic polytropic model adopted here, the total energy density is obtained from the EoS through
\begin{equation}
\varepsilon = \rho_b + \frac{p}{\Gamma - 1}.
\end{equation}
Thus, although the polytropic EoS is parametrized in terms of \(\rho_b\), the hydrostatic equilibrium equations in both GR and optimized \(f(R,T)\) gravity are integrated using \(\varepsilon\). In the main set of results presented below, we adopt \(K=100\) in geometrized units, and \(\Gamma=2\), while later in this section we also discuss the effect of varying \(\Gamma\) on the resulting equilibrium configurations.

The stellar equilibrium configurations are then obtained by numerically integrating the modified hydrostatic equilibrium equation, Eq.~(4.21), together with the associated modified mass equation, Eq.~(4.23).

\begin{figure}[h]
    \centering
    \includegraphics[width=0.8\textwidth]{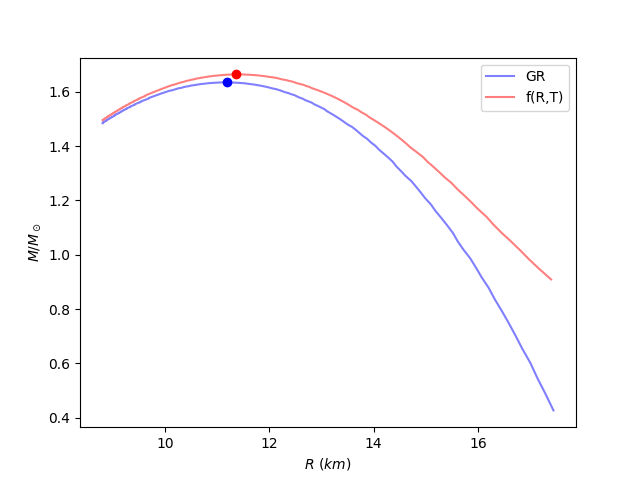}
    \caption{Mass--radius relation for neutron stars in General Relativity and optimized \(f(R,T)\) gravity. The markers indicate the maximum-mass configuration in each case.}
    \label{fig:optimized_mr}
\end{figure}

Figure \ref{fig:optimized_mr} shows the mass-radius relation obtained for NSs in GR and in the optimized \(f(R,T)\) gravity model. In both cases, the usual relativistic behavior is recovered: the stellar mass increases along the stable branch until a maximum value is reached, after which the sequence bends toward the decreasing branch associated with the onset of instability.

A direct comparison between the two curves reveals that the optimized $f(R,T)$ model predicts a slightly higher maximum mass than GR. In addition, for a comparable mass range, the modified-gravity sequence tends to yield somewhat larger radii. This indicates that the trace-dependent corrections introduced by the optimized functional modify the balance between gravitational compression and internal pressure support, allowing equilibrium configurations that are slightly more massive and more extended than their GR counterparts.

\begin{figure}[h]
    \centering
    \includegraphics[width=0.8\textwidth]{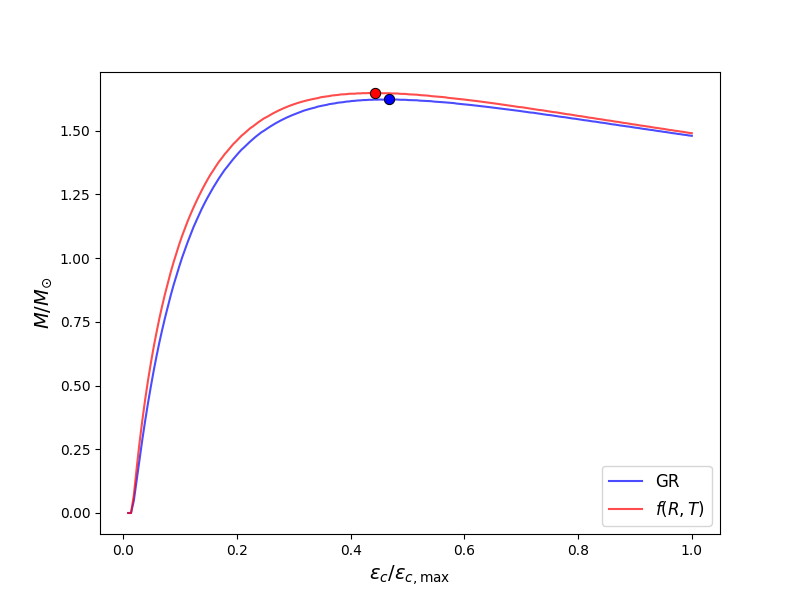}
    \caption{Stellar mass as a function of the normalized central energy density for neutron stars in General Relativity and optimized \(f(R,T)\) gravity.}
    \label{fig:optimized_mrho}
\end{figure}


Figure \ref{fig:optimized_mrho} shows the stellar mass as a function of the normalized central energy density. In both cases, the same qualitative behavior is recovered: the mass increases with the central density, reaches a maximum value, and then decreases along the branch usually associated with the onset of instability.


Therefore, the combined analysis of Figs. \ref{fig:optimized_mr} and \ref{fig:optimized_mrho} indicates that the optimized \(f(R,T)\) functional alters both the global observables and the central conditions associated with NS equilibrium. While the overall structure of the relativistic sequence is preserved, the modified theory allows slightly more massive configurations to be sustained under less extreme central conditions than in the GR case.

\begin{figure}[h]
    \centering
    \includegraphics[width=0.8\textwidth]{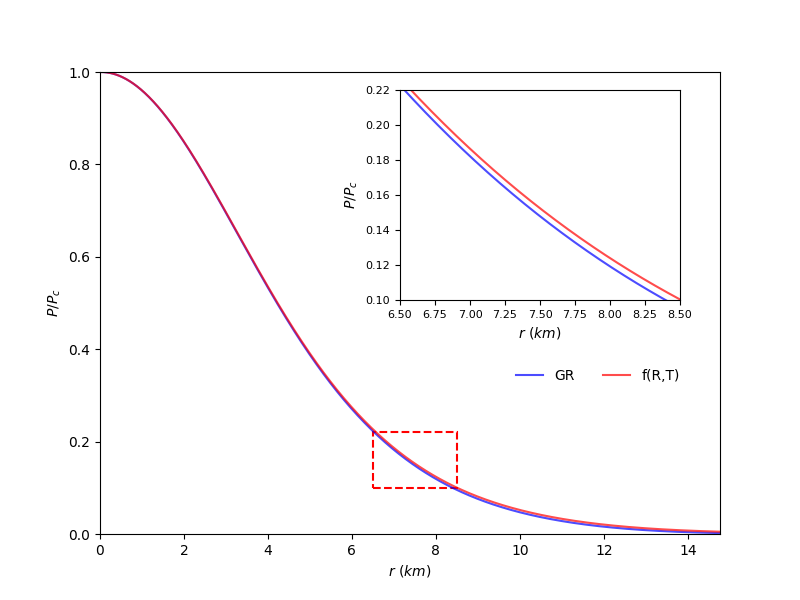}
    \caption{Normalized pressure profiles for neutron stars in General Relativity and optimized \(f(R,T)\) gravity. The inset highlights the small differences between the two curves in the outer layers of the star.}
    \label{fig:optimized_pressure}
\end{figure}

Figure \ref{fig:optimized_pressure} shows the normalized pressure profiles. The pressure decreases smoothly from the stellar center to the surface, as expected for physically acceptable equilibrium configurations. The two curves are very close to each other over the entire radial range, indicating that the optimized functional preserves the general internal structure predicted by GR.

A more careful inspection, however, reveals a small but systematic difference between the two profiles. As highlighted in the inset of Fig. \ref{fig:optimized_pressure}, the pressure in the optimized $f(R,T)$ model remains slightly higher than in GR over part of the outer region of the star. This indicates that the modified theory provides a modest enhancement of the pressure support against gravitational compression.

\begin{figure}[h]
    \centering
    \includegraphics[width=0.8\textwidth]{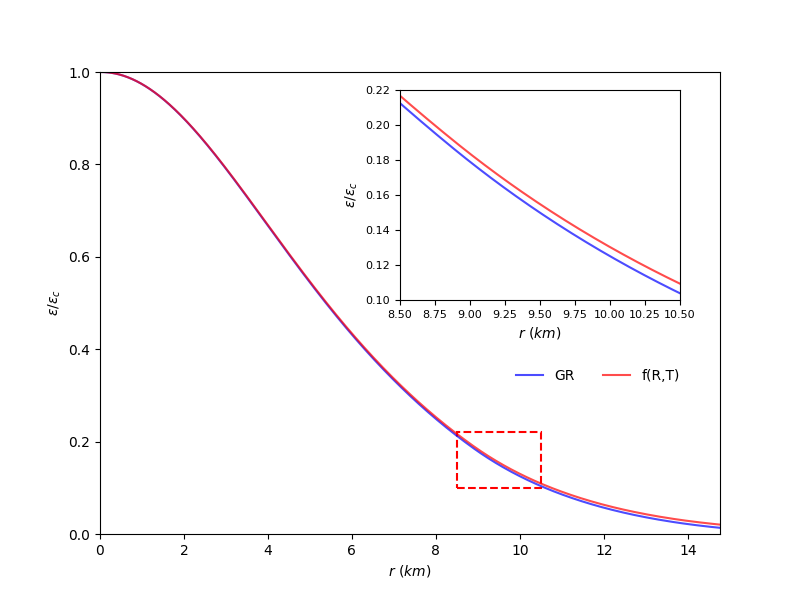}
    \caption{Normalized energy-density profiles for neutron stars in General Relativity and optimized \(f(R,T)\) gravity. The inset highlights the small differences between the two curves in the outer layers of the star.}
    \label{fig:optimized_density}
\end{figure}

In Figure \ref{fig:optimized_density} we have the normalized energy-density. The two curves are very similar, showing that the modified theory preserves the general density stratification of the stellar interior. Nevertheless, the inset makes clear that the $f(R,T)$ profile remains slightly above the GR one in part of the outer region, indicating a somewhat slower radial decay of the density. This suggests that the modified theory produces a slightly more extended matter distribution, which is consistent with the larger radii obtained in the mass-radius relation.

\begin{figure}[h]
    \centering
    \includegraphics[width=0.8\textwidth]{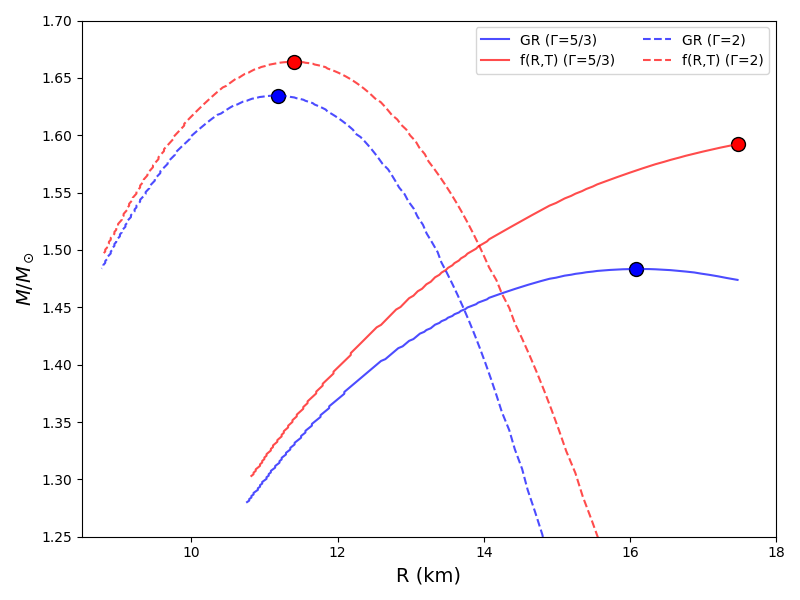}
    \caption{Mass--radius relations for neutron stars in General Relativity and optimized \(f(R,T)\) gravity for two values of the adiabatic index, \(\Gamma=2\) and \(\Gamma=5/3\). The markers indicate the maximum-mass configuration of each sequence.}
    \label{fig:gamma_comparison}
\end{figure}

The main results discussed so far were obtained for $\Gamma=2$, which corresponds to a relatively stiff fluid and leads to equilibrium configurations more compatible with the typical NS range. For comparison, Fig. \ref{fig:gamma_comparison} also shows the results obtained for $\Gamma=5/3$, corresponding to a softer fluid. In this case, the stellar sequences display lower maximum masses and noticeably larger radii.

This comparison indicates that, within the present polytropic description, the choice $\Gamma=2$ provides configurations that are more representative of NSs than those obtained with $\Gamma=5/3$. Therefore, the analysis makes clear that the stellar response depends not only on the underlying gravitational theory, but also on the effective stiffness of the matter sector.

\section{Deformed compact stars in $f(R,T)$ gravity}

In this section, we present, in a unified way, the two main structures explored in this
work to investigate massive compact stars: stellar deformation and modified gravity. The usual TOV equation provides a realistic and well-established description of static, isotropic, spherically symmetric compact stars in GR. However, astrophysical compact objects may in general depart from exact spherical symmetry due to effects such as rotation, magnetic fields, anisotropic stresses, or other internal mechanisms.

Since most TOV-like constructions in modified gravity are developed under the assumption
of spherical symmetry, the results presented here aim to go beyond this idealization by
effectively incorporating deviations from sphericity. In this sense, the goal is not to
replace the standard TOV framework, but to extend it phenomenologically in order to
explore how nonspherical effects, together with matter--geometry coupling, may influence
the equilibrium structure and maximum mass of compact objects.


The solutions discussed in this section follow the same strategy adopted for deformed compact objects in GR. The essential difference is that, in the present case, both the hydrostatic equilibrium equation and the mass equation defining the background configuration are modified, according to Eqs. (4.30) and (4.34). Within this framework, we again analyze the three equations of state considered in the previous sections — the polytropic model, the MIT Bag Model, and the GM1 parametrization — in order to investigate how the combined action of geometric deformation and $f(R,T)$ gravity affects the internal structure and global observables of compact stars.

In addition, the functional adopted in this section corresponds to the linear
trace-dependent case, \(f(R,T)=R+2\lambda T\), which, due to its mathematical simplicity
and phenomenological relevance, has been widely employed in the study of compact objects
in modified gravity \citep{moraes2016stellar}. In this model, \(\lambda\) is a coupling
constant that controls the strength of the correction associated with the explicit
dependence on the trace \(T\) of the energy-momentum tensor, thus allowing a controlled
investigation of matter--geometry interaction effects on stellar structure. Since both
\(R\) and \(T\) have dimensions of \(L^{-2}\) in geometrized units, the parameter
\(\lambda\) is dimensionless in the convention adopted here. Equivalently,
\(f_T=2\lambda\) acts as a dimensionless measure of the matter--geometry coupling in the
field equations.

\subsection{Polytropic equation of state}

For the NS configurations discussed in this subsection, we adopt a rest-mass polytropic EoS of the form $p = K \rho_b^\Gamma$, where $\rho_b$ is the baryonic (rest-mass) density. Following the setup implemented in this work, we take $\Gamma=2$ and \(K=150\,\mathrm{km}^{2}\). The stellar structure equations, however, are integrated in terms of the total energy density $\varepsilon$, obtained from the EoS through Eq. (5.8).

The equilibrium sequences are constructed by varying the central pressure on a logarithmic grid spanning
\(p_c \in [10^{-6},\,2\times10^{-2}]\,\mathrm{km}^{-2}\), and solving the deformed
\(f(R,T)=R+2\lambda T\) structure equations for the parameter sets
\(\mathcal{D}=0.9,1.0,1.1\) and \(\lambda=0,0.08,0.16\) \citep{moraes2016stellar}.
For each value of \(p_c\), the corresponding central energy density \(\varepsilon_c\) is
obtained from the polytropic EoS, rather than prescribed independently. Therefore, the
central-density sequences shown below are expressed in terms of the normalized energy
density \(\varepsilon_c/\varepsilon_{c,\max}\). The numerical integration is performed
through a fourth-order Runge--Kutta scheme with adaptive radial step, starting from a
small radius \(r_0\) with regular central conditions. The stellar surface is defined by
the condition \(p(R)=0\), and from the resulting solutions we extract the mass--radius
relation, the mass as a function of normalized central energy density, and the normalized
pressure and energy-density profiles.

\begin{figure}[H]
    \centering
    \includegraphics[width=0.8\textwidth]{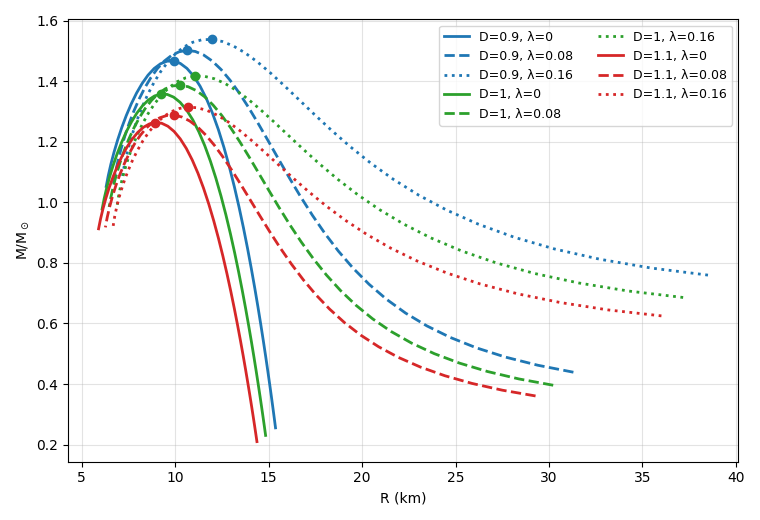}
    \caption{Mass--radius relations obtained for neutron stars described by the
    \(\mathcal{D}\)-TOV formalism in the \(f(R,T)=R+2\lambda T\) model, using the
    polytropic equation of state. The markers indicate the maximum-mass configuration
    of each sequence.}
    \label{fig:poly1}
\end{figure}

Figure~\ref{fig:poly1} shows the mass--radius relation for NSs described by the
\(\mathcal{D}\)-TOV formalism in the \(f(R,T)=R+2\lambda T\) model. The case
\(\mathcal{D}=1\) and \(\lambda=0\) recovers the usual relativistic solution for a
perfectly spherical star in GR. As discussed previously, geometric deformation modifies
the maximum mass that the star can support: values \(\mathcal{D}<1\) shift the sequences
toward higher masses, whereas values \(\mathcal{D}>1\) tend to reduce the maximum mass.

Before discussing the numerical trends, it is important to clarify the parameter space
explored in this figure. The parameter \(\mathcal{D}\) is a phenomenological measure of
global deformation, while \(\lambda\) controls the strength of the matter--geometry
coupling in the linear \(f(R,T)\) model. Since both parameters affect the hydrostatic
equilibrium, their effects are partially degenerate in the mass--radius diagram. In
particular an oblate-like deformation \((\mathcal{D}<1)\) and a positive coupling \((\lambda>0)\) both tend to shift the equilibrium sequences toward larger maximum masses. Consequently, different combinations of \(\mathcal{D}\) and \(\lambda\) may lead to similar changes in the global stellar properties.

For this reason, the values adopted here should be understood as an exploratory and
controlled subset of the parameter space, not as observationally inferred bounds. The
deformation parameter is restricted to moderate deviations around the spherical limit,
\(\mathcal{D}=0.9,1.0,1.1\). Similarly, the coupling parameter is restricted to positive
values \(\lambda=0,0.08,0.16\). Since, for the model \(f(R,T)=R+2\lambda T\), one has
\(f_T=2\lambda\), the largest value considered here corresponds to \(f_T=0.32\), which is
still small compared with the standard gravitational coupling scale \(8\pi\) in
geometrized units. Thus, the adopted range keeps the matter--geometry coupling moderate
while allowing its qualitative effects on the stellar structure to be identified.

A complete determination of the allowed ranges of \(\mathcal{D}\) and \(\lambda\) would
require additional physical and observational constraints, such as radial stability,
causality, tidal deformability, rotation, magnetic-field effects, and a consistent
treatment of the exterior spacetime for genuinely deformed configurations. Such a
complete parameter-space analysis lies beyond the scope of the present work. Here, the
goal is more limited: to investigate how the equilibrium sequences respond when geometric
deformation and matter--geometry coupling are varied in a controlled way.

When the matter--geometry coupling is switched on, that is, for \(\lambda\neq0\), an
additional displacement of the sequences toward higher maximum masses and larger
equatorial radii is observed. For each fixed value of \(\mathcal{D}\), increasing
\(\lambda\) systematically leads to more massive equilibrium configurations. Therefore,
Fig.~\ref{fig:poly1} shows that deformation and modified gravity may act jointly on the
stellar structure: for oblate-like configurations with \(\mathcal{D}<1\), the presence of
\(\lambda>0\) further enhances the increase in the maximum supported mass.

At the same time, the figure also illustrates the partial degeneracy between the two
parameters. A change in \(\lambda\) can partially mimic the effect of changing
\(\mathcal{D}\), and vice versa. Consequently, mass--radius curves alone are not
sufficient to uniquely separate the effect of geometric deformation from the effect of
the \(f(R,T)\) matter--geometry coupling. Table~\ref{tab:poly_maxima_frt} quantifies the
trends displayed in Fig.~\ref{fig:poly1}.

\begin{table}[H]
\centering
\caption{Maximum-mass configurations for the polytropic equation of state in the \(\mathcal{D}\)-TOV formalism with \(f(R,T)=R+2\lambda T\).}
\label{tab:poly_maxima_frt}
\begin{tabular}{cccc}
\hline
\(\mathcal{D}\) & \(\lambda\) & \(M_{\max}\,(M_\odot)\) & \(R_{\rm eq}\,(\mathrm{km})\) \\
\hline
0.9 & 0.00    & 1.467 & 9.92 \\
0.9 & 0.08 & 1.502 & 10.64 \\
0.9 & 0.16 & 1.538 & 11.97 \\
1.0 & 0.00    & 1.358 & 9.23 \\
1.0 & 0.08 & 1.387 & 10.27 \\
1.0 & 0.16 & 1.417 & 11.04 \\
1.1 & 0.00    & 1.264 & 8.94 \\
1.1 & 0.08 & 1.288 & 9.95 \\
1.1 & 0.16 & 1.314 & 10.70 \\
\hline
\end{tabular}
\end{table}

\begin{figure}[H]
    \centering
    \includegraphics[width=0.8\textwidth]{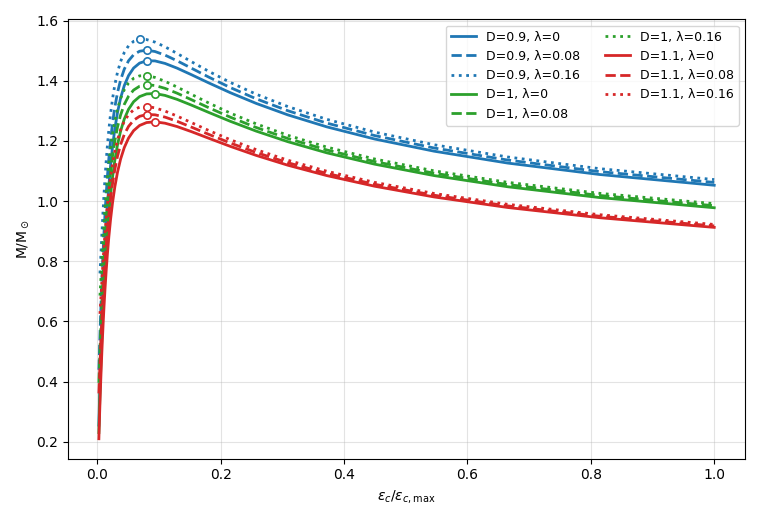}
    \caption{Stellar mass as a function of the normalized central energy density for neutron stars described by the \(\mathcal{D}\)-TOV formalism in the \(f(R,T)=R+2\lambda T\) model, using the polytropic equation of state.}
    \label{fig:poly2}
\end{figure}

\begin{figure}[H]
    \centering
    \includegraphics[width=0.8\textwidth]{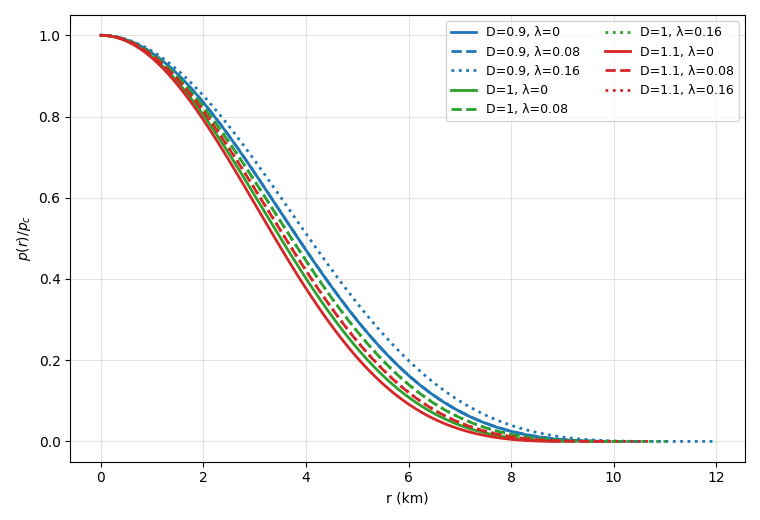}
    \caption{Normalized pressure profiles for the maximum-mass configurations obtained with the polytropic equation of state in the \(\mathcal{D}\)-TOV formalism with \(f(R,T)=R+2\lambda T\).}
    \label{fig:poly3}
\end{figure}

\begin{figure}[htbp]
    \centering
    \includegraphics[width=0.8\textwidth]{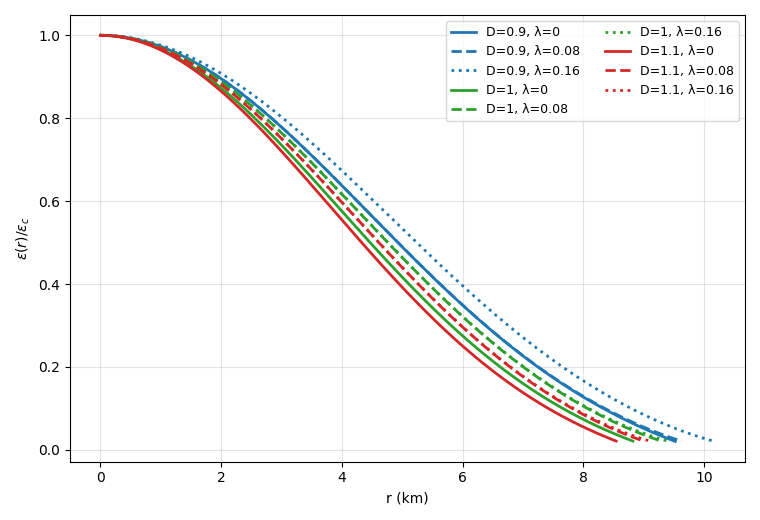}
    \caption{Normalized energy-density profiles for the maximum-mass configurations obtained with the polytropic equation of state in the \(\mathcal{D}\)-TOV formalism with \(f(R,T)=R+2\lambda T\).}
    \label{fig:poly4}
\end{figure}

Figures \ref{fig:poly2}--\ref{fig:poly4} provide the central-density sequences and the corresponding internal profiles for the same polytropic configurations discussed in Fig. \ref{fig:poly1}. Together, these results offer an internal-structure interpretation of the mass-radius behavior previously obtained.

Figure \ref{fig:poly2} shows the stellar mass as a function of the normalized central energy density. In all cases, the mass increases up to a maximum value and then decreases along the high-density branch, reproducing the usual turning-point structure of relativistic equilibrium sequences.

The pressure profiles displayed in Fig. \ref{fig:poly3} show that the configurations with $\mathcal{D}<1$ and larger $\lambda$ exhibit a slower radial decay of the normalized pressure. In other words, the hydrostatic support remains significant over a larger radial interval, which is consistent with the larger equatorial radii and larger maximum masses found in the mass-radius sequences. By contrast, for $\mathcal{D}>1$ and smaller $\lambda$, the pressure drops more rapidly, leading to more compact and less massive stars. A similar trend is observed in the normalized energy-density profiles shown in Fig. \ref{fig:poly4}. 

Taken together, Figs. \ref{fig:poly1}--\ref{fig:poly4} show that geometric deformation and trace-dependent modified gravity act in a consistent way within the present polytropic setup. Oblate-like configurations ($\mathcal{D}<1$) and positive matter--geometry coupling ($\lambda>0$) favor stars that are simultaneously more extended and able to sustain higher masses, while prolate-like configurations ($\mathcal{D}>1$) and weaker coupling lead to the opposite behavior.

\subsection{The MIT Bag Model solutions}

For the strange-star configurations discussed in this subsection, we adopt the MIT Bag Model with $B=60\,\mathrm{MeV/fm^3}$. In this case, the EoS can be written as
\begin{equation}
p = a(\varepsilon-4B),
\end{equation}
where $a=1/3$ in the massless quark approximation.

\begin{figure}[h]
    \centering
    \includegraphics[width=0.8\textwidth]{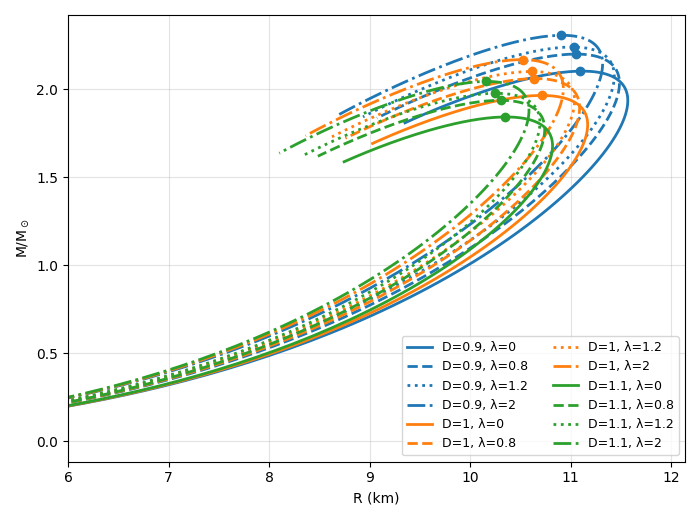}
    \caption{Mass--radius relations obtained for strange stars described by the \(\mathcal{D}\)-TOV formalism in the \(f(R,T)=R+2\lambda T\) model, using the massless MIT Bag Model with \(B=60\,\mathrm{MeV/fm^3}\). The markers indicate the maximum-mass configuration of each sequence.}
    \label{fig:mit1_frt}
\end{figure}

Figure \ref{fig:mit1_frt} shows a behavior qualitatively similar to that already observed in Fig. \ref{fig:mit_dtov_mr}, but now including the effect of the matter-geometry coupling parameter $\lambda$. In particular, the same ordering with respect to the deformation parameter is preserved: oblate-like configurations with $\mathcal{D}<1$ reach the highest masses, whereas prolate-like configurations with $\mathcal{D}>1$ lead to less massive stars.

The modified-gravity contribution introduces an additional systematic effect. For each fixed value of $\mathcal{D}$, increasing $\lambda$ shifts the sequences toward higher maximum masses. At the same time, the radius associated with the maximum-mass configuration decreases slightly. Therefore, in the MIT case, the matter-geometry coupling tends to produce more massive but also somewhat more compact limiting configurations. 

\begin{figure}[H]
    \centering
    \includegraphics[width=0.8\textwidth]{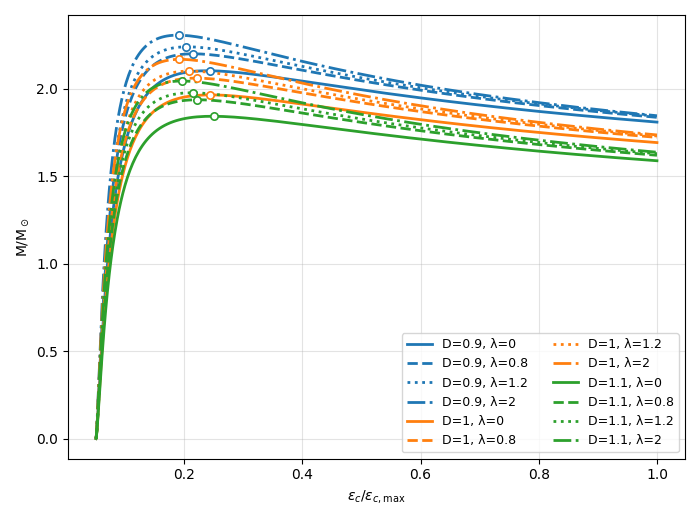}
    \caption{Stellar mass as a function of the normalized central energy density for strange stars described by the \(\mathcal{D}\)-TOV formalism in the \(f(R,T)=R+2\lambda T\) model, using the massless MIT Bag Model with \(B=60\,\mathrm{MeV/fm^3}\).}
    \label{fig:mit2_frt}
\end{figure}

\begin{figure}[H]
    \centering
    \includegraphics[width=0.8\textwidth]{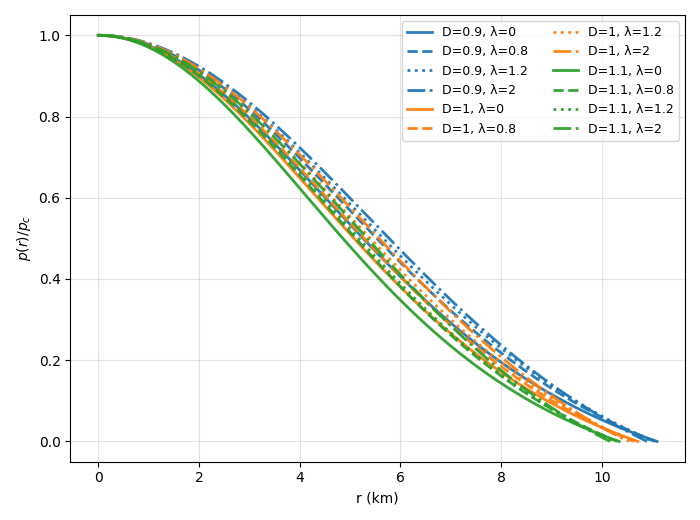}
    \caption{Normalized pressure profiles for the maximum-mass configurations obtained with the massless MIT Bag Model in the \(\mathcal{D}\)-TOV formalism with \(f(R,T)=R+2\lambda T\).}
    \label{fig:mit3_frt}
\end{figure}

\begin{figure}[H]
    \centering
    \includegraphics[width=0.8\textwidth]{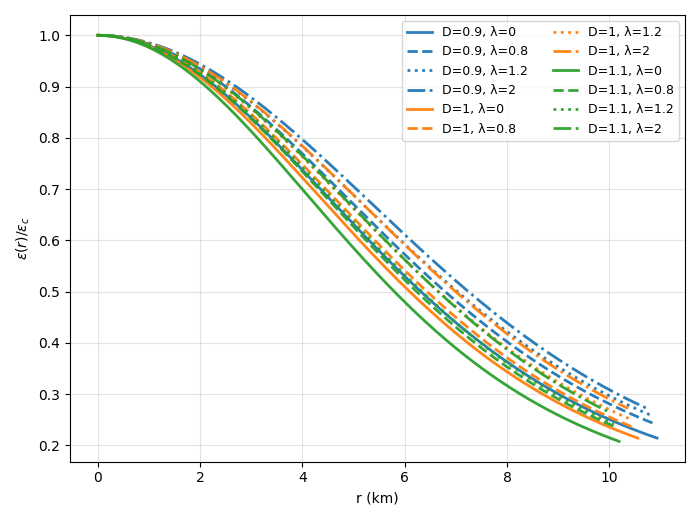}
    \caption{Normalized energy-density profiles for the maximum-mass configurations obtained with the massless MIT Bag Model in the \(\mathcal{D}\)-TOV formalism with \(f(R,T)=R+2\lambda T\).}
    \label{fig:mit4_frt}
\end{figure}

Figures \ref{fig:mit2_frt}--\ref{fig:mit4_frt} confirm, in terms of the central-density sequences and internal profiles, the same general trends already observed in the mass-radius relation of Fig. \ref{fig:mit1_frt}.

In Fig. \ref{fig:mit2_frt}, the ordering of the curves shows that, for fixed \(\mathcal{D}\), increasing \(\lambda\) systematically shifts the sequences toward higher masses, while for fixed \(\lambda\) the oblate-like branch \(\mathcal{D}<1\) remains the one associated with the largest maximum masses. As usual, the increasing branches up to the turning point correspond to the regime commonly associated with stable equilibrium configurations.

\begin{figure}[H]
    \centering
    \includegraphics[width=0.8\textwidth]{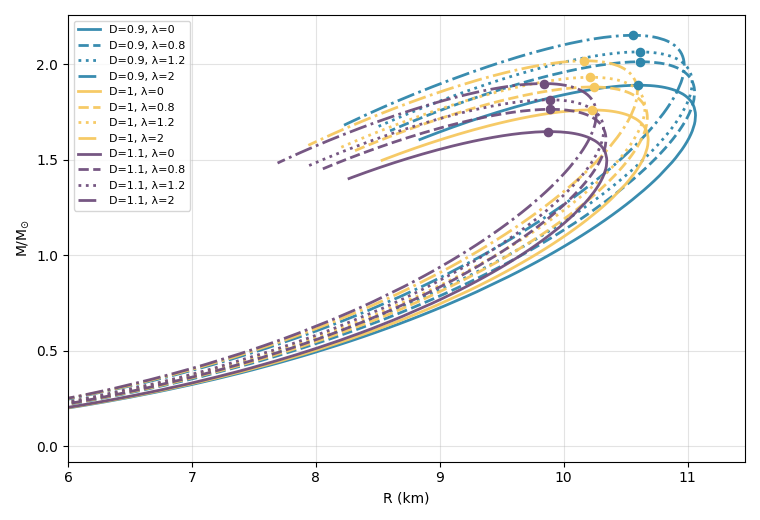}
    \caption{Mass--radius relations for strange stars in the \(\mathcal{D}\)-TOV formalism with \(f(R,T)=R+2\lambda T\), now including the finite strange-quark mass through the choice \(a=0.28\). The markers indicate the maximum-mass configuration of each sequence.}
    \label{fig:strange3}
\end{figure}

The normalized pressure profiles shown in Fig. \ref{fig:mit3_frt} indicate that the presence of the matter-geometry coupling leads to a slightly slower radial decay of the pressure, especially for larger values of \(\lambda\). This provides an internal-structure counterpart to the increase in the maximum supported mass.

A similar effect is seen in the normalized energy-density profiles of Fig. \ref{fig:mit4_frt}. The cases with larger \(\lambda\) and smaller \(\mathcal{D}\) exhibit a more extended matter distribution, with a less abrupt radial decrease of \(\varepsilon(r)/\varepsilon_c\). Therefore, even though the overall pattern remains similar to the deformed GR case, the modified-gravity contribution systematically reinforces the tendency toward more massive strange-star configurations.

Figure \ref{fig:strange3} shows the solutions obtained for the MIT Bag Model when the strange-quark mass, $m_s$, is taken into account, adopting \(a=0.28\) \citep{stergioulas2003rotating}. In comparison with the massless case, the inclusion of \(m_s\neq0\) makes the EoS softer, since the finite strange-quark mass reduces the kinetic contribution of the fluid and, consequently, the pressure available to support the star against gravitational compression. As a result, the mass-radius sequences are shifted toward lower maximum masses and, in general, slightly smaller radii.

Table \ref{tab:strange_massive} quantifies this effect by comparing the maximum-mass configurations with and without the inclusion of the strange-quark mass. It is seen that, for all values of \(\mathcal{D}\) and \(\lambda\), the case \(m_s\neq0\) yields lower maximum masses than the corresponding massless model. Even so, the overall hierarchy of the solutions is preserved: for fixed \(\lambda\), configurations with \(\mathcal{D}<1\) still produce the largest masses, while increasing \(\lambda\) continues to systematically raise the maximum mass along each branch.

\begin{table}[h]
\centering
\caption{Comparison between the maximum-mass configurations obtained for strange stars in the massless case (\(m_s=0\)) and in the case with finite strange-quark mass (\(m_s\neq0\)).}
\label{tab:strange_massive}
\begin{tabular}{cccccc}
\hline\hline
& & \multicolumn{2}{c}{\(m_s=0\)} & \multicolumn{2}{c}{\(m_s\neq0\)} \\
\cline{3-4} \cline{5-6}
\(\mathcal{D}\) & \(\lambda\) & \(M_{\max}\,(M_\odot)\) & \(R\,(\mathrm{km})\) & \(M_{\max}\,(M_\odot)\) & \(R\,(\mathrm{km})\) \\
\hline
0.9 & 0.0 & 2.102 & 11.097 & 1.885 & 10.610 \\
0.9 & 0.8 & 2.198 & 11.054 & 2.005 & 10.624 \\
0.9 & 1.2 & 2.238 & 11.029 & 2.063 & 10.632 \\
0.9 & 2.0 & 2.305 & 10.901 & 2.146 & 10.567 \\
1.0 & 0.0 & 1.964 & 10.714 & 1.749 & 10.245 \\
1.0 & 0.8 & 2.059 & 10.637 & 1.875 & 10.261 \\
1.0 & 1.2 & 2.100 & 10.613 & 1.927 & 10.220 \\
1.0 & 2.0 & 2.167 & 10.525 & 2.016 & 10.172 \\
1.1 & 0.0 & 1.842 & 10.345 & 1.644 & 9.882 \\
1.1 & 0.8 & 1.936 & 10.305 & 1.760 & 9.906 \\
1.1 & 1.2 & 1.976 & 10.246 & 1.807 & 9.898 \\
1.1 & 2.0 & 2.043 & 10.160 & 1.885 & 9.849 \\
\hline\hline
\end{tabular}
\end{table}

Table \ref{tab:strange_massive} provides a direct comparison between the maximum-mass configurations obtained in the massless case (\(m_s=0\)) and in the massive strange-quark case (\(m_s\neq0\)).

\subsection{GM1 equation of state}

\begin{figure}[H]
    \centering
    \includegraphics[width=0.8\textwidth]{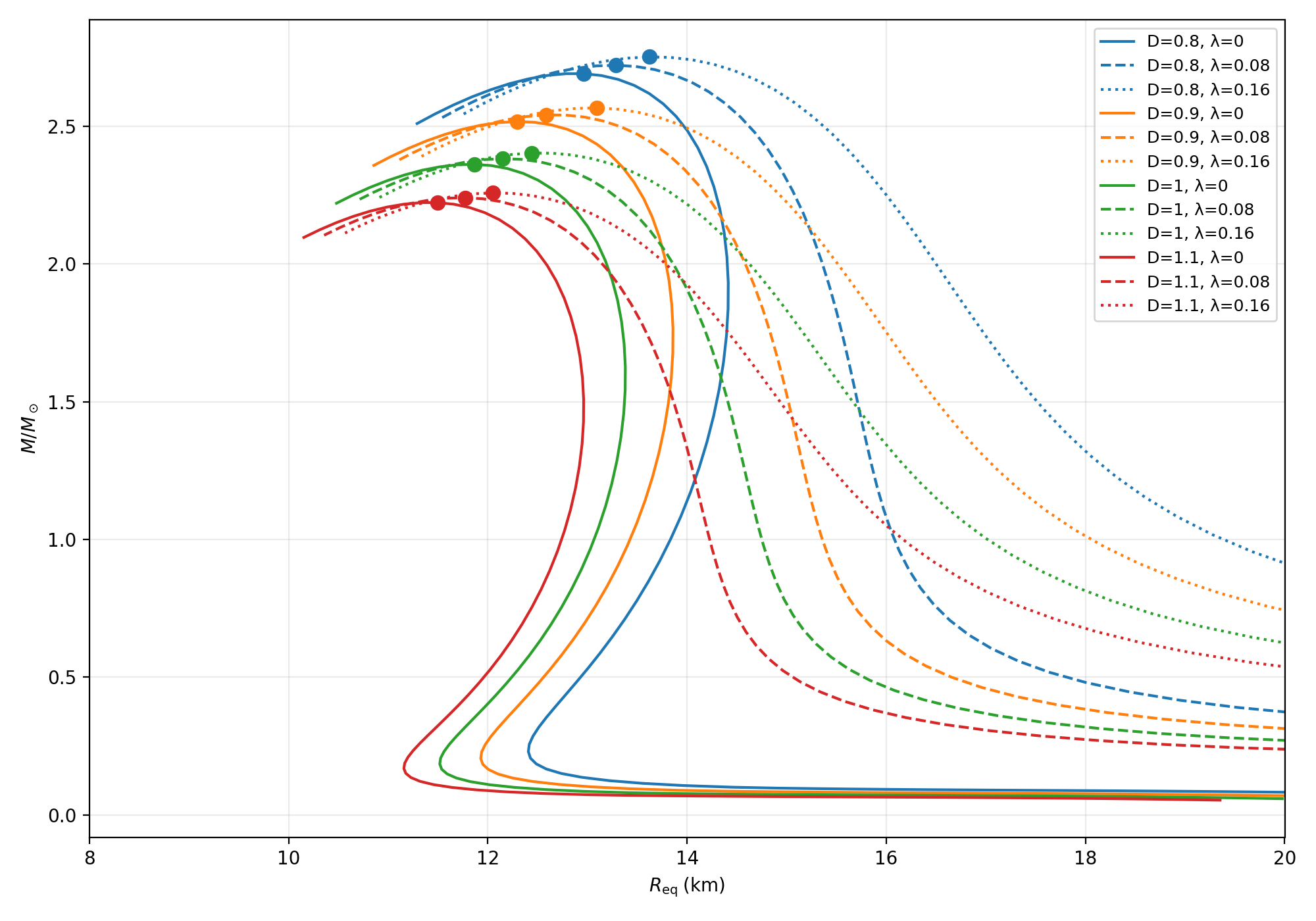}
    \caption{Mass--radius relations obtained for neutron stars described by the \(\mathcal{D}\)-TOV formalism in the \(f(R,T)=R+2\lambda T\) model, using the GM1 equation of state. The markers indicate the maximum-mass configuration of each sequence.}
    \label{fig:gm1_frt_mr}
\end{figure}

\begin{figure}[H]
    \centering
    \includegraphics[width=0.8\textwidth]{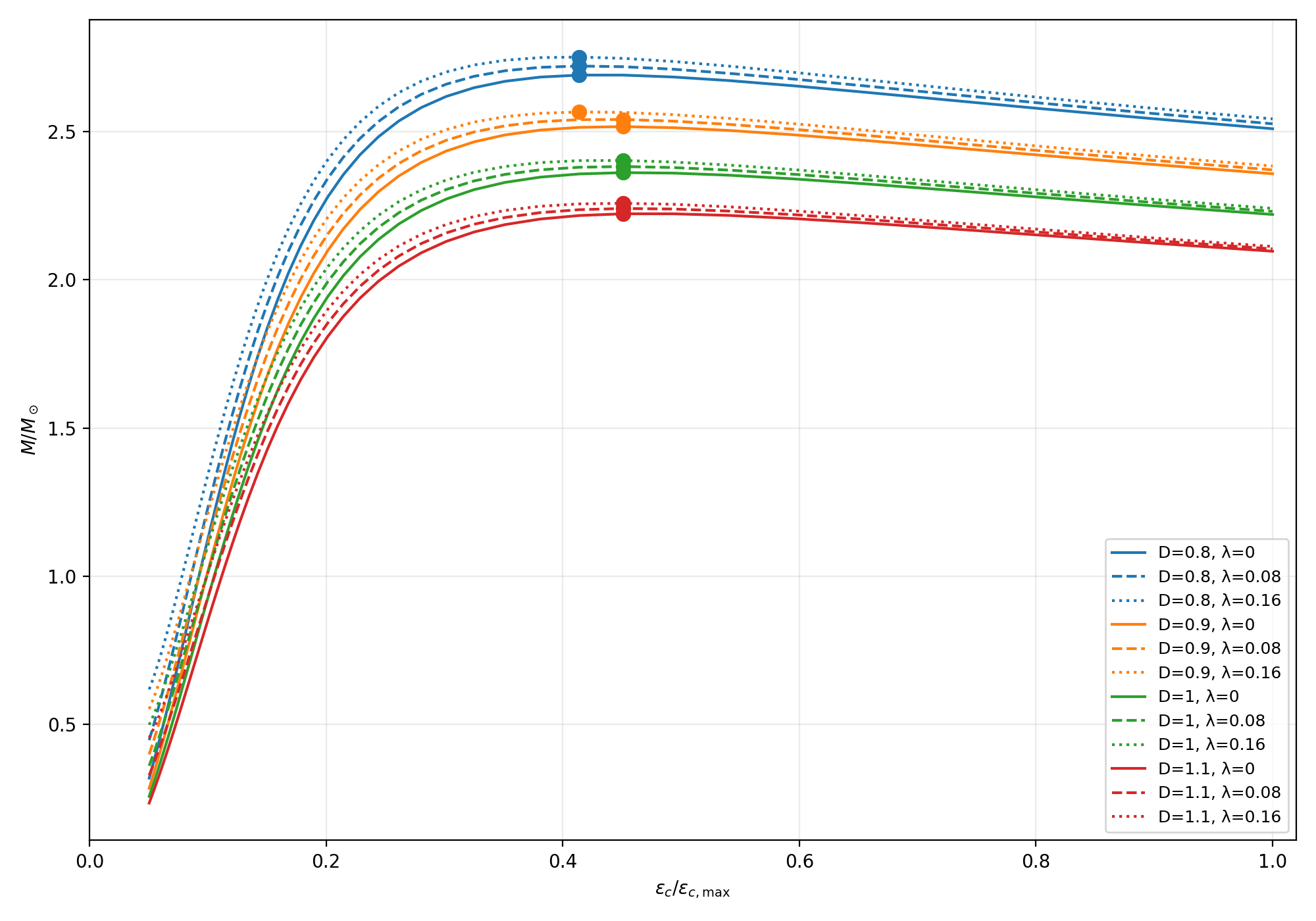}
    \caption{Stellar mass as a function of the normalized central energy density for neutron stars described by the \(\mathcal{D}\)-TOV formalism in the \(f(R,T)=R+2\lambda T\) model, using the GM1 equation of state.}
    \label{fig:gm1_frt_mrho}
\end{figure}

As an example of a realistic hadronic EoS for NSs, we present in Figs. \ref{fig:gm1_frt_mr} and \ref{fig:gm1_frt_mrho} the mass-radius and mass-central-density relations obtained with the GM1 parametrization. The overall behavior remains consistent with that found for the previous equations of state: the maximum mass is shifted toward higher values when oblate-like configurations ($\mathcal{D}<1$) and modified gravity effects ($\lambda>0$) are taken into account. In this sense, the GM1 results confirm, within a more realistic hadronic framework, the same qualitative trend already observed for the polytropic and MIT cases. The values listed in Table \ref{tab:gm1_maxima_frt} quantify the effect of the parameters $\mathcal{D}$ and $\lambda$ on the limiting equilibrium configurations. For each fixed value of $\mathcal{D}$, increasing $\lambda$ systematically raises both the maximum mass and the corresponding equatorial radius. Likewise, for fixed $\lambda$, the largest masses are always obtained in the oblate-like branch $\mathcal{D}<1$, while the prolate-like branch $\mathcal{D}>1$ leads to smaller maximum masses.

\begin{table}[H]
\centering
\caption{Maximum-mass configurations for the GM1 equation of state in the \(\mathcal{D}\)-TOV formalism with \(f(R,T)=R+2\lambda T\).}
\label{tab:gm1_maxima_frt}
\begin{tabular}{cccc}
\hline
\(\mathcal{D}\) & \(\lambda\) & \(M_{\max}\,(M_\odot)\) & \(R_{\rm eq}\,(\mathrm{km})\) \\
\hline
0.8 & 0.00    & 2.691 & 12.965 \\
0.8 & 0.08 & 2.722 & 13.288 \\
0.8 & 0.16 & 2.752 & 13.620 \\
0.9 & 0.00    & 2.517 & 12.291 \\
0.9 & 0.08 & 2.541 & 12.585 \\
0.9 & 0.16 & 2.567 & 13.096 \\
1.0 & 0.00    & 2.362 & 11.867 \\
1.0 & 0.08 & 2.383 & 12.151 \\
1.0 & 0.16 & 2.403 & 12.442 \\
1.1 & 0.00    & 2.223 & 11.496 \\
1.1 & 0.08 & 2.241 & 11.771 \\
1.1 & 0.16 & 2.259 & 12.053 \\
\hline
\end{tabular}
\end{table}

\section{Deformed hybrid stars: a preliminary study}

Hybrid stars constitute an interesting class of compact objects to be explored within the
present framework. In these configurations, a deconfined quark core is surrounded by
hadronic matter, allowing one to investigate the interplay between phase transitions,
dense-matter microphysics, and non-spherical stellar structure \citep{weber2007pulsars}. As a first exploratory result, we present a preliminary mass--radius analysis obtained by combining a GM1
hadronic phase with a quark phase described by the massless MIT Bag Model.

At the microphysical level, the hadron--quark phase transition is modeled through a
Maxwell construction. This construction is appropriate when the transition between the
hadronic and quark phases is assumed to be sharp, with local charge neutrality imposed
separately in each phase. In this case, no spatially extended mixed phase is included,
in contrast with a Gibbs construction, where global charge neutrality may allow both
phases to coexist over a finite pressure interval.

For cold matter in beta equilibrium, the relevant thermodynamic variable controlling the
phase transition is the baryon chemical potential \(\mu_B\). At zero temperature, the
thermodynamically favored phase at a given \(\mu_B\) is the one with the larger pressure.
The Maxwell transition point is therefore determined by the crossing of the hadronic and
quark pressures as functions of \(\mu_B\),
\begin{equation}
p_H(\mu_B^{\mathrm{trans}})
=
p_Q(\mu_B^{\mathrm{trans}})
\equiv
p_{\mathrm{trans}},
\label{eq:maxwell_pressure_equality}
\end{equation}
with
\begin{equation}
\mu_B^{H}
=
\mu_B^{Q}
\equiv
\mu_B^{\mathrm{trans}}.
\label{eq:maxwell_mu_equality}
\end{equation}
Equivalently, the transition occurs when the Gibbs free energy per baryon is the same in
the two phases. For cold beta-equilibrated matter, this quantity is identified with the
baryon chemical potential,
\begin{equation}
g = \mu_B = \frac{\varepsilon+p}{n_B},
\label{eq:gibbs_per_baryon}
\end{equation}
where \(n_B\) is the baryon number density. Thus, the Maxwell construction fixes a single
transition pressure \(p_{\mathrm{trans}}\). Across the phase interface, the pressure and
the baryon chemical potential are continuous, whereas the energy density is generally
discontinuous:
\begin{equation}
\Delta\varepsilon
=
\varepsilon_Q(p_{\mathrm{trans}})
-
\varepsilon_H(p_{\mathrm{trans}}).
\label{eq:energy_density_jump}
\end{equation}

In the stellar integration, this produces a hybrid barotropic EoS with a sharp transition:
for pressures below \(p_{\mathrm{trans}}\), the hadronic GM1 branch is used, while for
pressures above \(p_{\mathrm{trans}}\), the quark branch described by the massless MIT Bag
Model is adopted. The resulting EoS therefore contains an energy-density jump at constant
pressure, as expected for a first-order phase transition described by a Maxwell
construction \citep{contrera2017hybrid}.

It is important to emphasize, however, that in the present work the Maxwell construction
is performed at the microphysical EoS level and is not rederived self-consistently for
each value of the deformation parameter \(\mathcal{D}\). The parameter \(\mathcal{D}\)
enters only through the effective stellar-structure equations, modifying the global
hydrostatic equilibrium once the hybrid EoS has already been specified. Therefore, the
transition pressure, the baryon chemical potential at the transition, and the
energy-density jump are those obtained from the underlying Maxwell construction of the
hybrid EoS, not from a new deformation-dependent thermodynamic matching.

Accordingly, the analysis presented here should be understood as a preliminary and
phenomenological extension of the deformed stellar formalism to hybrid-star
configurations. A fully consistent treatment would require investigating how deformation
affects the phase equilibrium conditions, the stability of the interface, and the
matching between the internal and external geometries. Such an analysis lies beyond the
scope of the present thesis. The purpose of the present section is therefore more modest:
to provide an exploratory illustration of how a hybrid EoS, constructed through a
standard Maxwell prescription, behaves when inserted into the effective
\(\mathcal{D}\)-TOV framework.

\begin{figure}[H]
    \centering
    \includegraphics[width=0.8\textwidth]{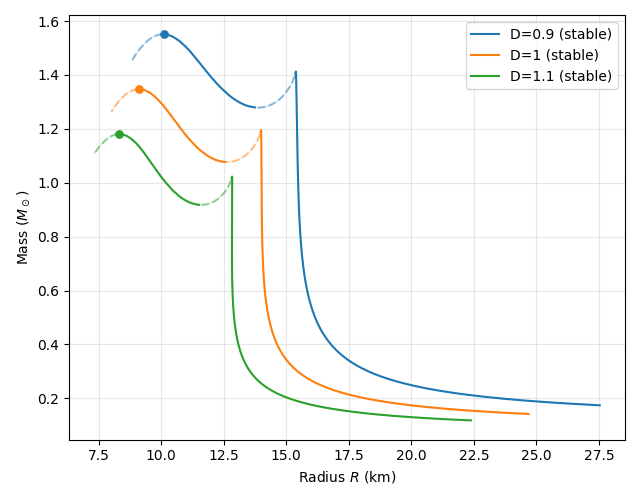}
    \caption{Preliminary mass--radius relations for hybrid stars obtained by employing a
    hybrid EoS built from a GM1 hadronic phase and a massless MIT Bag Model quark phase
    matched through a standard Maxwell construction. The resulting EoS is then used as an
    effective input in the \(\mathcal{D}\)-TOV equations for different values of the
    deformation parameter \(\mathcal{D}\). The solid segments indicate the branch
    identified as stable in the present analysis.}
    \label{fig:hybrid_mr}
\end{figure}

With this caveat in mind, the resulting mass--radius relation is shown in
Fig.~\ref{fig:hybrid_mr}. Within this preliminary treatment, deformation affects hybrid
stars in the same qualitative way found for NSs and strange stars: oblate-like
configurations are able to sustain larger masses, whereas more prolate solutions tend to
be less massive. The solid segments indicate the branch identified as stable in the
present analysis.

\section{Radial oscillations of deformed compact stars}

After constructing the equilibrium configurations, it is useful to investigate how the
deformation parameter affects the radial oscillation spectrum. Radial oscillations provide
a direct diagnostic of the internal structure and dynamical stability of compact stars,
since their eigenfrequencies depend on both the EoS and the equilibrium background
\citep{chandrasekhar1964dynamical,kokkotas2001radial,sagun2020asteroseismology}.

In the linear perturbative approach, the radial displacement and the pressure perturbation
are described by the dimensionless variables
\begin{equation}
\xi(r)=\frac{\Delta r}{r},
\qquad
\eta(r)=\frac{\Delta p}{p},
\end{equation}
so that the radial oscillation problem can be written as a first-order system. In the
present analysis, the perturbation equations are evaluated on top of the deformed
equilibrium background obtained from the \(\mathcal{D}\)-TOV formalism. Therefore, instead
of keeping the perturbation equations written only in terms of a generic radial metric
potential \(e^\lambda\), we explicitly substitute the effective radial metric component
of the \(\mathcal{D}\)-TOV background,
\begin{equation}
e^{\lambda(r)}
=
\left(1-\frac{2m(r)}{r}\right)^{-\mathcal{D}}.
\end{equation}
This prescription should be understood consistently with the effective character of the
\(\mathcal{D}\)-TOV model. It represents radial perturbations around a one-dimensional
deformed equilibrium background, rather than a full perturbative treatment of a genuinely
axisymmetric spacetime.

With these conventions, and denoting by \(\varepsilon\) the total energy density of the
stellar fluid, the system describing radial oscillations can be written as
\begin{equation}
\xi'(r)=-
\left(
\frac{3}{r}+\frac{p'}{\chi}
\right)\xi
-\frac{1}{\Gamma r}\eta,
\end{equation}
and
\begin{equation}
\begin{aligned}
\eta'(r)=\;&
\omega^2
\left[
r\left(1+\frac{\varepsilon}{p}\right)e^{-2\Phi}
\left(1-\frac{2m}{r}\right)^{-\mathcal{D}}
\right]\xi \\
&-
\left[
\frac{4p'}{p}
+8\pi \chi r \left(1-\frac{2m}{r}\right)^{-\mathcal{D}}
-\frac{r{p'}^2}{p\chi}
\right]\xi \\
&-
\left[
\frac{\varepsilon p'}{p\chi}
+4\pi \chi r \left(1-\frac{2m}{r}\right)^{-\mathcal{D}}
\right]\eta .
\end{aligned}
\end{equation}
where
\begin{equation}
\chi=\varepsilon+p,
\end{equation}
\(\omega\) is the oscillation frequency, \(\Gamma\) is the relativistic adiabatic index,
and \(\Phi(r)\) is the temporal metric potential of the equilibrium background.

In the numerical implementation, the EoS is introduced in tabulated form, except for the
MIT Bag Model, which is treated analytically. The energy density is reconstructed as a
function of pressure, \(\varepsilon=\varepsilon(p)\), and the relativistic adiabatic index
is computed consistently from the EoS through
\begin{equation}
c_s^2=\frac{dp}{d\varepsilon},
\qquad
\Gamma(r)=c_s^2\left(1+\frac{\varepsilon}{p}\right).
\end{equation}
The metric potential \(\Phi(r)\) is obtained by integrating the background relation
\begin{equation}
\Phi'(r)=-\frac{p'(r)}{\varepsilon(r)+p(r)}.
\end{equation}
The integration constant is fixed by imposing the usual surface normalization of the
temporal metric potential. For \(\mathcal{D}\neq1\), this matching should be interpreted
within the same effective framework adopted for the deformed equilibrium configurations.

At the stellar center, regularity requires the perturbation variables to satisfy
\begin{equation}
\xi(r_0)=r_0^3,
\qquad
\eta(r_0)=-3\Gamma_c\,\xi(r_0),
\end{equation}
where \(r_0\) is a small nonzero radius used to start the integration and \(\Gamma_c\) is
the central value of the relativistic adiabatic index. The stellar surface is defined
numerically by the condition \(p(R)=0\), and the oscillation eigenfrequencies are obtained
through a shooting procedure in \(\omega^2\). In practice, the correct eigenvalues are
those for which the surface residual vanishes \citep{sagun2020asteroseismology},
\begin{equation}
F(\omega^2)=
\eta_R-\xi_R
\left[
-4+\frac{1}{1-2M/R}
\left(
-\frac{M}{R}-\omega^2\frac{R^3}{M}
\right)
\right]
=0.
\end{equation}
Thus, the oscillation spectrum is obtained by scanning \(\omega^2\), identifying sign
changes in \(F(\omega^2)\), and refining the corresponding roots numerically.

The oscillation modes are determined for representative equilibrium configurations in
order to assess how the radial spectrum responds to changes in the deformation parameter
\(\mathcal{D}\). The results are summarized in Table~\ref{tab:radial_modes_preliminary},
where selected radial-mode frequencies are shown for the GM1, GM3, NL3 and MIT equations
of state.

\begin{table}[H]
\centering
\caption{Frequencies of selected radial modes (\(f\), in kHz) as a function of the
deformation parameter \(\mathcal{D}\), for representative compact-star configurations
described by the GM1, GM3, NL3 and MIT equations of state.}
\label{tab:radial_modes_preliminary}
\begin{tabular}{cccccc}
\hline
\(\mathcal{D}\) & Mode \(n\) & GM1 & GM3 & NL3 & MIT \\
\hline
0.9 & 0  & 3.96  & 4.76  & 3.09  & 4.62  \\
0.9 & 1  & 6.51  & 7.51  & 5.42  & 8.02  \\
0.9 & 7  & 18.04 & 19.46 & 16.08 & 26.22 \\
0.9 & 10 & 22.70 & 25.61 & 19.52 & 35.12 \\
\hline
1.0 & 0  & 4.44  & 5.29  & 3.45  & 5.17  \\
1.0 & 1  & 7.28  & 8.42  & 6.04  & 8.95  \\
1.0 & 7  & 20.71 & 22.23 & 18.15 & 29.23 \\
1.0 & 10 & 25.49 & 28.79 & 22.19 & 39.14 \\
\hline
\end{tabular}
\end{table}

The frequencies listed in Table~\ref{tab:radial_modes_preliminary} show that the radial
spectrum is sensitive to the deformation parameter \(\mathcal{D}\). For all equations of
state and for all representative modes shown here, the frequencies obtained for
\(\mathcal{D}=1.0\) are higher than those found for \(\mathcal{D}=0.9\). This indicates
that the deformation of the equilibrium background affects not only global quantities such
as the maximum mass and radius, but also the dynamical response of the stellar fluid.

These results provide a complementary stability diagnostic for the deformed compact-star
configurations studied in this thesis. Within the effective \(\mathcal{D}\)-TOV framework,
the radial oscillation spectrum confirms that changes in the global deformation parameter
leave a measurable imprint on the internal dynamics of the star. Therefore, radial modes
constitute an additional probe of the equilibrium configurations, complementing the
information obtained from the mass--radius relation and from the internal pressure and
energy-density profiles.

\chapter{Conclusions and perspectives}

The main motivations underlying the development of this work may be summarized as follows:
the search for a simple effective formulation capable of describing compact objects
beyond the idealized assumption of perfect spherical symmetry, and the investigation of
mechanisms that may allow such objects to attain higher maximum masses.


The motivation for extending the standard compact-star description becomes evident once
we examine the assumptions usually adopted in the derivation of the hydrostatic
equilibrium equation, namely that the object is perfectly spherical, isotropic, and
strictly static. These assumptions lead to the usual TOV equation and provide a realistic
and well-established description of non-rotating compact stars in GR. However,
astrophysical compact objects need not satisfy these idealizations exactly. From the
observational point of view, many NS constraints are obtained from pulsars,
which are rotating NSs. Rotation already breaks the assumption of a strictly static
configuration and, even in the slow-rotation regime, tends to induce departures from
perfect spherical symmetry. In addition, NSs may possess intense magnetic fields, which
can contribute to nonspherical structure and internal anisotropies. At supranuclear
densities, anisotropic stresses may also arise from the microphysics itself. Therefore,
although the spherical TOV solution remains a fundamental reference model, it is
physically motivated to investigate effective extensions capable of incorporating
deviations from exact spherical symmetry.


It is true that many models have already addressed more general compact-star
configurations by explicitly incorporating rotational effects, tidal deformations, strong
magnetic fields, and anisotropic matter distributions. As expected, however, such
analyses make the hydrostatic equilibrium problem considerably more complex, both from
the analytical and numerical points of view. This motivates the search for an alternative
description capable of preserving the essential simplicity of the standard TOV framework
while still allowing one to effectively describe departures from perfect spherical
symmetry.

This goal was pursued here through the introduction of a dimensionless parameter, \(\mathcal{D}\), defined as the ratio between the polar and equatorial radii of the star. In this way, \(\mathcal{D}\) provides a simple global measure of the stellar geometry: \(\mathcal{D}<1\) corresponds to an oblate object, flattened at the poles; \(\mathcal{D}>1\) describes a prolate configuration, stretched along the polar direction; and \(\mathcal{D}=1\) recovers the case of a perfectly spherical star.

It should be emphasized, however, that this formalism is also an approximation rather than a final description of compact objects. First, the parameter \(\mathcal{D}\) is treated here within an effective one-dimensional framework, whereas a fully multidimensional treatment would naturally provide a more realistic description of deformed stellar configurations. Second, \(\mathcal{D}\) does not encode the physical origin of the deformation itself. It does not specify whether the deviation from sphericity is caused by rotation, magnetic stresses, or anisotropic matter effects. Even so, the formalism remains physically relevant, since, despite its simplified character, it is able to generate equilibrium configurations compatible with observational constraints from pulsars.

It is also worth emphasizing that, in the construction of the deformed formalism, only one of the standard assumptions is explicitly relaxed, namely spherical symmetry. The star is still treated as static and effectively isotropic. Although this may at first seem contradictory, it actually highlights the usefulness of the model: by modifying only the geometric structure, the formalism can effectively mimic the global impact that rotation, magnetic fields, or anisotropic stresses may have on the stellar shape, without the need to introduce these ingredients explicitly into the hydrostatic equations.

At the same time, some care is required in the interpretation of the terms ``polar'' and ``equatorial'' in a configuration that is not explicitly rotating. In the present context, these directions should be understood geometrically rather than kinematically. The polar direction is defined by the symmetry axis of the deformation, while the equatorial direction corresponds to the orthogonal plane associated with the largest transverse extension of the star. In this sense, the equatorial radius represents the preferred direction along which the star would appear stretched under the action of effects such as rotation, strong magnetic fields, or anisotropic stresses with a privileged spatial orientation.

With regard to the description of more massive compact objects, we return to the discussion introduced at the beginning of this thesis concerning the secondary component of GW190814. While the primary object in that event is identified as a black hole, the nature of the secondary component remains uncertain, since the most reliably measured massive NSs are typically found around the two-solar-mass range. With an estimated mass between \(2.50\,M_\odot\) and \(2.67\,M_\odot\), the secondary object in GW190814 lies in an unusual regime and may represent either one of the most massive NSs ever inferred or, alternatively, a low-mass black hole.

Although some studies have succeeded in producing very massive NS configurations, this is often achieved through very stiff equations of state. Such equations of state may approach causality limits, and in any case the true behavior of matter at extremely high densities remains uncertain. Therefore, even though the GW190814 secondary is not necessarily inconsistent with the Rhoades--Ruffini upper bound, it still represents an object well outside the usual observational range and thus deserves careful theoretical investigation.

What has been shown in this thesis is that deformation also affects their maximum mass. In particular, oblate configurations, that is, those with \(\mathcal{D}<1\), are able to sustain larger masses for the same EoS when compared with perfectly spherical or prolate configurations. This behavior is qualitatively consistent with what would be expected if the deformation were associated with rotation, since rotational effects provide additional support against gravity and may therefore allow the existence of more massive stars.

An important point of caution in this formalism concerns the predictions associated with the canonical NS mass. Typical NSs are expected to have masses around \(1.4\,M_\odot\), with radii lying roughly in the range of \(10\)--\(15\) km. This information provides a useful criterion for constraining both the adopted EoS and the deformation parameter. In Fig.~\ref{fig:gm1_dtov_mr}, for instance, each point along a given curve represents a distinct stellar configuration. For \(\mathcal{D}=0.8\), the canonical-mass configuration is associated with a radius close to \(16\) km, whereas for \(\mathcal{D}=0.9\) the corresponding radius lies between \(14\) and \(15\) km. Therefore, if \(\mathcal{D}\) is interpreted as an effective parameter related to rotationally induced deformation, the comparison suggests that the model is more compatible with modest departures from spherical symmetry than with very large deformations. In this sense, the parameter \(\mathcal{D}\), together with the canonical-mass constraint, may also serve as a useful criterion for assessing the plausibility of a given EoS: combinations of EoS and deformation that place the canonical-mass radius well above the usual observational range should be regarded with caution.

Another framework adopted in this thesis to describe more massive compact objects was that of modified gravity, in particular the \(f(R,T)\) theory. In the functionals considered throughout this work, the geometric sector remains linear in \(R\), while the modification is introduced through an explicit dependence on the matter sector, encoded in the trace \(T\) of the energy-momentum tensor. Approaches of this kind, which may be written in the form \(f(R,T)=R+f(T)\), are often useful in preliminary studies of massive compact stars, since they preserve the basic geometric structure of GR while introducing additional matter-dependent contributions in the field equations.

These extra \(f(T)\) terms modify the effective hydrostatic balance and the internal structure of the star. As a consequence, depending on the adopted functional form and parameter values, the theory may support equilibrium configurations with higher maximum masses before the onset of gravitational instability.

In general, NS studies in modified gravity are still predominantly developed under the assumption of spherical symmetry. By introducing a simple deformation formalism and extending it to the \(f(R,T)\) theory, this thesis advances one of its main original contributions: the investigation of compact stars in a framework that is simultaneously beyond GR and geometrically more realistic.


Within the set of models considered here, the interplay between deformation and modified gravity was shown to favor the existence of even more massive equilibrium configurations for a fixed EoS, when compared with the corresponding $\lambda=0$ configurations within the same $f(R,T)$ formalism. Nevertheless, some caution is required when interpreting these results through the canonical-mass criterion alone. Unlike the effective General Relativistic case, the additional terms present in $f(R,T)$ gravity may significantly affect not only the maximum mass but also the stellar radius. For this reason, the canonical mass should be regarded as a useful first filter, especially for discarding parameter choices that predict excessively large radii.

In conclusion, some important caveats must be emphasized. At no point in this thesis do we claim that NSs, and in particular the secondary object in GW190814, are necessarily deformed. This statement cannot be made categorically since current observations do not yet allow the direct spatial resolution of these objects or a direct determination of their shape. It is also difficult, on a purely observational basis, to disentangle effects that may arise from stellar deformation from those that may instead be associated with an alternative gravitational framework such as \(f(R,T)\) gravity. What can be done, however, is to compare different models---separately and in combination, and for different equations of state---and assess which of them produce results that remain compatible with current observational constraints.

A more complete physical description of such compact objects would also require a deeper investigation of the stability of deformed configurations. This is expected to be pursued in future work through the proper derivation and analysis of the radial oscillation equations for these stars. Such a study would provide an important next step toward establishing which equilibrium solutions are not only mathematically consistent, but also physically stable.

Finally, it is worth noting that deformed compact stars could be described in even greater
detail by including additional nonspherical features. One possible extension would be to
consider small surface imperfections, or ``mountains,'' in the crust of NSs
\citep{gittins2024gravitational}. In rotating configurations, such irregularities may
produce gravitational-wave emission through a time-dependent quadrupole moment. It is
therefore reasonable to expect that the corresponding signatures could depend on whether
the underlying equilibrium background is spherical or already deformed. This provides a
natural direction for future work, aimed at connecting the effective deformation
formalism developed in this thesis with more detailed models of nonspherical compact
objects and their possible gravitational-wave signatures.


\cleardoublepage
\begin{otherlanguage*}{english}
\renewcommand{\bibname}{References}
\addcontentsline{toc}{chapter}{References}
\bibliography{ref}

@inproceedings{zubairi2017,
  title={Stellar Structure Models of Deformed Neutron Stars},
  author={Zubairi, Omair and Wigley, David and Weber, Fridolin},
  booktitle={International Journal of Modern Physics: Conference Series},
  volume={45},
  pages={1760029},
  year={2017},
  organization={World Scientific}
}

@article{ruderman1972pulsars,
  title={Pulsars: structure and dynamics},
  author={Ruderman, Malvin},
  journal={Annual Review of Astronomy and Astrophysics},
  volume={10},
  number={1},
  pages={427-476},
  year={1972},
  publisher={Annual Reviews 4139 El Camino Way, PO Box 10139, Palo Alto, CA 94303-0139, USA}
}

@article{abbott2020gw190814,
  title={GW190814: gravitational waves from the coalescence of a 23 solar mass black hole with a 2.6 solar mass compact object},
  author={Abbott, Richard and Abbott, TD and Abraham, S and Acernese, Fausto and Ackley, K and Adams, C and Adhikari, Rana X and Adya, VB and Affeldt, Christoph and Agathos, Michail and others},
  journal={The Astrophysical Journal Letters},
  volume={896},
  number={2},
  pages={L44},
  year={2020},
  publisher={IOP Publishing}
}

@article{huang2020possibility,
  title={The possibility of the secondary object in GW190814 as a neutron star},
  author={Huang, Kaixuan and Hu, Jinniu and Zhang, Ying and Shen, Hong},
  journal={The Astrophysical Journal},
  volume={904},
  number={1},
  pages={39},
  year={2020},
  publisher={IOP Publishing}
}

@article{vattis2020could,
  title={Could the 2.6 M$_\odot$ object in GW190814 be a primordial black hole?},
  author={Vattis, Kyriakos and Goldstein, Isabelle S and Koushiappas, Savvas M},
  journal={Physical Review D},
  volume={102},
  number={6},
  pages={061301},
  year={2020},
  publisher={APS}
}

@article{glendenning1991reconciliation,
  title={Reconciliation of neutron-star masses and binding of the $\Lambda$ in hypernuclei},
  author={Glendenning, NK and Moszkowski, SA},
  journal={Physical review letters},
  volume={67},
  number={18},
  pages={2414},
  year={1991},
  publisher={APS}
}

@article{baym1971ground,
  title={The ground state of matter at high densities: equation of state and stellar models},
  author={Baym, Gordon and Pethick, Christopher and Sutherland, Peter},
  journal={Astrophysical Journal, vol. 170, p. 299},
  volume={170},
  pages={299},
  year={1971}
}

@article{riley2021nicer,
  title={A NICER view of the massive pulsar PSR J0740+ 6620 informed by radio timing and XMM-Newton spectroscopy},
  author={Riley, Thomas E and Watts, Anna L and Ray, Paul S and Bogdanov, Slavko and Guillot, Sebastien and Morsink, Sharon M and Bilous, Anna V and Arzoumanian, Zaven and Choudhury, Devarshi and Deneva, Julia S and others},
  journal={The Astrophysical Journal Letters},
  volume={918},
  number={2},
  pages={L27},
  year={2021},
  publisher={IOP Publishing}
}

@article{cromartie2020relativistic,
  title={Relativistic Shapiro delay measurements of an extremely massive millisecond pulsar},
  author={Cromartie, H Thankful and Fonseca, Emmanuel and Ransom, Scott M and Demorest, Paul B and Arzoumanian, Zaven and Blumer, Harsha and Brook, Paul R and DeCesar, Megan E and Dolch, Timothy and Ellis, Justin A and others},
  journal={Nature Astronomy},
  volume={4},
  number={1},
  pages={72--76},
  year={2020},
  publisher={Nature Publishing Group UK London}
}

@article{antoniadis2013massive,
  title={A massive pulsar in a compact relativistic binary},
  author={Antoniadis, John and Freire, Paulo CC and Wex, Norbert and Tauris, Thomas M and Lynch, Ryan S and Van Kerkwijk, Marten H and Kramer, Michael and Bassa, Cees and Dhillon, Vik S and Driebe, Thomas and others},
  journal={Science},
  volume={340},
  number={6131},
  pages={1233232},
  year={2013},
  publisher={American Association for the Advancement of Science}
}

@article{demorest2010two,
  title={A two-solar-mass neutron star measured using Shapiro delay},
  author={Demorest, Paul B and Pennucci, Tim and Ransom, SM and Roberts, MSE and Hessels, JWT},
  journal={nature},
  volume={467},
  number={7319},
  pages={1081--1083},
  year={2010},
  publisher={Nature Publishing Group UK London}
}

@article{abbott2016observation,
  title={Observation of gravitational waves from a binary black hole merger},
  author={Abbott, Benjamin P and Abbott, Richard and Abbott, TDe and Abernathy, MR and Acernese, Fausto and Ackley, Kendall and Adams, Carl and Adams, Thomas and Addesso, Paolo and Adhikari, Rana X and others},
  journal={Physical review letters},
  volume={116},
  number={6},
  pages={061102},
  year={2016},
  publisher={APS}
}

@book{ryden2017introduction,
  title={Introduction to cosmology},
  author={Ryden, Barbara},
  year={2017},
  publisher={Cambridge University Press}
}

@article{adler1995vacuum,
  title={Vacuum catastrophe: An elementary exposition of the cosmological constant problem},
  author={Adler, Ronald J and Casey, Brendan and Jacob, Ovid C},
  journal={American Journal of Physics},
  volume={63},
  number={7},
  pages={620--626},
  year={1995},
  publisher={American Association of Physics Teachers}
}

@article{fortunato2024search,
  title={Search for the f (R, T) gravity functional form via gaussian processes},
  author={Fortunato, JAS and Moraes, PHRS and de Lima J{\'u}nior, JG and Brito, E},
  journal={The European Physical Journal C},
  volume={84},
  number={2},
  pages={198},
  year={2024},
  publisher={Springer}
}

@article{olmo2020stellar,
  title={Stellar structure models in modified theories of gravity: Lessons and challenges},
  author={Olmo, Gonzalo J and Rubiera-Garcia, Diego and Wojnar, Aneta},
  journal={Physics Reports},
  volume={876},
  pages={1--75},
  year={2020},
  publisher={Elsevier}
}

@article{astashenok2013further,
  title={Further stable neutron star models from f (R) gravity},
  author={Astashenok, Artyom V and Capozziello, Salvatore and Odintsov, Sergei D},
  journal={Journal of Cosmology and Astroparticle Physics},
  volume={2013},
  number={12},
  pages={040},
  year={2013},
  publisher={IOP Publishing}
}

@article{capozziello2009f,
  title={f (R) gravity constrained by PPN parameters and stochastic background of gravitational waves},
  author={Capozziello, Salvatore and De Laurentis, Mariafelicia and Nojiri, S and Odintsov, SD},
  journal={General Relativity and Gravitation},
  volume={41},
  pages={2313--2344},
  year={2009},
  publisher={Springer}
}

@article{harko2010f,
  title={f (R, L m) gravity},
  author={Harko, Tiberiu and Lobo, Francisco SN},
  journal={The European Physical Journal C},
  volume={70},
  pages={373--379},
  year={2010},
  publisher={Springer}
}

@article{haghani2021generalizing,
  title={Generalizing the coupling between geometry and matter: f R, L m, T gravity},
  author={Haghani, Zahra and Harko, Tiberiu},
  journal={The European Physical Journal C},
  volume={81},
  number={7},
  pages={615},
  year={2021},
  publisher={Springer}
}

@article{harko2011f,
  title={f (R, T) gravity},
  author={Harko, Tiberiu and Lobo, Francisco SN and Nojiri, Shin’ichi and Odintsov, Sergei D},
  journal={Physical Review D—Particles, Fields, Gravitation, and Cosmology},
  volume={84},
  number={2},
  pages={024020},
  year={2011},
  publisher={APS}
}

@article{liweber,
  title={Heavy baryons in compact stars},
  author={Sedrakian, Armen and Li, Jia Jie and Weber, Fridolin},
  journal={Progress in Particle and Nuclear Physics},
  volume={131},
  year={2023},
  publisher={Elsevier}
}

@book{glendenning2012compact,
  title={Compact stars: Nuclear physics, particle physics and general relativity},
  author={Glendenning, Norman K},
  year={2012},
  publisher={Springer Science \& Business Media}
}

@article{weberbrasil,
  title={Introdu{\c{c}}{\~a}o {\`a} Relatividade Geral e {\`a} F{\'\i}sica de Estrelas Compactas},
  author={Weber, Fridolin},
  journal={Livraria da F{\'\i}sica},
  year={2015}
}

@article{walecka1974nuclear,
  title={Nuclear hydrodynamics in a relativistic mean field theory},
  author={Walecka, JD},
  journal={Ann. Phys},
  volume={83},
  pages={491},
  year={1974}
}

@article{weber2007pulsars,
  title={Pulsars as astrophysical laboratories for nuclear and particle physics},
  author={Weber, Fridolin and Negreiros, R and Rosenfield, P and Stejner, M},
  journal={Progress in Particle and Nuclear Physics},
  volume={59},
  number={1},
  pages={94--113},
  year={2007},
  publisher={Elsevier}
}

@incollection{rajagopal2001condensed,
  title={The condensed matter physics of QCD},
  author={Rajagopal, Krishna and Wilczek, Frank},
  booktitle={At The Frontier of Particle Physics: Handbook of QCD (In 3 Volumes)},
  pages={2061--2151},
  year={2001},
  publisher={World Scientific}
}

@article{jaffe1984strange,
  title={Strange matter},
  author={Jaffe, RL and Farhi, E},
  journal={Phys. Rev. D},
  volume={30},
  pages={2379},
  year={1984}
}

@article{bodmer1971collapsed,
  title={Collapsed nuclei},
  author={Bodmer, Arnold R},
  journal={Physical Review D},
  volume={4},
  number={6},
  pages={1601},
  year={1971},
  publisher={APS}
}

@article{witten1984cosmic,
  title={Cosmic separation of phases},
  author={Witten, Edward},
  journal={Physical Review D},
  volume={30},
  number={2},
  pages={272},
  year={1984},
  publisher={APS}
}

@article{terezawa1989336,
  title={336 (Tokyo: Univ. Tokyo, INS),(1979); H. Terezawa},
  author={Terezawa, H and Rep, INS},
  journal={J. Phys. Soc. Japan},
  volume={58},
  pages={3555},
  year={1989}
}

@article{weber2005strange,
  title={Strange quark matter and compact stars},
  author={Weber, Fridolin},
  journal={Progress in Particle and Nuclear Physics},
  volume={54},
  number={1},
  pages={193--288},
  year={2005},
  publisher={Elsevier}
}

@article{wen2011properties,
  title={Properties of hyperon stars rotating at Keplerian frequency},
  author={Wen, De-Hua and Chen, Wei},
  journal={Chinese Physics B},
  volume={20},
  number={2},
  pages={029701},
  year={2011},
  publisher={IOP Publishing}
}

@inproceedings{rizaldy2018magnetized,
  title={Magnetized deformation of neutron stars},
  author={Rizaldy, R and Sulaksono, A},
  booktitle={Journal of Physics: Conference Series},
  volume={1080},
  number={1},
  pages={012031},
  year={2018},
  organization={IOP Publishing}
}

@article{ivanenko1965hypothesis,
  title={Hypothesis concerning quark stars},
  author={Ivanenko, DD and Kurdgelaidze, DF},
  journal={Astrophysics},
  volume={1},
  pages={251--252},
  year={1965},
  publisher={Springer}
}

@article{itoh1970hydrostatic,
  title={Hydrostatic equilibrium of hypothetical quark stars},
  author={Itoh, Naoki},
  journal={Progress of Theoretical Physics},
  volume={44},
  number={1},
  pages={291--292},
  year={1970},
  publisher={Oxford University Press}
}

@article{fritzsch197316th,
  title={16th Int. Conf. on High Energy Physics, Chicago (1972) H. Fritzsch, M. Gell-Mann and H. Leutwyler},
  author={Fritzsch, Harald},
  journal={Phys. Lett. B},
  volume={47},
  pages={365},
  year={1973}
}

@article{sagun2020asteroseismology,
  title={Asteroseismology: Radial oscillations of neutron stars with realistic equation of state},
  author={Sagun, V and Panotopoulos, G and Lopes, I},
  journal={Physical Review D},
  volume={101},
  number={6},
  pages={063025},
  year={2020},
  publisher={APS}
}

@article{kokkotas2001radial,
  title={Radial oscillations of relativistic stars},
  author={Kokkotas, KostasD and Ruoff, Johannes},
  journal={Astronomy \& Astrophysics},
  volume={366},
  number={2},
  pages={565--572},
  year={2001},
  publisher={EDP Sciences}
}

@article{chandrasekhar1964dynamical,
  title={Dynamical instability of gaseous masses approaching the Schwarzschild limit in general relativity},
  author={Chandrasekhar, S},
  journal={Physical Review Letters},
  volume={12},
  number={4},
  pages={114},
  year={1964},
  publisher={APS}
}

@book{haensel2007neutron,
  title={Neutron stars 1},
  author={Haensel, Pawe{\l} and Potekhin, Aleksander Yu and Yakovlev, Dmitry G},
  year={2007},
  publisher={Springer}
}

@article{oertel2017equations,
  title={Equations of state for supernovae and compact stars},
  author={Oertel, Micaela and Hempel, Matthias and Kl{\"a}hn, Thomas and Typel, Stefan},
  journal={Reviews of Modern Physics},
  volume={89},
  number={1},
  pages={015007},
  year={2017},
  publisher={APS}
}

@article{quartuccio2025equilibrium,
  title={The equilibrium configurations of neutron stars in the optimized $ f (R, T) $ gravity},
  author={Quartuccio, JT and Moraes, PHRS and Zeminiani, GN and Lapola, MM},
  journal={Astrophysics and Space Science},
  volume={370},
  number={4},
  pages={1--7},
  year={2025},
  publisher={Springer}
}

@article{quartuccio2025deformed,
  title={Deformed Compact Objects},
  author={Quartuccio, JT and Moraes, PHRS and Arbanil, JDV},
  journal={International Journal of Theoretical Physics},
  volume={64},
  number={2},
  pages={23},
  year={2025},
  publisher={Springer}
}

@article{moraes2016stellar,
  title={Stellar equilibrium configurations of compact stars in f (R, T) theory of gravity},
  author={Moraes, PHRS and Arba{\~n}il, Jos{\'e} DV and Malheiro, M},
  journal={Journal of Cosmology and Astroparticle Physics},
  volume={2016},
  number={06},
  pages={005},
  year={2016},
  publisher={IOP Publishing}
}

@article{orsaria2014quark,
  title={Quark deconfinement in high-mass neutron stars},
  author={Orsaria, M and Rodrigues, H and Weber, Fridolin and Contrera, GA},
  journal={Physical Review C},
  volume={89},
  number={1},
  pages={015806},
  year={2014},
  publisher={APS}
}

@book{carroll2017introduction,
  title={An introduction to modern astrophysics},
  author={Carroll, Bradley W and Ostlie, Dale A},
  year={2017},
  publisher={Cambridge University Press}
}

@book{shapiro2024black,
  title={Black holes, white dwarfs and neutron stars: the physics of compact objects},
  author={Shapiro, Stuart L and Teukolsky, Saul A},
  year={1983},
  publisher={John Wiley \& Sons}
}

@article{lattimer2000nuclear,
  title={Nuclear matter and its role in supernovae, neutron stars and compact object binary mergers},
  author={Lattimer, James M and Prakash, Madappa},
  journal={Physics Reports},
  volume={333},
  pages={121--146},
  year={2000},
  publisher={Elsevier}
}

@article{sagert2006compact,
  title={Compact stars for undergraduates},
  author={Sagert, Irina and Hempel, Matthias and Greiner, Carsten and Schaffner-Bielich, J{\"u}rgen},
  journal={European journal of physics},
  volume={27},
  number={3},
  pages={577},
  year={2006},
  publisher={IOP Publishing}
}

@article{chamel2008physics,
  title={Physics of neutron star crusts},
  author={Chamel, Nicolas and Haensel, Pawel},
  journal={Living Reviews in relativity},
  volume={11},
  number={1},
  pages={1--182},
  year={2008},
  publisher={Springer}
}

@article{yamada2024physical,
  title={Physical mechanism of core-collapse supernovae that neutrinos drive},
  author={Yamada, Shoichi and Nagakura, Hiroki and Akaho, Ryuichiro and Harada, Akira and Furusawa, Shun and Iwakami, Wakana and Okawa, Hirotada and Matsufuru, Hideo and Sumiyoshi, Kohsuke},
  journal={Proceedings of the Japan Academy, Series B},
  volume={100},
  number={3},
  pages={190--233},
  year={2024},
  publisher={The Japan Academy}
}

@article{kumar2025constraints,
  title={Constraints on maximum neutron star mass from protoneutron star evolution},
  author={Kumar, Deepak and Malik, Tuhin and Mishra, Hiranmaya and Provid{\^e}ncia, Constan{\c{c}}a},
  journal={Physical Review D},
  volume={112},
  number={6},
  pages={063042},
  year={2025},
  publisher={APS}
}

@article{lattimer2014neutron,
  title={Neutron stars},
  author={Lattimer, James M},
  journal={General Relativity and Gravitation},
  volume={46},
  number={5},
  pages={1713},
  year={2014},
  publisher={Springer}
}

@article{motta2020delta,
  title={Do Delta baryons play a role in neutron stars?},
  author={Motta, TF and Thomas, AW and Guichon, Pierre AM},
  journal={Physics Letters B},
  volume={802},
  pages={135266},
  year={2020},
  publisher={Elsevier}
}

@article{ravenhall1983structure,
  title={Structure of matter below nuclear saturation density},
  author={Ravenhall, DG and Pethick, CJ and Wilson, JR},
  journal={Physical Review Letters},
  volume={50},
  number={26},
  pages={2066},
  year={1983},
  publisher={APS}
}

@incollection{serot1992relativistic,
  title={Relativistic nuclear many-body theory},
  author={Serot, Brian D and Walecka, John Dirk},
  booktitle={Recent Progress in Many-Body Theories: Volume 3},
  pages={49--92},
  year={1992},
  publisher={Springer}
}

@book{fetter2012quantum,
  title={Quantum theory of many-particle systems},
  author={Fetter, Alexander L and Walecka, John Dirk},
  year={2012},
  publisher={Courier Corporation}
}

@article{day1967elements,
  title={Elements of the Brueckner-Goldstone theory of nuclear matter},
  author={Day, BD},
  journal={Reviews of Modern Physics},
  volume={39},
  number={4},
  pages={719},
  year={1967},
  publisher={APS}
}

@techreport{hui1982relativistic,
  title={Relativistic mean field model of nuclear matter},
  author={Hui, Gao Shang and Rong, Liang Shao and Zao, Ge Yun and others},
  year={1982},
  institution={PRE-25435}
}

@article{ring1996relativistic,
  title={Relativistic mean field theory in finite nuclei},
  author={Ring, Peter},
  journal={Progress in Particle and Nuclear Physics},
  volume={37},
  pages={193--263},
  year={1996},
  publisher={Elsevier}
}

@article{yakovlev2004neutron,
  title={Neutron star cooling},
  author={Yakovlev, Dima G and Pethick, CJ},
  journal={Annu. Rev. Astron. Astrophys.},
  volume={42},
  number={1},
  pages={169--210},
  year={2004},
  publisher={Annual Reviews}
}

@article{lattimer1991direct,
  title={Direct URCA process in neutron stars},
  author={Lattimer, James M and Pethick, CJ and Prakash, Madappa and Haensel, Pawel},
  journal={Physical review letters},
  volume={66},
  number={21},
  pages={2701},
  year={1991},
  publisher={APS}
}

@article{friman1979neutrino,
  title={Neutrino emissivities of neutron stars},
  author={Friman, BL and Maxwell, OV},
  journal={Astrophysical Journal, Part 1, vol. 232, Sept. 1, 1979, p. 541-557.},
  volume={232},
  pages={541--557},
  year={1979}
}

@article{haskell2019superfluidity,
  title={Superfluidity and superconductivity in neutron stars},
  author={Haskell, Brynmor and Sedrakian, Armen},
  journal={The Physics and Astrophysics of Neutron Stars},
  pages={401--454},
  year={2019},
  publisher={Springer}
}

@article{oppenheimer1939massive,
  title={On massive neutron cores},
  author={Oppenheimer, J Robert and Volkoff, George M},
  journal={Physical Review},
  volume={55},
  number={4},
  pages={374},
  year={1939},
  publisher={APS}
}

@article{miller2021radius,
  title={The radius of PSR J0740+ 6620 from NICER and XMM-Newton data},
  author={Miller, M Coleman and Lamb, FK and Dittmann, AJ and Bogdanov, S and Arzoumanian, Z and Gendreau, KC and Guillot, S and Ho, WCG and Lattimer, JM and Loewenstein, M and others},
  journal={The Astrophysical Journal Letters},
  volume={918},
  number={2},
  pages={L28},
  year={2021},
  publisher={IOP Publishing}
}

@article{bowers1974anisotropic,
  title={Anisotropic spheres in general relativity},
  author={Bowers, Richard L and Liang, EPT},
  journal={Astrophysical Journal, Vol. 188, p. 657 (1974)},
  volume={188},
  pages={657},
  year={1974}
}

@article{stergioulas2003rotating,
  title={Rotating stars in relativity},
  author={Stergioulas, Nikolaos},
  journal={Living Reviews in Relativity},
  volume={6},
  number={1},
  pages={1--109},
  year={2003},
  publisher={Springer}
}

@article{terrero2019modeling,
  title={Modeling anisotropic magnetized white dwarfs with $\gamma$ metric},
  author={Terrero, D Alvear and Mederos, V Hern{\'a}ndez and P{\'e}rez, S L{\'o}pez and Paret, D Manreza and Mart{\'\i}nez, A P{\'e}rez and Angulo, G Quintero},
  journal={Physical Review D},
  volume={99},
  number={2},
  pages={023011},
  year={2019},
  publisher={APS}
}

@inproceedings{gendreau2016neutron,
  title={The neutron star interior composition explorer (NICER): design and development},
  author={Gendreau, Keith C and Arzoumanian, Zaven and Adkins, Phillip W and Albert, Cheryl L and Anders, John F and Aylward, Andrew T and Baker, Charles L and Balsamo, Erin R and Bamford, William A and Benegalrao, Suyog S and others},
  booktitle={Space telescopes and instrumentation 2016: Ultraviolet to gamma ray},
  volume={9905},
  pages={420--435},
  year={2016},
  organization={SPIE}
}

@misc{lam2014peeling,
  author       = {Lam, Anson},
  title        = {Peeling apart a neutron star},
  howpublished = {Astrobites},
  year         = {2014},
  month        = aug,
  note         = {Accessed: 2026-03-09. Available at: \url{https://astrobites.org/2014/08/11/peeling-apart-a-neutron-star/}}
}

@article{weber2014properties,
  author       = {Weber, Fridolin and Contrera, Gustavo A. and Orsaria, Milva G. and Spinella, William and Zubairi, Omair},
  title        = {Properties of High-Density Matter in Neutron Stars},
  journal      = {Modern Physics Letters A},
  volume       = {29},
  number       = {23},
  pages        = {1430022},
  year         = {2014},
  eprint       = {1408.0079},
  archivePrefix= {arXiv},
  primaryClass = {astro-ph.SR}
}

@incollection{weber2009neutron,
  title={Neutron star interiors and the equation of state of superdense matter},
  author={Weber, Fridolin and Negreiros, Rodrigo and Rosenfield, Philip},
  booktitle={Neutron Stars and Pulsars},
  pages={213--245},
  year={2009},
  publisher={Springer}
}

@article{einstein1915feldgleichungen,
  title={Die feldgleichungen der gravitation},
  author={Einstein, Albert},
  journal={Sitzungsberichte der K{\"o}niglich Preu{\ss}ischen Akademie der Wissenschaften},
  pages={844--847},
  year={1915}
}

@article{minazzoli2018rethinking,
  title={Rethinking the link between matter and geometry},
  author={Minazzoli, Olivier},
  journal={Physical Review D},
  volume={98},
  number={12},
  pages={124020},
  year={2018},
  publisher={APS}
}

@book{poisson2004relativist,
  title={A relativist's toolkit: the mathematics of black-hole mechanics},
  author={Poisson, Eric},
  year={2004},
  publisher={Cambridge university press}
}

@article{zubairi2014solutions,
  title={Solutions of Einstein's field equation modified for the cosmological constant},
  author={Zubairi, O and Weber, F},
  journal={Astronomische Nachrichten},
  volume={335},
  number={6-7},
  pages={593--598},
  year={2014},
  publisher={Wiley Online Library}
}

@article{menezes2022introduccao,
  title={Introdu{\c{c}}{\~a}o {\`a} F{\'\i}sica Nuclear e de H{\'a}drons},
  author={Menezes, DP and Marquez, KD and da Silva, TJN},
  journal={S{\~a}o Paulo: Editora Livraria da F{\'\i}sica},
  volume={2},
  year={2022}
}

@article{rhoades1974maximum,
  title={Maximum mass of a neutron star},
  author={Rhoades Jr, Clifford E and Ruffini, Remo},
  journal={Physical Review Letters},
  volume={32},
  number={6},
  pages={324},
  year={1974},
  publisher={APS}
}

@article{tooper1964general,
  title={General Relativistic Polytropic Fluid Spheres.},
  author={Tooper, Robert F},
  journal={Astrophysical Journal, vol. 140, p. 434},
  volume={140},
  pages={434},
  year={1964}
}

@inproceedings{contrera2017hybrid,
  title={Hybrid stars in the framework of njl models},
  author={Contrera, Gustavo A and Orsaria, Milva and Ranea-Sandoval, Ignacio Francisco and Weber, Fridolin},
  booktitle={International Journal of Modern Physics: Conference Series},
  volume={45},
  pages={1760026},
  year={2017},
  organization={World Scientific}
}

@article{gittins2024gravitational,
  title={Gravitational waves from neutron-star mountains},
  author={Gittins, Fabian},
  journal={Classical and Quantum Gravity},
  volume={41},
  number={4},
  pages={043001},
  year={2024},
  publisher={IOP Publishing}
}

@article{ozel2016masses,
  title={Masses, radii, and the equation of state of neutron stars},
  author={{\"O}zel, Feryal and Freire, Paulo},
  journal={Annual Review of Astronomy and Astrophysics},
  volume={54},
  pages={401--440},
  year={2016},
  publisher={Annual Reviews}
}

@article{steiner2013neutron,
  title={The neutron star mass--radius relation and the equation of state of dense matter},
  author={Steiner, Andrew W and Lattimer, James M and Brown, Edward F},
  journal={The Astrophysical Journal Letters},
  volume={765},
  number={1},
  pages={L5},
  year={2013},
  publisher={The American Astronomical Society}
}

@article{tolman1939static,
  title={Static solutions of Einstein's field equations for spheres of fluid},
  author={Tolman, Richard C},
  journal={Physical Review},
  volume={55},
  number={4},
  pages={364},
  year={1939},
  publisher={APS}
}

@article{quartuccio2026deformed,
  title={Deformed compact objects in modified gravity},
  author={Quartuccio, JT and Moraes, PHRS},
  journal={The European Physical Journal Plus},
  volume={141},
  number={4},
  pages={447},
  year={2026},
  publisher={Springer}
}

@article{abbott2017gw170817,
  title={GW170817: observation of gravitational waves from a binary neutron star inspiral},
  author={Abbott, Benjamin P and Abbott, Rich and Abbott, Thomas D and Acernese, Fausto and Ackley, Kendall and Adams, Carl and Adams, Thomas and Addesso, Paolo and Adhikari, Rana X and Adya, Vaishali B and others},
  journal={Physical review letters},
  volume={119},
  number={16},
  pages={161101},
  year={2017},
  publisher={APS}
}

@article{abbott2020gw190425,
  title={GW190425: Observation of a compact binary coalescence with total $\sim 3.4\,M_{\odot}$},
  author={Abbott, Benjamin P and Abbott, Robert and Abbott, TD and Abraham, S and Acernese, Fausto and Ackley, K and Adams, C and Adhikari, RX and Adya, VB and Affeldt, Christoph and others},
  journal={The Astrophysical journal letters},
  volume={892},
  number={1},
  pages={L3},
  year={2020},
  publisher={The American Astronomical Society}
}

@misc{freire2024pulsar,
  title={Pulsar mass measurements and tests of general relativity},
  author={Freire, P},
  year={2024}
}

@article{fonseca2021refined,
  title={Refined mass and geometric measurements of the high-mass PSR J0740+ 6620},
  author={Fonseca, Emmanuel and Cromartie, H Thankful and Pennucci, Timothy T and Ray, Paul S and Kirichenko, A Yu and Ransom, Scott M and Demorest, Paul B and Stairs, Ingrid H and Arzoumanian, Zaven and Guillemot, Lucas and others},
  journal={The Astrophysical Journal Letters},
  volume={915},
  number={1},
  pages={L12},
  year={2021},
  publisher={The American Astronomical Society}
}

@article{buchdahl1970non,
  title={Non-linear Lagrangians and cosmological theory},
  author={Buchdahl, Hans A},
  journal={Monthly Notices of the Royal Astronomical Society},
  volume={150},
  number={1},
  pages={1--8},
  year={1970},
  publisher={Oxford University Press Oxford, UK}
}

@article{starobinsky1980new,
  title={A new type of isotropic cosmological models without singularity},
  author={Starobinsky, Alexei A},
  journal={Physics Letters B},
  volume={91},
  number={1},
  pages={99--102},
  year={1980},
  publisher={Elsevier}
}
\end{otherlanguage*}

\begin{appendices}
\chapter{Einstein's field equations}
The total action for gravity can be write in the form
\begin{equation}
	S = S_c + S_m,
\end{equation}
where $S_c$ is the curvature part and $S_m$ is the matter part. We write
\begin{equation}
	S_c = \int \mathcal{L}_c \sqrt{-g}d^4x,
\end{equation}
where the minus sign is from the fact that in GR the determinant of the metric is always negative. The action of curvature is integrated throughout four-dimensional space. Since the action must be a scalar, then $\mathcal{L}_c \propto R$, where $R$ is the Ricci scalar. Taking another scalar quantity arbitrarily
\begin{equation}
	\mathcal{L}_c = \alpha(R - 2\Lambda),
\end{equation}
we can write the total action in the form
\begin{equation}
	S = \alpha \int (R - 2\Lambda)\sqrt{-g}d^4x + S_m.
\end{equation}
Equation (A.4) describes the Einstein-Hilbert action. Now we can apply the principle of least action to the Einstein-Hilbert action. This principle tell us that the correct field equations that we are looking for should be produced when the value of the action is minimized (stationary). Mathematically, this means that $\delta S = 0$. As $R$ and $S$ are functions of the metric tensor, the variation of the action corresponds to the derivative of the functional with respect to the metric. The derivative of the functional is defined as
\begin{equation}
	\delta g^{\mu \nu} \frac{\delta}{\delta g^{\mu \nu}},
\end{equation}
and so
\begin{equation*}
	\delta S = \alpha \delta \int (R - 2\Lambda)\sqrt{-g}d^4x + \delta S_m
\end{equation*}
\begin{equation}
	 = \alpha \int \frac{\delta}{\delta g^{\mu \nu}}\left[(R - 2\Lambda)\sqrt{-g}\right] \delta g^{\mu \nu}d^4x + \delta S_m,
\end{equation}
so
\begin{equation*}
	\delta S = \alpha \int \left[ \frac{\delta(\sqrt{-g}R)}{\delta g^{\mu \nu}} - 2\Lambda \frac{\delta (\sqrt{-g})}{\delta g^{\mu \nu}}\right]\delta g^{\mu \nu}d^4x + \delta S_m
\end{equation*}
\begin{equation*}
	 = \alpha \int \left[ R\frac{\delta(\sqrt{-g})}{\delta g^{\mu \nu}} + \sqrt{-g} \frac{\delta R}{\delta g^{\mu \nu}} - 2\Lambda \frac{\delta (\sqrt{-g})}{\delta g^{\mu \nu}}\right]\delta g^{\mu \nu}d^4x + \delta S_m.
\end{equation*}
The derivative of $\sqrt{-g}$ is
\begin{equation}
	\frac{\delta(\sqrt{-g})}{\delta g^{\mu \nu}} = - \frac{1}{2\sqrt{-g}}\frac{\delta g}{\delta g^{\mu \nu}},
\end{equation}
and therefore
\begin{equation*}
	 \delta S = \alpha \int \left[ - \frac{R}{2\sqrt{-g}} \frac{\delta g}{\delta g^{\mu \nu}} + \sqrt{-g} \frac{\delta R}{\delta g^{\mu \nu}} + \frac{\Lambda}{\sqrt{-g}} \frac{\delta g}{\delta g^{\mu \nu}}\right]\delta g^{\mu \nu}d^4x + \delta S_m,
\end{equation*}
and applying the principle of least action
\begin{equation}
	 \alpha \int \left[ - \frac{R}{2\sqrt{-g}} \frac{\delta g}{\delta g^{\mu \nu}} + \sqrt{-g} \frac{\delta R}{\delta g^{\mu \nu}} + \frac{\Lambda}{\sqrt{-g}} \frac{\delta g}{\delta g^{\mu \nu}}\right]\delta g^{\mu \nu}d^4x =- \delta S_m.
\end{equation}

Now we must calculate the right side of (A.8). Let us write the action $S_m$ as an integral over spacetime in the form
\begin{equation}
	S_m = \int \mathcal{L}_m\sqrt{-g}d^4 x,
\end{equation}
and therefore
\begin{equation*}
	\delta S_m = \int \frac{\delta}{\delta g^{\mu \nu}}(\mathcal{L}_m\sqrt{-g})\delta g^{\mu \nu}d^4x
\end{equation*}
\begin{equation*}
	 = \int \left( \sqrt{-g}\frac{\delta \mathcal{L}_m}{\delta g^{\mu\nu}} + \mathcal{L}_m \frac{\delta \sqrt{-g}}{\delta g^{\mu \nu}}\right) \delta g^{\mu \nu}d^4 x,
\end{equation*}
using equation (A.7)
\begin{equation}
	 \delta S_m = \int \left( \sqrt{-g}\frac{\delta \mathcal{L}_m}{\delta g^{\mu\nu}} - \frac{\mathcal{L}_m}{2\sqrt{-g}}\frac{\delta g}{\delta g^{\mu \nu}}\right) \delta g^{\mu \nu}d^4 x.
\end{equation}

Equation (A.8) yields
\begin{equation}
	 \alpha \int \left[ - \frac{R}{2\sqrt{-g}} \frac{\delta g}{\delta g^{\mu \nu}} + \sqrt{-g} \frac{\delta R}{\delta g^{\mu \nu}} + \frac{\Lambda}{\sqrt{-g}} \frac{\delta g}{\delta g^{\mu \nu}}\right]\delta g^{\mu \nu}d^4x =- \int \left( \sqrt{-g}\frac{\delta \mathcal{L}_m}{\delta g^{\mu\nu}} - \frac{\mathcal{L}_m}{2\sqrt{-g}}\frac{\delta g}{\delta g^{\mu \nu}}\right) \delta g^{\mu \nu}d^4 x.
\end{equation}

Both sides of (A.11) involve integrals over the same region and therefore
\begin{equation}
	 \int\left[ \alpha \left( - \frac{R}{2\sqrt{-g}} \frac{\delta g}{\delta g^{\mu \nu}} + \sqrt{-g} \frac{\delta R}{\delta g^{\mu \nu}} + \frac{\Lambda}{\sqrt{-g}} \frac{\delta g}{\delta g^{\mu \nu}}\right) + \sqrt{-g}\frac{\delta \mathcal{L}_m}{\delta g^{\mu\nu}} - \frac{\mathcal{L}_m}{2\sqrt{-g}}\frac{\delta g}{\delta g^{\mu \nu}}\right] \delta g^{\mu \nu}d^4 x = 0.
\end{equation}
The solution of this integral is obtained when the integrant goes to zero, and so
\begin{equation}
	 \alpha \left( - \frac{R}{2\sqrt{-g}} \frac{\delta g}{\delta g^{\mu \nu}} + \sqrt{-g} \frac{\delta R}{\delta g^{\mu \nu}} + \frac{\Lambda}{\sqrt{-g}} \frac{\delta g}{\delta g^{\mu \nu}}\right) + \sqrt{-g}\frac{\delta \mathcal{L}_m}{\delta g^{\mu\nu}} - \frac{\mathcal{L}_m}{2\sqrt{-g}}\frac{\delta g}{\delta g^{\mu \nu}} = 0.
\end{equation}
If we multiply (A.13) by $\frac{2}{\sqrt{-g}}$, we obtain
\begin{equation}
	\alpha \left( \frac{R}{g} \frac{\delta g}{\delta g^{\mu \nu}} + 2 \frac{\delta R}{\delta g^{\mu \nu}} - \frac{2\Lambda}{g}\frac{\delta g}{\delta g^{\mu \nu}}\right) = - \frac{\mathcal{L}_m}{g}\frac{\delta g}{\delta g^{\mu \nu}} - 2 \frac{\delta \mathcal{L}_m}{\delta g^{\mu \nu}},
\end{equation}
where we use $\sqrt{-g} \cdot \sqrt{-g} = (\sqrt{-g})^2 = -g$.

The Lagrangian $\mathcal{L}_m$ is the Lagrangian density for any matter present in spacetime. The right-hand of (A.14) is a tensor with two indices that represents all matter and gravity sources in spacetime. In other words, the right-hand defines the energy-momentum tensor,
\begin{equation}
	T_{\mu \nu} = - \frac{\mathcal{L}_m}{g}\frac{\delta g}{\delta g^{\mu \nu}} - 2 \frac{\delta \mathcal{L}_m}{\delta g^{\mu \nu}} = - \frac{2}{\sqrt{-g}}\frac{\delta(\sqrt{-g}\mathcal{L}_m)}{\delta g^{\mu \nu}},
\end{equation}
and then equation (A.14)  becomes
\begin{equation*}
	\alpha \left( \frac{R}{g} \frac{\delta g}{\delta g^{\mu \nu}} + 2 \frac{\delta R}{\delta g^{\mu \nu}} - \frac{2\Lambda}{g}\frac{\delta g}{\delta g^{\mu \nu}}\right) = T_{\mu \nu},
\end{equation*}
and dividing both sides by $2\alpha$
\begin{equation}
	\frac{\delta R}{\delta g^{\mu \nu}} + \frac{R}{2g}\frac{\delta g}{\delta g^{\mu \nu}} - \frac{\Lambda}{g} \frac{\delta g}{\delta g^{\mu \nu}} = \frac{1}{2\alpha}T_{\mu \nu}.
\end{equation}

Now we must calculate the variation of the metric determinant and the variation of the Ricci scalar. First we will calculate the variation of $g$ using a linear algebra identity applied to a matrix $A$,
\begin{equation}
	\ln (\det A) = \text{Tr}(\ln A),
\end{equation}
and taking the variation on both sides
\begin{equation*}
	\delta \ln (\det A) = \delta \text{Tr}(\ln A),
\end{equation*}
\begin{equation*}
	\frac{1}{\det A} \delta \det A = \text{Tr}(\delta \ln A) = \text{Tr}(A^{-1} \delta A),
\end{equation*}
\begin{equation}
	\delta \det A = \text{Tr}(A^{-1} \delta A)\det A.
\end{equation}
Now we can substitute $A = g_{\mu \nu}$, $A^{-1} = g^{\mu \nu}$ and $\det A = g$ to get
\begin{equation}
	\delta g = \text{Tr}(g^{\mu \nu} \delta g_{\alpha \beta})g.
\end{equation}
The trace is found taking a simple contraction
\begin{equation}
	\delta g = \text{Tr}(g^{\mu \nu}\delta g_{\alpha \beta})g = g g^{\mu \nu}\delta g_{\mu \nu}.
\end{equation}
The contraction of the metric always returns the dimension of the space, so
\begin{equation}
	g^{\mu \nu}g_{\mu \nu} = 4.
\end{equation}
Now we can take the variation on both sides of (A.21)
\begin{equation*}
	\delta(g^{\mu \nu}g_{\mu \nu}) = \delta(4),
\end{equation*}
\begin{equation*}
	g_{\mu \nu}\delta g^{\mu \nu} + g^{\mu \nu}\delta g_{\mu \nu} = 0,
\end{equation*}
\begin{equation}
	g^{\mu \nu}\delta g_{\mu \nu} = -g_{\mu \nu}\delta g^{\mu \nu}.
\end{equation}
Using (A.22) in (A.20)
\begin{equation*}
	\delta g = -gg_{\mu \nu}\delta g^{\mu \nu},
\end{equation*}
and then
\begin{equation}
	\frac{\delta g}{\delta g^{\mu \nu}} = -gg_{\mu \nu}.
\end{equation}
Using this result in (A.16) we obtain
\begin{equation*}
	\frac{\delta R}{\delta g^{\mu \nu}} + \frac{R}{2g}(-gg_{\mu \nu}) - \frac{\Lambda}{g} (-gg_{\mu \nu}) = \frac{1}{2\alpha}T_{\mu \nu},
\end{equation*}
\begin{equation}
	\frac{\delta R}{\delta g^{\mu \nu}} - \frac{R}{2}g_{\mu \nu} + \Lambda g_{\mu \nu} = \frac{1}{2 \alpha}T_{\mu \nu}.
\end{equation}

Now it is necessary to calculate the variation of the Ricci scalar. We know that
\begin{equation}
	R = g^{\mu \nu}R_{\mu \nu},
\end{equation}
and taking the variation
\begin{equation}
	\delta R = R_{\mu \nu}\delta g^{\mu \nu} + g^{\mu \nu}\delta R_{\mu \nu}.
\end{equation}

The Ricci tensor is defined as
\begin{equation}
	R_{\mu \nu} = \partial_\alpha \Gamma^\alpha_{  \mu \nu} - \partial_\nu \Gamma^\alpha_{  \mu \alpha} + \Gamma^\alpha_{  \mu \nu}\Gamma^\beta_{  \alpha \beta} - \Gamma^\alpha_{  \mu \beta}\Gamma^\beta_{  \nu \alpha},
\end{equation}
and its variation
\begin{equation}
	\delta R_{\mu \nu} = \partial_\alpha \delta \Gamma^\alpha_{ \mu \nu} - \partial_\nu \delta \Gamma^\alpha_{ \mu \alpha} + \Gamma^\alpha_{ \mu \nu}\delta \Gamma^\beta_{ \alpha \beta} + \Gamma^\beta_{ \alpha \beta}\delta \Gamma^{\alpha}_{ \mu \nu} - \Gamma^{\alpha}_{ \mu \beta} \delta \Gamma^{\beta}_{ \nu \alpha} - \Gamma^\beta_{ \nu \alpha}\delta \Gamma^\alpha_{ \mu \beta}.
\end{equation}

In general, for a tensor $A^\lambda_{ \mu \nu}$, its covariant derivative is given by
\begin{equation}
	\nabla_\alpha A^\lambda_{ \mu \nu} = \partial_\alpha A^\lambda_{ \mu \nu} + \Gamma^\beta_{ \alpha \beta}A^\lambda_{ \mu \nu} - \Gamma^\lambda_{ \mu \beta}A^\beta_{ \nu \alpha} - \Gamma^\beta_{ \nu \alpha}A^\lambda_{ \mu \beta}.
\end{equation}
Although Christoffel's symbols are not tensors, their variations are. Therefore
\begin{equation}
	\nabla_\alpha \delta \Gamma^\alpha_{ \mu \nu} = \partial_\alpha \delta \Gamma^{\alpha}_{ \mu \nu} + \Gamma^\beta_{ \alpha \beta}\delta \Gamma^{\alpha}_{ \mu \nu} - \Gamma^{\alpha}_{ \mu \beta}\delta \Gamma^\beta_{ \nu \alpha} - \Gamma^\beta_{ \nu \alpha}\delta \Gamma^{\alpha}_{ \mu \beta},
\end{equation}
\begin{equation}
	\nabla_\nu \delta \Gamma^\alpha_{ \mu \alpha} = \partial_\nu \delta \Gamma^{\alpha}_{ \mu \alpha} + \Gamma^\beta_{ \alpha \nu}\delta \Gamma^{\alpha}_{ \mu \beta} - \Gamma^{\alpha}_{ \mu \nu}\delta \Gamma^\beta_{ \alpha \beta} - \Gamma^\beta_{ \alpha \nu}\delta \Gamma^{\alpha}_{ \mu \beta} = \partial_\nu \delta \Gamma^\alpha_{ \mu \alpha} - \Gamma^\alpha_{ \mu \nu}\delta \Gamma^\beta_{ \alpha \beta}.
\end{equation}
Looking again at $\delta R$ we can use Equations (A.30) and (A.31) and rewrite (A.28) as
\begin{equation}
	\delta R_{\mu \nu} = \nabla_\alpha \delta \Gamma^{\alpha}_{ \mu \nu} - \nabla_\nu \delta \Gamma^\alpha_{ \mu \alpha}.
\end{equation}
Equation (A.32) defines the Palatini Identity. Now we can look again at Equation (A.26) and rewrite it as
\begin{equation}
	\delta R = R_{\mu \nu}\delta g^{\mu \nu} + g^{\mu \nu}\delta R_{\mu \nu} = R_{\mu \nu} \delta g^{\mu \nu} + g^{\mu \nu}\left( \nabla_\alpha \delta \Gamma^\alpha_{ \mu \nu} - \nabla_\nu \delta \Gamma^\alpha_{ \mu \alpha}\right),
\end{equation}
\begin{equation}
	\delta R = R_{\mu \nu}\delta g^{\mu \nu} + \nabla_\alpha \left(\delta \Gamma^\alpha_{ \mu \nu} g^{\mu \nu}\right) - \nabla_\nu \left(\delta \Gamma^\alpha_{ \mu \alpha} g^{\mu \nu}\right).
\end{equation}
Since the indices in the last term of (A.34) are fictitious, we will make a change. Let's start by exchanging $\alpha$ for $\beta$.
\begin{equation*}
	\delta R = R_{\mu \nu}\delta g^{\mu \nu} + \nabla_\alpha \left(\delta \Gamma^\alpha_{ \mu \nu} g^{\mu \nu}\right) - \nabla_\nu \left(\delta \Gamma^\beta_{ \mu \beta} g^{\mu \nu}\right),
\end{equation*}
and changing, in the last term, $\nu$ to $\alpha$
\begin{equation*}
	\delta R = R_{\mu \nu}\delta g^{\mu \nu} + \nabla_\alpha \left(\delta \Gamma^\alpha_{ \mu \nu} g^{\mu \nu}\right) - \nabla_\alpha \left(\delta \Gamma^\beta_{ \mu \beta} g^{\mu \alpha}\right),
\end{equation*}
and therefore
\begin{equation}
	\delta R = R_{\mu \nu}\delta g^{\mu \nu} + \nabla_\alpha \left(\delta \Gamma^\alpha_{ \mu \nu}g^{\mu \nu} - \delta \Gamma^\beta_{ \mu \beta}g^{\mu \alpha}\right).
\end{equation}
The quantity inside the parentheses is a tensor object that we will denote by $V^\alpha$, so
\begin{equation}
	\delta R = R_{\mu \nu}\delta g^{\mu \nu} + \nabla_\alpha V^\alpha.
\end{equation}
The second term gives the total divergence. We are still calculating the variation in action that relates to an integral. The integral is in the form
\begin{equation}
	S = \int g^{\mu \nu} \delta R_{\mu \nu} \sqrt{-g}d^4 x = \int \nabla_\alpha V^\alpha \sqrt{-g}d^4 x.
\end{equation}
According to the divergence theorem, the integral gives us a three-dimensional boundary term
\begin{equation}
	\int \nabla_\alpha V^\alpha \sqrt{-g}d^4x = \int V^\alpha n_\alpha \sqrt{-\gamma}d^3 x,
\end{equation}
where $n_\alpha$ is the vector normal to the surface and $\gamma$ is the induced metric. Therefore
\begin{equation}
	\int V^\alpha n_\alpha \sqrt{-\gamma}d^3 x = \int \left( \delta \Gamma^\alpha_{\mu \nu}g^{\mu \nu} - \delta \Gamma^\nu_{\mu \nu} g^{\mu \alpha}\right)n_\alpha\sqrt{-\gamma}d^3x.
\end{equation}
An assumption we can make about the gravitational field is to say that the variations $\delta \Gamma$ go to zero at infinity, so that
\begin{equation}
	S = \int \nabla_\alpha V^\alpha \sqrt{-g}d^4x = 0.
\end{equation}
Since this term does not contribute to the action, it does not affect the field equations, and therefore the variation of the Ricci scalar in this case simply returns
\begin{equation}
	\delta R = R_{\mu \nu}\delta g^{\mu \nu},
\end{equation}
and so
\begin{equation}
	\frac{\delta R}{\delta g^{\mu \nu}} = R_{\mu \nu}.
\end{equation}
This result is telling us that the variation of the gravitational field goes to zero at infinity, which means that the variation of the Christoffel symbols goes to zero at infinity. Equation (A.24) is rewritten as
\begin{equation}
	R_{\mu \nu} - \frac{1}{2}g_{\mu \nu}R + \Lambda g_{\mu \nu} = \frac{1}{2\alpha}T_{\mu \nu}.
\end{equation}
Taking the term $1/2\alpha$ as a constant $k$
\begin{equation}
	R_{\mu \nu} - \frac{1}{2}g_{\mu \nu}R + \Lambda g_{\mu \nu} = kT_{\mu \nu},
\end{equation}
where $k = \frac{8\pi G}{c^4}$. Therefore, we obtain
\begin{equation}
	R_{\mu \nu} - \frac{1}{2}g_{\mu \nu}R + \Lambda g_{\mu \nu} = \frac{8\pi G}{c^4}T_{\mu \nu},
\end{equation}
or in a compact way
\begin{equation}
	G_{\mu \nu} + \Lambda g_{\mu \nu} = 8\pi T_{\mu \nu},
\end{equation}
where we assume $G = c = 1$.  

\end{appendices}

\begin{appendices}
\chapter{Derivation of the Tolman-Oppenheimer-Volkoff Equations}

The derivation presented in this Appendix follows, in part, the approach developed in \citet{menezes2022introduccao}, with adaptations in notation and presentation for the purposes of this thesis.

The most general line element describing a static and spherically symmetric spacetime is given by
\begin{equation}
	ds^2 = B(r)\,dt^2 - A(r)\,dr^2 - r^2(d\theta^2 + \sin^2\theta\,d\varphi^2),
\end{equation}
which can be written in matrix form as
\begin{equation}
	g_{\mu \nu} =
	\begin{pmatrix}
		B(r)&0&0&0\\
		0&-A(r)&0&0\\
		0&0&-r^2&0\\
		0&0&0&-r^2 \sin^2 \theta
	\end{pmatrix}.
\end{equation}
Therefore, the nonvanishing covariant components of the metric are
\begin{eqnarray}
	&&g_{00} = B(r), \nonumber\\
	&&g_{11} = -A(r),\\
	&&g_{22} = -r^2, \nonumber\\
	&&g_{33} = -r^2\sin^2\theta. \nonumber
\end{eqnarray}
The corresponding contravariant metric is
\begin{equation}
	g^{\mu \nu} =
	\begin{pmatrix}
		\dfrac{1}{B(r)}&0&0&0\\
		0&-\dfrac{1}{A(r)}&0&0\\
		0&0&-\dfrac{1}{r^2}&0\\
		0&0&0&-\dfrac{1}{r^2 \sin^2 \theta}
	\end{pmatrix},
\end{equation}
or equivalently,
\begin{eqnarray}
	&&g^{00} = \dfrac{1}{B(r)}, \nonumber\\
	&&g^{11} = -\dfrac{1}{A(r)},\\
	&&g^{22} = -\dfrac{1}{r^2}, \nonumber\\
	&&g^{33} = -\dfrac{1}{r^2\sin^2\theta}. \nonumber
\end{eqnarray}

In the Einstein field equations, $T_{\mu\nu}$ denotes the energy-momentum tensor. In the stellar context, this tensor describes the matter distribution inside the star. Assuming that the stellar interior can be modeled as a perfect fluid, one has
\begin{equation}
	T_{\mu \nu} = (\varepsilon + p)u_\mu u_\nu - pg_{\mu \nu},
\end{equation}
where $\varepsilon$ is the energy density and $p$ is the pressure, both depending only on the radial coordinate $r$. The quantity $u_\mu \equiv dx_\mu/ds$ is the four-velocity of the fluid element and satisfies the normalization condition $u_\mu u^\mu = 1$.

In the comoving frame, the energy-momentum tensor is diagonal. In the coordinate basis associated with the metric above, its nonvanishing covariant components are
\begin{eqnarray}
	&&T_{00} = \varepsilon B(r), \nonumber\\
	&&T_{11} = pA(r),\\
	&&T_{22} = pr^2, \nonumber\\
	&&T_{33} = pr^2\sin^2\theta. \nonumber
\end{eqnarray}
The corresponding contravariant components are
\begin{eqnarray}
	&&T^{00} = \dfrac{\varepsilon}{B(r)}, \nonumber\\
	&&T^{11} = \dfrac{p}{A(r)},\\
	&&T^{22} = \dfrac{p}{r^2}, \nonumber\\
	&&T^{33} = \dfrac{p}{r^2\sin^2\theta}. \nonumber
\end{eqnarray}

We now compute the Christoffel symbols. Since the metric is diagonal, all terms of the form $g_{ij,k}$ with $i\neq j$ vanish. The nonvanishing Christoffel symbols are
\begin{equation}
	\Gamma^{0}_{10} = \Gamma^{0}_{01} = \dfrac{g^{00}}{2}(g_{10,0} + g_{00,1} - g_{10,0}) = \dfrac{1}{2B}B' = \dfrac{B'}{2B},
\end{equation}
\begin{equation}
	\Gamma^{1}_{00} = \dfrac{g^{11}}{2}(g_{01,0} + g_{01,0} - g_{00,1}) = -\dfrac{1}{2A}(-B') = \dfrac{B'}{2A},
\end{equation}
\begin{equation}
	\Gamma^{1}_{11} = \dfrac{g^{11}}{2}(g_{11,1} + g_{11,1} - g_{11,1}) = -\dfrac{1}{2A}(-A') = \dfrac{A'}{2A},
\end{equation}
\begin{equation}
	\Gamma^{1}_{22} = \dfrac{g^{11}}{2}(g_{21,2} + g_{21,2} - g_{22,1}) = -\dfrac{1}{2A}(2r) = -\dfrac{r}{A},
\end{equation}
\begin{equation}
	\Gamma^{1}_{33} = \dfrac{g^{11}}{2}(g_{31,3} + g_{31,3} - g_{33,1}) = -\dfrac{1}{2A}(2r\sin^2\theta) = -\dfrac{r\sin^2\theta}{A},
\end{equation}
\begin{equation}
	\Gamma^{2}_{12} = \Gamma^{2}_{21} = \dfrac{g^{22}}{2}(g_{12,2} + g_{22,1} - g_{12,2}) = -\dfrac{1}{2r^2}(-2r) = \dfrac{1}{r},
\end{equation}
\begin{equation}
	\Gamma^{2}_{33} = \dfrac{g^{22}}{2}(g_{32,3} + g_{32,3} - g_{33,2}) = -\dfrac{1}{2r^2}(2r^2\cos\theta\sin\theta) = -\sin\theta\cos\theta,
\end{equation}
\begin{equation}
	\Gamma^{3}_{13} = \Gamma^{3}_{31} = \dfrac{g^{33}}{2}(g_{13,3} + g_{33,1} - g_{13,3}) = -\dfrac{1}{2r^2\sin^2\theta}(-2r\sin^2\theta) = \dfrac{1}{r},
\end{equation}
\begin{equation}
	\Gamma^{3}_{23} = \Gamma^{3}_{32} = \dfrac{g^{33}}{2}(g_{23,3} + g_{33,2} - g_{23,3}) = -\dfrac{1}{2r^2\sin^2\theta}(-2r^2\sin\theta\cos\theta) = \cot\theta.
\end{equation}

Having obtained the Christoffel symbols, we may now compute the Ricci tensor. Since the full derivation is lengthy, only the first two components will be presented explicitly. For the $R_{00}$ component, one obtains
\begin{equation*}
	\begin{split}
		R_{00} = \Gamma^{0}_{00,0} - \Gamma^{0}_{00,0} + \{ \Gamma^{0}_{00}\Gamma^{0}_{00} + \Gamma^{0}_{10}\Gamma^{1}_{00} + \Gamma^{0}_{20}\Gamma^{2}_{00} + \Gamma^{0}_{30}\Gamma^{3}_{00}\} \\
		- \{ \Gamma^{0}_{00}\Gamma^{0}_{00} + \Gamma^{0}_{10}\Gamma^{1}_{00} + \Gamma^{0}_{20}\Gamma^{2}_{00} + \Gamma^{0}_{30}\Gamma^{3}_{00} \} \\
		+ \Gamma^{1}_{00,1} - \Gamma^{1}_{01,0} +  \{ \Gamma^{1}_{01}\Gamma^{0}_{00} + \Gamma^{1}_{11}\Gamma^{1}_{00} + \Gamma^{1}_{21}\Gamma^{2}_{00} + \Gamma^{1}_{31}\Gamma^{3}_{00} \} \\
		- \{ \Gamma^{1}_{00}\Gamma^{0}_{01} + \Gamma^{1}_{10}\Gamma^{1}_{01} + \Gamma^{1}_{20}\Gamma^{2}_{01} + \Gamma^{1}_{30}\Gamma^{3}_{01} \} \\
		+ \Gamma^{2}_{00,2} - \Gamma^{2}_{02,0} + \{\Gamma^{2}_{02}\Gamma^{0}_{00} +\Gamma^{2}_{12}\Gamma^{1}_{00} + \Gamma^{2}_{22}\Gamma^{2}_{00} + \Gamma^{2}_{32}\Gamma^{3}_{00} \} \\
		- \{\Gamma^{2}_{00}\Gamma^{0}_{02} + \Gamma^{2}_{10}\Gamma^{1}_{02} + \Gamma^{2}_{20}\Gamma^{2}_{02} + \Gamma^{2}_{30}\Gamma^{3}_{02} \} \\
		+ \Gamma^{3}_{00,3} - \Gamma^{3}_{03,3} + \{ \Gamma^{3}_{03}\Gamma^{0}_{00} + \Gamma^{3}_{13}\Gamma^{1}_{00} + \Gamma^{3}_{23}\Gamma^{2}_{00} + \Gamma^{3}_{33}\Gamma^{3}_{00}\} \\
		- \{\Gamma^{3}_{00}\Gamma^{0}_{03} + \Gamma^{3}_{10}\Gamma^{1}_{03} + \Gamma^{3}_{20}\Gamma^{2}_{03} + \Gamma^{3}_{30}\Gamma^{3}_{03} \}.
	\end{split}
\end{equation*}
Substituting the nonvanishing Christoffel symbols, we find
\begin{equation*}
	\begin{split}
	R_{00} = 0 - 0 + \Big\{  0\cdot 0 + \dfrac{B'}{2B} \cdot \dfrac{B'}{2A} + 0\cdot 0 + 0 \cdot 0 \Big\} - \Big\{0\cdot 0 + \dfrac{B'}{2B}\cdot\dfrac{B'}{2A} + 0 \cdot 0 + 0\cdot 0  \Big \} \\
	+ \dfrac{B''}{2A} - \dfrac{A'B'}{2A^2} - 0 + \Big\{0\cdot0 + \dfrac{A'}{2A}\cdot \dfrac{B'}{2A} + 0\cdot0 + 0\cdot0   \Big\} - \Big\{\dfrac{B'}{2A}\cdot \dfrac{B'}{2B} + 0\cdot 0 + 0\cdot 0 + 0\cdot 0  \Big\} \\
	+ 0 - 0 + \Big\{0 \cdot0 + \dfrac{1}{r}\cdot \dfrac{B'}{2A} + 0\cdot 0 + 0 \cdot 0 \Big \} - \Big \{ 0\cdot 0 + 0 \cdot 0 + 0\cdot 0 + 0\cdot 0  \Big\} \\
	+ 0 - 0 + \Big\{0\cdot0 + \dfrac{1}{r}\cdot \dfrac{B'}{2A} + \cot \theta \cdot 0 + 0 \cdot 0   \Big\} - \Big\{0\cdot 0 + 0 \cdot0 + 0 \cdot 0 + 0 \cdot 0 \Big\}.
	\end{split}
\end{equation*}
Therefore,
\begin{equation*}
	R_{00} = \dfrac{B'^2}{4AB} - \dfrac{B'^2}{4AB} + \dfrac{B''}{2A} - \dfrac{A'B'}{2A^2} + \dfrac{A'B'}{4A^2} - \dfrac{B'^2}{4AB} + \dfrac{B'}{2rA} + \dfrac{B'}{2rA}
\end{equation*}
which yields
\begin{equation*}
	R_{00} = \dfrac{B''}{2A} - \dfrac{A'B'}{4A^2} - \dfrac{B'^2}{4AB} + \dfrac{B'}{rA}.
\end{equation*}
It should be noted that, in the term $\Gamma^{1}_{00,1}$, the quotient rule has been used, since both $B'$ and $A$ depend on $r$.

Proceeding analogously, the $R_{11}$ component is obtained from
\begin{equation*}
	\begin{split}
	R_{11} = \Gamma^0_{11,0} - \Gamma^{0}_{10,1} + \Big\{\Gamma^{0}_{00}\Gamma^{0}_{11} + \Gamma^{0}_{10}\Gamma^{1}_{11} + \Gamma^{0}_{20}\Gamma^{2}_{11} + \Gamma^{0}_{30}\Gamma^{3}_{11}   \Big\} \\
	- \Big\{\Gamma^{0}_{01}\Gamma^{0}_{10} + \Gamma^{0}_{11}\Gamma^{1}_{10} + \Gamma^{0}_{21}\Gamma^{2}_{10} + \Gamma^{0}_{31}\Gamma^{3}_{10}   \Big\} \\
	+ \Gamma^{1}_{11,1} - \Gamma^{1}_{11,1} + \Big\{ \Gamma^{1}_{01}\Gamma^{0}_{11} + \Gamma^{1}_{11}\Gamma^{1}_{11} + \Gamma^{1}_{21}\Gamma^{2}_{11} + \Gamma^{1}_{31}\Gamma^{3}_{11}  \Big\} \\
	- \Big\{\Gamma^{1}_{01}\Gamma^{0}_{11} + \Gamma^{1}_{11}\Gamma^{1}_{11} + \Gamma^{1}_{21}\Gamma^{2}_{11} + \Gamma^{1}_{31}\Gamma^{3}_{11}\Big\} \\
	+ \Gamma^{2}_{11,2} - \Gamma^{2}_{12,1} + \Big\{\Gamma^{2}_{02}\Gamma^{0}_{11} + \Gamma^{2}_{12}\Gamma^{1}_{11} + \Gamma^{2}_{22}\Gamma^{2}_{11} + \Gamma^{2}_{32}\Gamma^{3}_{11} \Big\} \\
	- \Big\{\Gamma^{2}_{01}\Gamma^{0}_{12} + \Gamma^{2}_{11}\Gamma^{1}_{12} + \Gamma^{2}_{21}\Gamma^{2}_{12} + \Gamma^{2}_{31}\Gamma^{3}_{12}  \Big\} \\
	+ \Gamma^{3}_{11,3} - \Gamma^{3}_{13,1} + \Big\{\Gamma^{3}_{03}\Gamma^{0}_{11} + \Gamma^{3}_{13}\Gamma^{1}_{11} + \Gamma^{3}_{23}\Gamma^{2}_{11} + \Gamma^{3}_{33}\Gamma^{3}_{11} \Big\} \\
	- \Big\{\Gamma^{3}_{01}\Gamma^{0}_{13} + \Gamma^{3}_{11}\Gamma^{1}_{13} + \Gamma^{3}_{21}\Gamma^{2}_{13} + \Gamma^{3}_{31}\Gamma^{3}_{13}\Big\}.
	\end{split}
\end{equation*}
Substituting the corresponding Christoffel symbols,
\begin{equation*}
	\begin{split}
	R_{11} = 0 - \dfrac{B''}{2B} + \dfrac{B'^2}{2B^2} + \Big\{0\cdot 0 + \dfrac{B'}{2B}\cdot \dfrac{A'}{2A} + 0\cdot 0 + 0 \cdot 0 \Big\} - \Big\{\dfrac{B'}{2B}\cdot \dfrac{B'}{2B} + 0 \cdot 0 + 0\cdot 0 + 0 \cdot 0 \Big\} \\
	+ \dfrac{A''}{2A} - \dfrac{A'^2}{2A^2} - \dfrac{A''}{2A} + \dfrac{A'^2}{2A^2} + \Big\{0 \cdot 0 + \dfrac{A'}{2A}\cdot \dfrac{A'}{2A} + 0\cdot0 + 0\cdot0\Big\} - \Big\{0\cdot0 + \dfrac{A'}{2A}\cdot \dfrac{A'}{2A} + 0\cdot0 + 0\cdot0 \Big\} \\
	+0 - \Big(- \dfrac{1}{r^2} \Big) + \Big\{0\cdot0 + \dfrac{1}{r}\cdot \dfrac{A'}{2A} + 0\cdot0 + 0\cdot0 \Big\} - \Big\{0\cdot0 + 0\cdot0 + \dfrac{1}{r}\cdot \dfrac{1}{r} + 0\cdot0  \Big\} \\
	+ 0 - \Big(- \dfrac{1}{r^2} \Big) + \Big\{0\cdot0 + \dfrac{1}{r}\cdot \dfrac{A'}{2A} + \cot \theta \cdot 0 + 0 \cdot 0 \Big\} - \Big\{0\cdot0 + 0\cdot0 + 0\cdot0 + \dfrac{1}{r}\cdot \dfrac{1}{r} \Big\}.
	\end{split}
\end{equation*}
Thus,
\begin{equation*}
	R_{11} = - \dfrac{B''}{2B} + \dfrac{B'^2}{2B^2} + \dfrac{A'B'}{4AB} - \dfrac{B'^2}{4B^2} + \dfrac{A'^2}{4A^2} - \dfrac{A'^2}{4A^2} + \dfrac{1}{r^2} + \dfrac{A'}{2rA} - \dfrac{1}{r^2} + \dfrac{1}{r^2} + \dfrac{A'}{2rA} - \dfrac{1}{r^2},
\end{equation*}
which simplifies to
\begin{equation*}
	R_{11} = \dfrac{A'}{rA} + \dfrac{A'B'}{4AB} + \dfrac{B'^2}{4B^2} - \dfrac{B''}{2B}.
\end{equation*}

The four nonvanishing components of the Ricci tensor are therefore
\begin{eqnarray}
	&&R_{00} = \dfrac{B''}{2A} - \dfrac{A'B'}{4A^2} - \dfrac{B'^2}{4AB} + \dfrac{B'}{rA}, \nonumber\\
	&&R_{11} = \dfrac{A'}{rA} + \dfrac{A'B'}{4AB} + \dfrac{B'^2}{4B^2} - \dfrac{B''}{2B},\\
	&&R_{22} = 1 - \dfrac{1}{A} + \dfrac{rA'}{2A^2} - \dfrac{rB'}{2AB}, \nonumber\\
	&&R_{33} = R_{22}\sin^2\theta. \nonumber
\end{eqnarray}

The curvature scalar is then obtained as
\begin{equation}
	R = g^{\mu \nu}R_{\mu \nu} = g^{00}R_{00} + g^{11}R_{11} + g^{22}R_{22} + g^{33}R_{33}.
\end{equation}
Substituting the previous results,
\begin{equation*}
	\begin{split}
	R = \dfrac{1}{B} \Big( \dfrac{B''}{2A} - \dfrac{A'B'}{4A^2} - \dfrac{B'^2}{4AB} + \dfrac{B'}{rA} \Big) - \dfrac{1}{A} \Big(\dfrac{A'}{rA} + \dfrac{A'B'}{4AB} + \dfrac{B'^2}{4B^2} - \dfrac{B''}{2B} \Big) \\
	- \dfrac{1}{r^2} \Big( 1 - \dfrac{1}{A} + \dfrac{rA'}{2A^2} - \dfrac{rB'}{2AB}\Big) - \dfrac{1}{r^2 \sin^2 \theta}\Bigg[\Big(1 - \dfrac{1}{A} + \dfrac{rA'}{2A^2} - \dfrac{rB'}{2AB} \Big) \sin^2 \theta \Bigg].
	\end{split}
\end{equation*}
After simplification,
\begin{equation*}
	\begin{split}
	R = \dfrac{B''}{2AB} - \dfrac{A'B'}{4A^2B} - \dfrac{B'^2}{4AB^2} + \dfrac{B'}{rAB} - \dfrac{A'}{rA^2} - \dfrac{A'B'}{4A^2B} - \dfrac{B'^2}{4AB^2} + \dfrac{B''}{2AB}\\
	-\dfrac{1}{r^2} + \dfrac{1}{r^2A} - \dfrac{A'}{2rA^2} + \dfrac{B'}{2rAB} -\dfrac{1}{r^2} + \dfrac{1}{r^2A} - \dfrac{A'}{2rA^2} + \dfrac{B'}{2rAB},
	\end{split}
\end{equation*}
which yields
\begin{equation*}
	R = \dfrac{B''}{AB} - \dfrac{A'B'}{2A^2B} - \dfrac{B'^2}{2AB^2} + \dfrac{2B'}{rAB} -\dfrac{2A'}{rA^2} + \dfrac{2}{r^2A} - \dfrac{2}{r^2}.
\end{equation*}
A more compact form is
\begin{equation}
	R = \dfrac{B''}{AB} - \dfrac{B'}{2AB}\Big( \dfrac{A'}{A} + \dfrac{B'}{B} \Big) - \dfrac{2}{rA}\Big( \dfrac{A'}{A} - \dfrac{B'}{B}\Big) - \dfrac{2}{r^2}\Big(1 - \dfrac{1}{A}\Big).
\end{equation}

We can now determine the Einstein tensor from
\begin{equation}
	G_{\mu \nu} = R_{\mu \nu} - \dfrac{1}{2}g_{\mu \nu}R.
\end{equation}
For the first component, one finds
\begin{equation*}
	\begin{split}
	G_{00} = R_{00} - \dfrac{1}{2}g_{00}R = \dfrac{B''}{2A} - \dfrac{A'B'}{4A^2} - \dfrac{B'^2}{4AB} + \dfrac{B'}{rA} - \dfrac{1}{2}B\Bigg[\dfrac{B''}{AB} - \dfrac{B'}{2AB}\Big(\dfrac{A'}{A} + \dfrac{B'}{B} \Big) \\
	- \dfrac{2}{rA}\Big( \dfrac{A'}{A} - \dfrac{B'}{B} \Big) - \dfrac{2}{r^2}\Big( 1 - \dfrac{1}{A}\Big)\Bigg].
	\end{split}
\end{equation*}
Expanding and simplifying,
\begin{equation*}
	\begin{split}
	G_{00} = \dfrac{B''}{2A} - \dfrac{A'B'}{4A^2} - \dfrac{B'^2}{4AB} + \dfrac{B'}{rA} - \dfrac{B''}{2A} + \dfrac{B'}{4A}\Big(\dfrac{A'}{A} + \dfrac{B'}{B} \Big) + \dfrac{B}{rA}\Big(\dfrac{A'}{A} - \dfrac{B'}{B}\Big) + \dfrac{B}{r^2}\Big(1 - \dfrac{1}{A} \Big) \\
	= -\dfrac{A'B'}{4A^2} - \dfrac{B'^2}{4AB} + \dfrac{B'}{rA} + \dfrac{A'B'}{4A^2} + \dfrac{B'^2}{4AB} + \dfrac{A'B}{rA^2} - \dfrac{B'}{rA} + \dfrac{B}{r^2}\Big(1 - \dfrac{1}{A} \Big),
	\end{split}
\end{equation*}
so that
\begin{equation}
	G_{00} = \dfrac{A'B}{rA^2} + \dfrac{B}{r^2}\Big(1 - \dfrac{1}{A} \Big).
\end{equation}
The remaining components are obtained in the same manner, yielding
\begin{eqnarray}
	&&G_{00} =  \dfrac{A'B}{rA^2} + \dfrac{B}{r^2}\Big(1 - \dfrac{1}{A} \Big), \nonumber\\
	&&G_{11} = \dfrac{B'}{rB} - \dfrac{A}{r^2}\Big(1 - \dfrac{1}{A}\Big),\\
	&&G_{22} = \dfrac{r^2B''}{2AB} - \dfrac{r^2B'}{4AB}\Big(\dfrac{A'}{A} + \dfrac{B'}{B} \Big) - \dfrac{r}{2A}\Big(\dfrac{A'}{A} - \dfrac{B'}{B} \Big), \nonumber\\
	&&G_{33} = G_{22}\sin^2 \theta. \nonumber
\end{eqnarray}

Since $G_{\mu\nu}$ is diagonal, the above form is consistent with a static, isotropic perfect fluid in a spherically symmetric spacetime. Using $G_{\mu \nu} = kT_{\mu \nu}$ and adopting $k = 8 \pi G$, one obtains
\begin{eqnarray}
	&&\dfrac{A'B}{rA^2} + \dfrac{B}{r^2}\Big(1 - \dfrac{1}{A} \Big) = 8\pi G \varepsilon B, \nonumber\\
	&&\dfrac{B'}{rB} - \dfrac{A}{r^2}\Big(1 - \dfrac{1}{A}\Big) = 8 \pi G pA,\\
	&&\dfrac{r^2B''}{2AB} - \dfrac{r^2B'}{4AB}\Big(\dfrac{A'}{A} + \dfrac{B'}{B} \Big) - \dfrac{r}{2A}\Big(\dfrac{A'}{A} - \dfrac{B'}{B} \Big) = 8 \pi G pr^2. \nonumber
\end{eqnarray}
The $G_{33}$ equation is not independent, since it is directly related to the $G_{22}$ component by a factor of $\sin^2\theta$.

Rewriting the first of the equations above,
\begin{equation*}
	\begin{split}
	\dfrac{A'}{rA^2} + \dfrac{1}{r^2}\Big(1 - \dfrac{1}{A}\Big) = 8\pi G \varepsilon \\
	\dfrac{rA'}{A^2} + \Big( 1 - \dfrac{1}{A}\Big) = 8\pi G \varepsilon r^2 \\
	\dfrac{rA'}{A^2} - \dfrac{1}{A} + 1 = 8 \pi G \varepsilon r^2 \\
	1 - 8 \pi G r^2 \varepsilon = \dfrac{1}{A} - \dfrac{rA'}{A^2}.
	\end{split}
\end{equation*}
The integration of the right-hand side gives $r/A$. Hence,
\begin{equation*}
	\begin{split}
	\int (1 - 8\pi G r'^2 \varepsilon)\,dr' = \dfrac{r}{A} \\
	r - \int 8\pi G r'^2 \varepsilon \,dr' = \dfrac{r}{A} \\
	1 - \dfrac{1}{r}\int 8\pi G r'^2 \varepsilon \,dr' = \dfrac{1}{A} \\
	A(r) = \Big(1 - \dfrac{1}{r}\int 8 \pi G r'^2 \varepsilon(r') \,dr' \Big)^{-1}.
	\end{split}
\end{equation*}
This may be written in the more compact form
\begin{equation}
	A(r) = \Big(1 - \dfrac{2GM(r)}{r}\Big)^{-1},
\end{equation}
where
\begin{equation}
	M(r) = \int_{0}^{r} 4 \pi r'^2 \varepsilon(r') \, dr'.
\end{equation}

The integration of $\varepsilon(r)$ is performed over the stellar interior and therefore gives the mass enclosed within the radius $r$. Equation (B.27) is the mass-continuity equation. In the Newtonian case, the corresponding dependence involves the mass density alone. In the relativistic regime, however, one must work with the energy density $\varepsilon$. If desired, this quantity may be decomposed into a rest-mass contribution plus an internal-energy term. At the stellar surface, located at $r=R$, one has
\begin{equation}
	p(R)=0.
\end{equation}

We may now substitute the result of Eq.~(B.25) into the second of Eqs.~(B.24). Defining
\begin{equation}
	B(r)=e^{2\Phi(r)},
\end{equation}
we obtain an expression for the gravitational potential:
\begin{equation*}
	\begin{split}
	\dfrac{B'}{rB} - \dfrac{A}{r^2}\Big(1 - \dfrac{1}{A}\Big) = 8\pi G p A\\
	\dfrac{B'}{rB} - \dfrac{A}{r^2}\Big(1 -  \Big(1 - \dfrac{2GM}{r}\Big)\Big) = 8\pi G p A\\
	\dfrac{B'}{B} - \dfrac{A}{r}\Big(1 -\Big(1 - \dfrac{2GM}{r}\Big)\Big) = 8\pi G rp A.
	\end{split}
\end{equation*}
Since $B=e^{2\Phi}$, it follows that $B' = 2\Phi' B$, and therefore $\Phi' = B'/(2B)$. Thus,
\begin{equation*}
	\begin{split}
	\dfrac{2\Phi'B}{B} - \dfrac{A}{r} + \dfrac{A}{r}\Big(1 - \dfrac{2GM}{r}\Big) = 8\pi G rp A\\
	2\Phi' = 8\pi G rp A + \dfrac{A}{r} - \dfrac{A}{r}\Big(1 - \dfrac{2GM}{r}\Big)\\
	\Phi' = 4\pi G rp A + \dfrac{A}{2r} - \dfrac{A}{2r}\Big(1 - \dfrac{2GM}{r}\Big)\\
	\Phi' = \Big(4\pi G rp + \dfrac{1}{2r} - \dfrac{1}{2r}\Big(1 - \dfrac{2GM}{r}\Big)\Big)A\\
	\Phi' = \Big(4 \pi G rp + \dfrac{1}{2r} - \dfrac{1}{2r} + \dfrac{GM}{r^2}\Big)A\\
	\Phi' = \dfrac{GM}{r^2}\Big(1 + \dfrac{4\pi r^3 p}{M}\Big)A.
	\end{split}
\end{equation*}
Using the expression for $A(r)$ given in Eq.~(B.25), we obtain
\begin{equation}
	 \Phi'(r) = \dfrac{GM(r)}{r^2}\Big(1 + \dfrac{4\pi r^3 p(r)}{M(r)}\Big)\Big(1 - \dfrac{2GM(r)}{r}\Big)^{-1}.
\end{equation}

A second expression for $\Phi'$ follows from the covariant conservation of the energy-momentum tensor. In GR, the pressure gradient in a gravitational field described by the potential $\Phi$ satisfies
\begin{equation}
	\Phi'(r) = - \dfrac{p'(r)}{\varepsilon(r) + p(r)}.
\end{equation}
This result is obtained from the conservation law $\nabla^\mu T_{\mu\nu}=0$. Comparing Eqs.~(B.29) and (B.30), one obtains
\begin{equation*}
	- \dfrac{p'(r)}{\varepsilon(r) + p(r)} = \dfrac{GM(r)}{r^2}\Big(1 + \dfrac{4\pi r^3 p(r)}{M(r)}\Big)\Big(1 - \dfrac{2GM(r)}{r}\Big)^{-1}.
\end{equation*}
Rearranging the above expression yields
\begin{equation}
	\dfrac{dp(r)}{dr} = - \dfrac{GM(r)\varepsilon(r)}{r^2} \Big(1 + \dfrac{p(r)}{\varepsilon(r)}\Big) \Big(1 + \dfrac{4\pi r^3 p(r)}{M(r)} \Big) \Big(1 - \dfrac{2GM(r)}{r}\Big)^{-1}.
\end{equation}

Equations (B.26) and (B.31) correspond to the reduction of Einstein’s field equations inside a static, isotropic, and spherically symmetric star. These two equations describe hydrostatic equilibrium in GR and are known as the Tolman-Oppenheimer-Volkoff equations. Once an equation of state relating pressure and energy density is specified, they allow one to determine observable stellar properties such as mass and radius.

It is straightforward to write Eq.~(B.31) in a more compact form. Setting $G=1$, one obtains
\begin{equation}
	\frac{dp(r)}{dr} = - \frac{[\varepsilon(r) + p(r)][M(r) + 4 \pi r^3 p(r)]}{r[r - 2M(r)]}.
\end{equation}

\end{appendices}

\begin{appendices}
\chapter{Derivation of the $\mathcal{D}$-TOV Equations}

In this appendix, we present the detailed derivation of the $\mathcal{D}$-TOV equations. We begin with the metric
\begin{equation}
\label{dtov1}
ds^2 = e^{2\Phi(r)}dt^2 - \left( 1 - \dfrac{2m(r)}{r}\right)^{-\mathcal{D}}dr^2 - r^2\left( d\theta^2 + \sin^2 \theta\, d\varphi^2 \right).
\end{equation}

The first step consists in determining the Christoffel symbols, defined by
\begin{equation}
\Gamma^{\alpha}_{\beta\gamma} = \frac{g^{\alpha\delta}}{2}\left(g_{\beta\delta,\gamma} + g_{\gamma\delta,\beta} - g_{\beta\gamma,\delta}\right).
\end{equation}
For the metric \eqref{dtov1}, the nonvanishing Christoffel symbols are
\begin{eqnarray}
	&&\Gamma^{t}_{tr} = \Gamma^{t}_{rt} = \Phi', 
	\qquad 
	\Gamma^{r}_{tt} = e^{2\Phi}\left(1 - \dfrac{2m}{r}\right)^{\mathcal{D}}\Phi', \nonumber\\
	&&\Gamma^{r}_{rr} = \dfrac{\mathcal{D}(rm' - m)}{r(r - 2m)}, 
	\qquad
	\Gamma^{r}_{\theta \theta} = -r \left(1 - \dfrac{2m}{r}\right)^{\mathcal{D}}, \nonumber\\
	&&\Gamma^{r}_{\varphi \varphi} = -r \left(1 - \dfrac{2m}{r}\right)^{\mathcal{D}}\sin^2\theta,
	\qquad
	\Gamma^{\theta}_{r \theta} = \Gamma^{\theta}_{\theta r} = \dfrac{1}{r}, \\
	&&\Gamma^{\theta}_{\varphi \varphi} = -\sin \theta \cos \theta,
	\qquad
	\Gamma^{\varphi}_{r \varphi} = \Gamma^{\varphi}_{\varphi r} = \dfrac{1}{r}, \nonumber\\
	&&\Gamma^{\varphi}_{\theta \varphi} = \Gamma^{\varphi}_{\varphi \theta} = \cot \theta. \nonumber
\end{eqnarray}

Once the Christoffel symbols have been obtained, the Ricci tensor is calculated from
\begin{equation}
R_{\mu \nu} = \Gamma^\lambda_{\mu \nu,\lambda} - \Gamma^\lambda_{\mu \lambda,\nu}
+ \Gamma^\lambda_{\sigma \lambda}\Gamma^\sigma_{\mu \nu}
- \Gamma^\lambda_{\sigma \nu}\Gamma^\sigma_{\mu \lambda}.
\end{equation}

Its nonvanishing components are
\begin{eqnarray}
&& R_{tt} = \frac{e^{2\Phi}a^{\mathcal{D}}}{r(r-2m)}
\left[-\Phi' \mathcal{D} (rm' - m) - 2rm\Phi'' - 2rm\Phi'^2 - 4m\Phi' \right] \nonumber\\
&&\hspace{1.6cm}
+ e^{2\Phi}a^{\mathcal{D}}
\left[ \frac{r^2\Phi'' + r^2\Phi'^2 + 2r\Phi'}{r(r-2m)} \right], \\
&&R_{rr} = -\Phi'^2 - \dfrac{\mathcal{D} m(2 + r\Phi')}{r^2(r - 2m)}
+ \dfrac{\mathcal{D} m'(2 + r\Phi')}{r(r-2m)} - \Phi'', \nonumber\\
&&R_{\theta \theta} = 1 - a^{\mathcal{D}} + \dfrac{\mathcal{D} a^\mathcal{D} (-m + rm')}{r - 2m} - ra^\mathcal{D} \Phi', \nonumber\\
&&R_{\varphi \varphi} = \sin^2\theta\, R_{\theta \theta}, \nonumber
\end{eqnarray}
where, for convenience, we have defined
\begin{equation}
a = \left(1 - \frac{2m}{r}\right).
\end{equation}

Having determined the Ricci tensor, the Ricci scalar follows from
\begin{equation}
R = g^{\mu \nu}R_{\mu \nu},
\end{equation}
which yields
\begin{eqnarray}
&&R = -\dfrac{2}{r^2(r-2m)}\Bigg( r - 2m - \Big( 1 - \dfrac{2m}{r}\Big)^{\mathcal{D}}
\nonumber\\
&&\times \Big[r + r(2 + r\Phi')(-\mathcal{D} m' + r\Phi') + r^3\Phi'' 
\nonumber\\
&&\qquad\qquad
+ m\Big(2(\mathcal{D} - 1) + r\big(\Phi'(\mathcal{D} - 4 - 2r\Phi') - 2r\Phi''\big)\Big)\Big]\Bigg).
\end{eqnarray}

We may now write the Einstein tensor as
\begin{equation}
G_{\mu \nu} = R_{\mu \nu} - \frac{1}{2}g_{\mu \nu}R,
\end{equation}
which appears in Einstein's field equations,
\begin{equation}
G_{\mu \nu} = 8\pi T_{\mu \nu},
\end{equation}
where $T_{\mu \nu}$ denotes the energy-momentum tensor. Once the line element \eqref{dtov1} and the energy-momentum tensor are specified, the relevant components of the Einstein tensor, namely $G_{tt}$ and $G_{rr}$, are given by
\begin{eqnarray}
&&G_{tt} =
\dfrac{e^{2\Phi}\Bigg[-2m\Big(1 - \Big(1 - \dfrac{2m}{r}\Big)^{\mathcal{D}} + \mathcal{D}\Big(1 - \dfrac{2m}{r} \Big)^\mathcal{D}\Big)}{r^2(r-2m)} \nonumber\\
&&\hspace{1.5cm}
+\frac{e^{2\Phi}r^2\Big(1 - \Big(1 - \dfrac{2m}{r}\Big)^{\mathcal{D}}\Big)+2\mathcal{D}m' \Big(1 - \dfrac{2m}{r}\Big)^\mathcal{D}\Bigg]}{r^2(r-2m)}, \\
&&G_{rr} = \dfrac{1 - \Big(1 - \dfrac{2m}{r}\Big)^{-\mathcal{D}}+2r\Phi'}{r^2}. \nonumber
\end{eqnarray}

From the $G_{rr}$ component, the $\mathcal{D}$-TOV equation can be obtained by imposing $G_{rr} = 8\pi T_{rr}$. Since, for the metric \eqref{dtov1},
\[
T_{rr}=-p\,g_{rr}=p\left(1-\frac{2m}{r}\right)^{-\mathcal{D}},
\]
one then finds
\begin{equation}
\Phi' = \dfrac{-\dfrac{1}{2}\Big(1 - \dfrac{2m}{r}\Big)^\mathcal{D} + 4\pi r^2 p + \dfrac{1}{2}}{r\Big(1 - \dfrac{2m}{r}\Big)^\mathcal{D}}.
\end{equation}
Equating this result to the expression obtained from the covariant conservation of the energy-momentum tensor, one arrives at
\begin{equation}
\dfrac{dp}{dr}
=
- \dfrac{(\varepsilon + p)\left[8\pi r^2 p -\left(1 - \dfrac{2m}{r}\right)^\mathcal{D} + 1\right]}
{2\left(1 - \dfrac{2m}{r}\right)^{\mathcal{D}}r}.
\end{equation}

The mass-continuity equation may be obtained from the $G_{tt}$ component. For this purpose, it is convenient to rewrite this component in the form
\begin{equation}
G_{tt} = \dfrac{A'B}{rA^2} + \dfrac{B}{r^2}\left(1 - \dfrac{1}{A} \right),
\end{equation}
where $A$ denotes the positive radial metric coefficient and $B$ the corresponding temporal metric coefficient. This compact form can be readily verified for a general line element of the form
\begin{equation}
ds^2 = B(r)dt^2 - A(r)dr^2 - r^2(d\theta^2 + \sin^2 \theta\, d\varphi^2).
\end{equation}

For simplicity, let us first consider the case $\mathcal{D}=1$, and only afterwards generalize the result to arbitrary $\mathcal{D}$. Using $G_{tt} = 8\pi T_{tt}$, we obtain
\begin{equation}
\dfrac{A'B}{rA^2} + \dfrac{B}{r^2}\left(1 - \dfrac{1}{A} \right) = 8\pi \varepsilon B.
\end{equation}
Integrating the above equation gives
\begin{equation}
A = \left(1 - \dfrac{1}{r}\int 8 \pi r^2 \varepsilon \, dr \right)^{-1},
\end{equation}
or, equivalently,
\begin{equation}
A = \left( 1 - \frac{2m(r)}{r}\right)^{-1},
\end{equation}
where the mass function is defined by
\begin{equation}
m(r) = \int_0^{r} 4\pi \tilde{r}^2 \varepsilon(\tilde{r})\, d\tilde{r}.
\end{equation}

The above expression corresponds to the standard spherically symmetric case. In order to generalize it to the deformed configuration considered in this work, we adopt the geometric prescription that the spherical radial factor $r^2$ appearing in the mass integral is replaced by the spheroidal combination $rz$. By defining the deformation through
\begin{equation}
z = \mathcal{D}r,
\end{equation}
the mass function can still be written in terms of the radial coordinate $r$, yielding
\begin{equation}
m(r) = \int_0^r 4\pi \tilde{r}^{\,2}\mathcal{D}\,\varepsilon(\tilde{r})\, d\tilde{r}.
\end{equation}
Equivalently, one obtains the mass-continuity equation
\begin{equation}
\frac{dm(r)}{dr}=4\pi r^2\mathcal{D}\,\varepsilon(r).
\end{equation}
This expression constitutes the mass-continuity equation for the deformed configuration described by the $\mathcal{D}$-dependent metric.
\end{appendices}

\begin{appendices}
\chapter{TOV equation in $f(R,T)$ Gravity}
In order to derive the Tolman-Oppenheimer-Volkoff equation in the $f(R,T)$ formalism, an additional contribution must be taken into account. We begin with the action
\begin{equation}
	S = \int \left(\frac{1}{16 \pi G} f(R,T)\sqrt{-g}\right)d^4x + S_m,
\end{equation}
where $S_m$ is the matter action defined. Varying the action and imposing $\delta S = 0$, we obtain
\begin{equation}
	\delta S = \int \frac{\delta}{\delta g^{\mu \nu}}\left( \frac{1}{16 \pi G} f(R,T)\sqrt{-g} \right)\delta g^{\mu \nu}d^4 x + \int \frac{\delta}{\delta g^{\mu \nu}}\left( \mathcal{L}_m\sqrt{-g}\right)\delta g^{\mu \nu}d^4 x = 0,
\end{equation}
where the functional derivative given in Eq.~(A.5) has been used. Therefore we have
\begin{equation}
	\frac{1}{16 \pi G}\int \left( \sqrt{-g} \frac{\delta f(R,T)}{\delta g^{\mu \nu}} + f(R,T) \frac{\delta(\sqrt{-g})}{\delta g^{\mu \nu}}\right)\delta g^{\mu \nu}d^4 x = - \int \frac{\delta}{\delta g^{\mu \nu}}\left(\mathcal{L}_m\sqrt{-g}\right)\delta g^{\mu \nu}d^4x.
\end{equation}

We now use the result of Eq.~(A.7), while for the matter sector on the right-hand side we use Eq.~(A.10). Since both integrands are defined over spacetime, they may be written under a single integral:
\begin{equation}
	\int \left[\frac{1}{16 \pi G} \left(\sqrt{-g} \frac{\delta f(R,T)}{\delta g^{\mu \nu}} + f(R,T)\frac{\delta \sqrt{-g}}{\delta g^{\mu \nu}}\right) + \sqrt{-g} \frac{\delta \mathcal{L}_m}{\delta g^{\mu \nu}} - \frac{\mathcal{L}_m}{2\sqrt{-g}}\frac{\delta g}{\delta g^{\mu \nu}}\right]\delta g^{\mu \nu}d^4 x = 0.
\end{equation}
Since the variation $\delta g^{\mu\nu}$ is arbitrary, the integrand must vanish, yielding
\begin{equation}
	\frac{1}{16 \pi G}\left( \sqrt{-g} \frac{\delta f(R,T)}{\delta g^{\mu \nu}} - \frac{f(R,T)}{2\sqrt{-g}}\frac{\delta g}{\delta g^{\mu \nu}}\right) = \frac{\mathcal{L}_m}{2\sqrt{-g}}\frac{\delta g}{\delta g^{\mu \nu}} - \sqrt{-g}\frac{\delta \mathcal{L}_m}{\delta g^{\mu \nu}},
\end{equation}
where Eq.~(A.7) was used in the second term inside parentheses. Multiplying both sides by $2/\sqrt{-g}$, we obtain
\begin{eqnarray*}
	&& \frac{1}{8\pi G}\left( \frac{\delta f(R,T)}{\delta g^{\mu \nu}} + \frac{f(R,T)}{2g} \frac{\delta g}{\delta g^{\mu \nu}}\right) = - \frac{\mathcal{L}_m}{g}\frac{\delta g}{\delta g^{\mu \nu}} - 2 \frac{\delta \mathcal{L}_m}{\delta g^{\mu \nu}},\\
	&& \frac{1}{8\pi G}\left( \frac{\delta f(R,T)}{\delta g^{\mu \nu}} + \frac{f(R,T)}{2g} \frac{\delta g}{\delta g^{\mu \nu}}\right) = - \frac{2}{\sqrt{-g}}\frac{\delta(\sqrt{-g}\mathcal{L}_m)}{\delta g^{\mu \nu}},\\
	&& \frac{1}{8\pi G}\left( \frac{\delta f(R,T)}{\delta g^{\mu \nu}} + \frac{1}{2g}f(R,T) \frac{\delta g}{\delta g^{\mu \nu}}\right) = T_{\mu \nu},
\end{eqnarray*}
where, on the right-hand side, we have used the definition of the energy-momentum tensor given in Eq.~(A.15). Using Eq.~(A.23), we can write
\begin{equation}
	\frac{1}{8 \pi G}\left( \frac{\delta f(R,T)}{\delta g^{\mu \nu}} - \frac{1}{2}f(R,T)g_{\mu \nu}\right) = T_{\mu \nu},
\end{equation}
and therefore
\begin{equation}
	\frac{\delta f(R,T)}{\delta g^{\mu \nu}} - \frac{1}{2}f(R,T)g_{\mu \nu} = 8 \pi G T_{\mu \nu}. \label{fieldequationfRT}
\end{equation}

At this stage, we have obtained a generalized form of the Einstein field equations. We now decompose the variation of $f(R,T)$ into two parts:
\begin{equation}
	\frac{\delta R}{\delta g^{\mu \nu}} \frac{\partial f}{\partial R} + \frac{\delta T}{\delta g^{\mu \nu}}\frac{\partial f}{\partial T} - \frac{1}{2}f(R,T)g_{\mu \nu} = 8\pi G T_{\mu \nu}.
\end{equation}

Defining $\partial f/\partial R \equiv f_R(R,T)$ and $\partial f/\partial T \equiv f_T(R,T)$,  one finds
\begin{equation*}
	\frac{\delta R}{\delta g^{\mu \nu}} = \left( \frac{R_{\mu \nu}\delta g^{\mu \nu} + g_{\mu \nu}\Box \delta g^{\mu \nu} - \nabla_\mu \nabla_\nu \delta g^{\mu \nu}}{\delta g^{\mu \nu}}\right)f_R(R,T),
\end{equation*}
that is,
\begin{equation}
	\frac{\delta R}{\delta g^{\mu \nu}} = \left(R_{\mu \nu} + g_{\mu \nu}\Box - \nabla_\mu \nabla_\nu \right)f_R(R,T).
\end{equation}
Hence,
\begin{equation}
	R_{\mu \nu}f_R(R,T) + \left(g_{\mu \nu}\Box - \nabla_\mu \nabla_\nu \right)f_R(R,T) + \frac{\delta T}{\delta g^{\mu \nu}}f_T(R,T) - \frac{1}{2}f(R,T)g_{\mu \nu} = 8\pi G T_{\mu \nu}. \label{fRT}
\end{equation}

The trace of the energy-momentum tensor is defined as
\begin{equation}
	T = g^{\alpha \beta}T_{\alpha \beta},
\end{equation}
and its variation is
\begin{equation}
	\delta T = T_{\alpha \beta}\delta g^{\alpha \beta} + g^{\alpha \beta}\delta T_{\alpha \beta}.
\end{equation}
Applying this result to Eq.~(\ref{fRT}), we obtain
\begin{equation}
	\frac{\delta T}{\delta g^{\mu \nu}} = \frac{T_{\alpha \beta}\delta g^{\mu \nu}}{\delta g^{\mu \nu}} + \frac{g^{\alpha \beta}\delta T_{\alpha \beta}}{\delta g^{\mu \nu}} = T_{\mu \nu} + \Theta_{\mu \nu}, \label{deltaT}
\end{equation}
where
\begin{equation}
	\Theta_{\mu \nu} = g^{\alpha \beta}\frac{\delta T_{\alpha \beta}}{\delta g^{\mu \nu}} = -2T_{\mu \nu} - pg_{\mu \nu}.
\end{equation}

Substituting Eq.~(\ref{deltaT}) into Eq.~(\ref{fRT}), we obtain
\begin{equation*}
	R_{\mu \nu}f_R(R,T) + \frac{\delta T}{\delta g^{\mu \nu}}f_T(R,T) - \frac{1}{2}f(R,T)g_{\mu \nu} = 8\pi G T_{\mu \nu},
\end{equation*}
\begin{equation*}
	R_{\mu \nu}f_R(R,T) + f_T(R,T)(T_{\mu \nu} + \Theta_{\mu \nu}) - \frac{1}{2}f(R,T)g_{\mu \nu} = 8\pi G T_{\mu \nu},
\end{equation*}
which may be rewritten as
\begin{equation}
	R_{\mu \nu}f_R(R,T) - \frac{1}{2}f(R,T)g_{\mu \nu} = 8\pi G T_{\mu \nu} - f_T(R,T)T_{\mu \nu} - f_T(R,T)\Theta_{\mu \nu}.
\end{equation}

For the specific model considered in this work, it is convenient to write
\begin{equation}
	f(R,T) = R + f(T),
\end{equation}
so that $f_R(R,T)=1$ and the derivative terms involving $f_R$ vanish. Therefore,
\begin{equation*}
	R_{\mu \nu} - \frac{1}{2}Rg_{\mu \nu} - \frac{1}{2}f(T)g_{\mu \nu} = 8\pi G T_{\mu \nu} - f_T(R,T)T_{\mu \nu} - f_T(R,T)\Theta_{\mu \nu},
\end{equation*}
or equivalently,
\begin{equation}
	G_{\mu \nu} - \frac{1}{2}f(T)g_{\mu \nu} = 8\pi G T_{\mu \nu} - f_T(R,T)T_{\mu \nu} - f_T(R,T)\Theta_{\mu \nu}. \label{GfRT}
\end{equation}

We now apply the covariant derivative to Eq.~(\ref{GfRT}):
\begin{equation*}
	\nabla^\mu G_{\mu \nu} - \frac{1}{2}f_T\nabla^\mu Tg_{\mu \nu} = 8\pi G \nabla^\mu T_{\mu \nu} - T_{\mu \nu}(\nabla^\mu f_T) - f_T\nabla^\mu T_{\mu \nu} - \Theta_{\mu \nu}(\nabla^\mu f_T) -f_T \nabla^\mu \Theta_{\mu \nu},
\end{equation*}
and, setting $G=1$ and using $\nabla^\mu G_{\mu\nu}=0$, this becomes
\begin{equation*}
	- \frac{1}{2}f_T\nabla^\mu Tg_{\mu \nu} = (8\pi - f_T)\nabla^\mu T_{\mu \nu} - \nabla^\mu f_T(T_{\mu \nu} + \Theta_{\mu \nu}) - f_T \nabla^\mu \Theta_{\mu \nu}.
\end{equation*}
Therefore,
\begin{equation*}
	(8\pi - f_T)\nabla^\mu T_{\mu \nu} = \nabla^\mu f_T(T_{\mu \nu} + \Theta_{\mu \nu}) + f_T \nabla^\mu \Theta_{\mu \nu} - \frac{1}{2}f_T\nabla^\mu Tg_{\mu \nu},
\end{equation*}
which yields
\begin{equation}
	\nabla^\mu T_{\mu \nu} = \frac{f_T}{8\pi - f_T} \left[(T_{\mu \nu} + \Theta_{\mu \nu})\nabla^\mu \ln f_T + \nabla^\mu \Theta_{\mu \nu} - \frac{1}{2}\nabla^\mu Tg_{\mu \nu}\right],
\end{equation}
where we have used $\nabla^\mu f_T = f_T \nabla^\mu \ln f_T$.

Using
\begin{equation}
	\nabla^\mu \ln f_T = \frac{f_{TT}}{f_T}\nabla^\mu T,
\end{equation}
where $f_{TT}$ denotes the second derivative of $f(T)$, we obtain
\begin{equation*}
	\nabla^\mu T_{\mu \nu} = \frac{f_T}{8\pi - f_T} \left[(T_{\mu \nu} + \Theta_{\mu \nu}) \frac{f_{TT}}{f_T}\nabla^\mu T + \nabla^\mu \Theta_{\mu \nu} - \frac{1}{2}\nabla^\mu Tg_{\mu \nu}\right].
\end{equation*}

Assuming
\begin{equation}
\Theta_{\mu \nu} = -2T_{\mu \nu} - pg_{\mu \nu},
\end{equation}
it follows that
\begin{equation}
T_{\mu \nu} + \Theta_{\mu \nu} = -T_{\mu \nu} - pg_{\mu \nu}.
\end{equation}
For a perfect fluid, the trace is
\begin{equation}
T = \varepsilon - 3p.
\end{equation}
Thus,
\begin{equation*}
	\nabla^\mu T_{\mu \nu} = \frac{f_T}{8\pi - f_T} \left[(T_{\mu \nu} + \Theta_{\mu \nu}) \frac{f_{TT}}{f_T}\nabla^\mu(\varepsilon - 3p) - 2\nabla^\mu T_{\mu \nu} - \nabla_\nu p - \frac{1}{2}(\varepsilon' - 3p')\right],
\end{equation*}
where we have used $\nabla^\mu(pg_{\mu \nu}) = \nabla_\nu p$. Therefore,
\begin{equation*}
	\nabla^\mu T_{\mu \nu} = \frac{f_T}{8\pi - f_T} \left[(- T_{\mu \nu} - pg_{\mu \nu})\frac{f_{TT}}{f_T}(\varepsilon' - 3p') - p' - \frac{1}{2}\varepsilon' + \frac{3}{2}p'\right] - \frac{2f_T}{8\pi - f_T}\nabla^\mu T_{\mu \nu},
\end{equation*}
which may be rearranged as
\begin{eqnarray*}
	&&\nabla^\mu T_{\mu \nu} +  \frac{2f_T}{8\pi - f_T}\nabla^\mu T_{\mu \nu} = \frac{f_T}{8\pi - f_T} \left[(- T_{\mu \nu} - pg_{\mu \nu})\frac{f_{TT}}{f_T}(\varepsilon' - 3p') - p' - \frac{1}{2}\varepsilon' + \frac{3}{2}p'\right],\\
	&&\nabla^\mu T_{\mu \nu}\left(1 +  \frac{2f_T}{8\pi - f_T}\right) = \frac{f_T}{8\pi - f_T} \left[(- T_{\mu \nu} - pg_{\mu \nu})\frac{f_{TT}}{f_T}(\varepsilon' - 3p') - p' - \frac{1}{2}\varepsilon' + \frac{3}{2}p'\right],\\
	&&\nabla^\mu T_{\mu \nu}\left(\frac{8\pi + f_T}{8\pi - f_T}\right) = \frac{f_T}{8\pi - f_T} \left[(- T_{\mu \nu} - pg_{\mu \nu})\frac{f_{TT}}{f_T}(\varepsilon' - 3p') - p' - \frac{1}{2}\varepsilon' + \frac{3}{2}p'\right].
\end{eqnarray*}
Hence,
\begin{equation*}
	\nabla^\mu T_{\mu \nu} = \frac{f_T}{8\pi + f_T} \left[(- T_{\mu \nu} - pg_{\mu \nu})\frac{f_{TT}}{f_T}(\varepsilon' - 3p') - p' - \frac{1}{2}\varepsilon' + \frac{3}{2}p'\right],
\end{equation*}
or, more compactly,
\begin{equation}
	\nabla^\mu T_{\mu \nu} = \frac{f_T}{8\pi + f_T} \left[(- T_{\mu \nu} - pg_{\mu \nu})\frac{f_{TT}}{f_T}(\varepsilon' - 3p') + \frac{1}{2}p' - \frac{1}{2}\varepsilon'\right].
\end{equation}

Now consider the spherically symmetric metric
\begin{equation}
	ds^2 = e^{\upsilon(r)}dt^2 - e^{\omega(r)}dr^2 - r^2(d\theta^2 + \sin^2\theta d\varphi^2),
\end{equation}
for which
\begin{equation*}
	\nabla^\mu T_{\mu \nu} \equiv p' +  (\varepsilon + p)\frac{\upsilon'}{2}.
\end{equation*}
Therefore,
\begin{equation}
	p' + (\varepsilon + p)\frac{\upsilon'}{2} = \frac{f_T}{8\pi + f_T} \left[(- T_{\mu \nu} - pg_{\mu \nu})\frac{f_{TT}}{f_T}(\varepsilon' - 3p') + \frac{1}{2}p' - \frac{1}{2}\varepsilon'\right]. \label{plinha2}
\end{equation}

From Eq.~(\ref{GfRT}), we may also write
\begin{equation}
	G_{\mu \nu} = 8\pi T_{\mu \nu} - f_T(R,T)T_{\mu \nu} - f_T(R,T)\Theta_{\mu \nu} + \frac{1}{2}f(T)g_{\mu \nu},
\end{equation}
and using $\Theta_{\mu \nu} = -2T_{\mu \nu} - pg_{\mu \nu}$, this becomes
\begin{equation*}
	G_{\mu \nu} = 8\pi T_{\mu \nu} - f_TT_{\mu \nu} + 2f_T T_{\mu \nu}+ pf_T g_{\mu \nu} + \frac{1}{2}f(T)g_{\mu \nu},
\end{equation*}
that is,
\begin{equation*}
	G_{\mu \nu} = 8\pi T_{\mu \nu} + f_TT_{\mu \nu} + pf_T g_{\mu \nu} + \frac{1}{2}f(T)g_{\mu \nu},
\end{equation*}
or equivalently,
\begin{equation}
	G_{\mu \nu} = 8\pi T_{\mu \nu} + f_T(T_{\mu \nu} + p g_{\mu \nu}) + \frac{1}{2}f(T)g_{\mu \nu}. \label{GmunufRT}
\end{equation}

To obtain the modified TOV equation, we consider the $rr$ component,
\begin{equation}
	G_{rr} = \frac{1-e^\omega + r\upsilon'}{r^2},
\end{equation}
while
\begin{equation}
	T_{rr} = (\varepsilon + p)0\cdot0 - p(-e^\omega) = pe^\omega,
\end{equation}
since $g_{rr}=-e^\omega$. Therefore,
\begin{equation*}
	\frac{1-e^\omega + r\upsilon'}{r^2} = 8\pi pe^\omega + f_T(pe^\omega - p e^\omega) - \frac{1}{2}f(T)e^\omega,
\end{equation*}
and thus
\begin{equation*}
	\frac{1-e^\omega + r\upsilon'}{r^2} = 8\pi pe^\omega  - \frac{1}{2}f(T)e^\omega.
\end{equation*}
Multiplying through by $r^2$, we obtain
\begin{equation*}
	1-e^\omega + r\upsilon' = 8\pi pr^2e^\omega  - \frac{1}{2}f(T)r^2e^\omega,
\end{equation*}
so that
\begin{equation}
	\upsilon' = 8\pi pre^\omega  - \frac{1}{2}f(T)re^\omega +\frac{e^\omega - 1}{r}. \label{upsilonlinha}
\end{equation}

Substituting Eq.~(\ref{upsilonlinha}) into Eq.~(\ref{plinha2}), we obtain
\begin{equation*}
	p' + (\varepsilon + p)\frac{1}{2}\left[ 8\pi pre^\omega  - \frac{1}{2}f(T)re^\omega + \frac{e^\omega - 1}{r}\right] = \frac{f_T}{8\pi + f_T} \left[(- T_{\mu \nu} - pg_{\mu \nu})\frac{f_{TT}}{f_T}(\varepsilon' - 3p') + \frac{1}{2}p' - \frac{1}{2}\varepsilon'\right].
\end{equation*}
For the $rr$ component, one has
\begin{equation*}
	(- T_{\mu \nu} - pg_{\mu \nu}) = (-pe^\omega + pe^\omega)=0,
\end{equation*}
and therefore
\begin{equation*}
	p' + (\varepsilon + p)\frac{1}{2}\left[ 8\pi pre^\omega  - \frac{1}{2}f(T)re^\omega +\frac{e^\omega - 1}{r}\right] = \frac{f_T}{8\pi + f_T} \left[\frac{1}{2}p' - \frac{1}{2}\varepsilon'\right].
\end{equation*}
Rearranging,
\begin{equation*}
	p' =- (\varepsilon + p)\left[ 4\pi pre^\omega  - \frac{1}{4}f(T)re^\omega +\frac{e^\omega - 1}{2r}\right] +\frac{f_T}{2(8\pi + f_T)}[p' - \varepsilon'].
\end{equation*}

If we assume
\begin{equation}
	\varepsilon' = \left(\frac{d\varepsilon}{dp}\right)p',
\end{equation}
then
\begin{equation*}
	p' -\frac{f_T}{2(8\pi + f_T)}\left[p' - \frac{d\varepsilon}{dp} p'\right]=- (\varepsilon + p)\left[ 4\pi pre^\omega  - \frac{1}{4}f(T)re^\omega +\frac{e^\omega - 1}{2r}\right],
\end{equation*}
that is,
\begin{equation*}
	p'\left(1 -\frac{f_T}{2(8\pi + f_T)}\left(1 - \frac{d\varepsilon}{dp}\right)\right)=- (\varepsilon + p)\left[ 4\pi pre^\omega  - \frac{1}{4}f(T)re^\omega +\frac{e^\omega-1}{2r}\right].
\end{equation*}
Therefore,
\begin{equation*}
	p'=- (\varepsilon + p)\frac{\left[ 4\pi pre^\omega  - \frac{1}{4}f(T)re^\omega +\frac{e^\omega - 1}{2r}\right]}{1 -\frac{f_T}{2(8\pi + f_T)}\left(1 - \frac{d\varepsilon}{dp}\right)}.
\end{equation*}

Finally, using
\begin{equation}
	e^\omega = \left(1 - \frac{2m}{r}\right)^{-1} = \frac{r}{r-2m},
\end{equation}
we obtain
\begin{equation}
	p'=- (\varepsilon + p)\frac{\left[ \frac{4\pi pr^2}{r - 2m}  - \frac{f(T)r^2}{4(r-2m)} + \frac{m}{r(r-2m)}\right]}{1 -\frac{f_T}{2(8\pi + f_T)}\left(1 - \frac{d\varepsilon}{dp}\right)},
\end{equation}
or, equivalently,
\begin{equation}
\frac{dp}{dr} = -(\varepsilon + p) \frac{\left[4\pi pr - \frac{f(T)r}{4} + \frac{m}{r^2} \right]}{\left(1 - \frac{2m}{r}\right)\left[1 - \frac{f_T}{2(8\pi + f_T)}\left(1 - \frac{d\varepsilon}{dp}\right)\right]}. \label{TOVfRT2}
\end{equation}

To obtain the mass equation, we need the $G_{tt}$ component,
\begin{equation}
	G_{tt} = \frac{e^{\upsilon - \omega}(-1 + e^\omega + r\omega')}{r^2},
\end{equation}
and using $T_{tt} = \varepsilon e^{\upsilon}$ and $g_{tt} = e^\upsilon$, we obtain
\begin{equation*}
	G_{tt} = 8\pi T_{tt} + f_T(T_{tt} + p g_{tt}) + \frac{1}{2}f(T)g_{tt},
\end{equation*}
\begin{equation*}
	\frac{e^{\upsilon - \omega}(-1 + e^\omega + r\omega')}{r^2} = 8\pi \varepsilon e^\upsilon + f_T(\varepsilon e^\upsilon + p e^\upsilon) + \frac{1}{2}f(T)e^\upsilon,
\end{equation*}
\begin{equation*}
	\frac{e^\upsilon}{e^\omega}\frac{(-1 + e^\omega + r\omega')}{r^2} = 8\pi \varepsilon e^\upsilon + f_T(\varepsilon e^\upsilon + p e^\upsilon) + \frac{1}{2}f(T)e^\upsilon,
\end{equation*}
\begin{equation*}
	\frac{1}{e^\omega}\frac{(-1 + e^\omega + r\omega')}{r^2} = 8\pi \varepsilon + f_T(\varepsilon + p) + \frac{1}{2}f(T),
\end{equation*}
\begin{equation*}
	\frac{(-1 + e^\omega + r\omega')}{e^\omega} = 8\pi r^2 \varepsilon + f_T(\varepsilon + p)r^2 + \frac{1}{2}f(T)r^2,
\end{equation*}
\begin{equation*}
	-1 + e^\omega + r\omega'=e^\omega r^2\left[8\pi \varepsilon + f_T(\varepsilon + p) + \frac{1}{2}f(T)\right],
\end{equation*}
\begin{equation*}
	 r\omega'=e^\omega r^2\left[8\pi \varepsilon + f_T(\varepsilon + p) + \frac{1}{2}f(T)\right] +1 - e^\omega,
\end{equation*}
\begin{equation}
	 \omega'=e^\omega r\left[8\pi \varepsilon + f_T(\varepsilon + p) + \frac{1}{2}f(T)\right] +\frac{1 - e^\omega}{r},
\end{equation}
and since 
\begin{equation}
	e^\omega = \frac{r}{r-2m},
\end{equation}
we have
\begin{equation}
	\omega' = \frac{2rm' - 2m}{r(r-2m)} = \frac{2rm'}{r(r-2m)} - \frac{2m}{r(r-2m)},
\end{equation}
and so
\begin{equation*}
	 \omega'=e^\omega r\left[8\pi \varepsilon + f_T(\varepsilon + p) + \frac{1}{2}f(T)\right] -\frac{2m}{r(r-2m)},
\end{equation*}
\begin{equation*}
	 \frac{2rm'}{r(r-2m)} - \frac{2m}{r(r-2m)}= \frac{r}{r-2} r\left[8\pi \varepsilon + f_T(\varepsilon + p) + \frac{1}{2}f(T)\right] -\frac{2m}{r(r-2m)},
\end{equation*}
\begin{equation*}
	 \frac{2rm'}{r(r-2m)} =\frac{r^2}{r-2m}\left[8\pi \varepsilon + f_T(\varepsilon + p) + \frac{1}{2}f(T)\right] ,
\end{equation*}
\begin{equation*}
	 \frac{2rm'}{r} =r^2\left[8\pi \varepsilon + f_T(\varepsilon + p) + \frac{1}{2}f(T)\right] ,
\end{equation*}
\begin{equation*}
	 2m' =r^2\left[8\pi \varepsilon + f_T(\varepsilon + p) + \frac{1}{2}f(T)\right] ,
\end{equation*}
\begin{equation*}
	 2m' =8\pi r^2 \varepsilon + r^2\left[f_T(\varepsilon+ p) + \frac{1}{2}f(T)\right] ,
\end{equation*}
and  finally
\begin{equation}
		m' = 4\pi r^2 \varepsilon + r^2\left[ \frac{f_T(\varepsilon + p)}{2} + \frac{f(T)}{4}\right].
\end{equation}
\end{appendices}

\begin{appendices}

\chapter[MIT Bag Model and Walecka-type RMF model]{MIT Bag Model and Walecka-type relativistic mean-field model}
\label{app:mit_walecka}

In this appendix, we present the microphysical background of two equations of state used, directly or indirectly, throughout this thesis: the MIT Bag Model for deconfined quark matter and the Walecka-type relativistic mean-field model for dense hadronic matter. The purpose is not to provide an exhaustive review of dense-matter theory, but rather to collect the main equations and physical assumptions needed to understand the EoS inputs adopted in the stellar-structure calculations.

Both models enter the present work through the barotropic relation
\begin{equation}
p=p(\varepsilon),
\end{equation}
which closes the hydrostatic equilibrium equations. Here \(p\) is the isotropic pressure and \(\varepsilon\) is the total energy density. This notation is important: whenever an auxiliary baryonic or rest-mass density appears, it will be denoted explicitly by \(\rho_b\) or \(n_B\), in order to avoid confusion with the total energy density \(\varepsilon\) that enters the TOV, \(\mathcal{D}\)-TOV, and modified-gravity structure equations.

\section{MIT Bag Model}
\label{app:MIT_bag_model}

The MIT Bag Model provides a phenomenological description of deconfined quark matter. In this picture, quarks are treated as quasi-free particles inside a finite region, the ``bag'', while confinement is modeled through a constant vacuum energy density \(B\), known as the bag constant. This constant contributes positively to the total energy density and negatively to the pressure. Therefore, it acts as an external pressure that confines the quark gas.

The model is particularly useful in the context of strange quark matter and strange stars. According to the strange-matter hypothesis, bulk matter composed of up, down, and strange quarks may have an energy per baryon lower than that of ordinary nuclear matter at zero pressure. If this condition is satisfied, strange quark matter could be absolutely stable, and compact objects made predominantly of deconfined quark matter could exist as self-bound stars \citep{bodmer1971collapsed, witten1984cosmic, jaffe1984strange}.

At zero temperature, the thermodynamic quantities of a free Fermi gas of quarks can be written as sums over the quark flavors. For each flavor \(f\), with Fermi momentum \(k_{F,f}\), chemical potential \(\mu_f\), and current mass \(m_f\), the number density is
\begin{equation}
n_f = \frac{k_{F,f}^{3}}{\pi^2},
\end{equation}
where the color and spin degeneracies have already been included. The corresponding energy density and pressure are
\begin{equation}
\varepsilon_f
=
\frac{3}{\pi^2}
\int_0^{k_{F,f}}
\sqrt{k^2+m_f^2}\,k^2\,dk,
\end{equation}
and
\begin{equation}
p_f
=
\frac{1}{\pi^2}
\int_0^{k_{F,f}}
\frac{k^4}{\sqrt{k^2+m_f^2}}\,dk.
\end{equation}
The total energy density and pressure of quark matter in the MIT Bag Model are then
\begin{equation}
\varepsilon
=
\sum_f \varepsilon_f + B,
\end{equation}
and
\begin{equation}
p
=
\sum_f p_f - B.
\end{equation}
Thus, the bag constant shifts the zero of the pressure and energy density. Physically, this means that even when the kinetic pressure of the quarks vanishes at the surface, the energy density remains finite.

In the massless and noninteracting limit,
\begin{equation}
m_u=m_d=m_s=0,
\end{equation}
one has, for each quark flavor,
\begin{equation}
p_f = \frac{1}{3}\varepsilon_f.
\end{equation}
Consequently, for three-flavor massless quark matter, the total pressure and energy density satisfy the simple linear relation
\begin{equation}
p = \frac{1}{3}\left(\varepsilon-4B\right),
\label{eq:MIT_p_epsilon}
\end{equation}
or, equivalently,
\begin{equation}
\varepsilon = 3p+4B.
\label{eq:MIT_epsilon_p}
\end{equation}
This is the form used in the numerical integrations of the massless MIT Bag Model throughout this thesis. The surface of the star is defined by the condition
\begin{equation}
p(R)=0.
\end{equation}
From Eq.~\eqref{eq:MIT_epsilon_p}, however, this implies
\begin{equation}
\varepsilon(R)=4B.
\end{equation}
Therefore, strange stars described by the MIT Bag Model are self-bound objects with a finite surface energy density. This behavior differs from ordinary hadronic stars, for which the density decreases continuously toward zero at the surface.

The parameter \(B\) controls the stiffness of the EoS. Larger values of \(B\) increase the vacuum contribution to the energy density and reduce the pressure at fixed \(\varepsilon\), leading to a softer EoS. As a result, equilibrium configurations tend to support smaller maximum masses and smaller radii. Conversely, smaller values of \(B\) produce a stiffer relation between pressure and energy density, allowing more massive and more extended strange-star configurations.

The speed of sound in the massless MIT Bag Model is constant:
\begin{equation}
c_s^2 = \frac{dp}{d\varepsilon}=\frac{1}{3},
\end{equation}
in units where \(c=1\). This value is below the causal limit \(c_s^2=1\), and is characteristic of a gas of ultrarelativistic noninteracting particles. The simplicity of Eq.~\eqref{eq:MIT_p_epsilon} makes the MIT Bag Model especially useful for isolating the role of the gravitational framework and of the deformation parameter in strange-star configurations.

For numerical stellar-structure calculations, it is convenient to use Eq.~\eqref{eq:MIT_epsilon_p}, since the integration is usually performed by evolving the pressure outward from a prescribed central value \(p_c\). Once \(p(r)\) is known at each integration step, the corresponding total energy density is obtained directly from
\begin{equation}
\varepsilon(r)=3p(r)+4B.
\end{equation}
The central energy density is therefore
\begin{equation}
\varepsilon_c = 3p_c+4B.
\end{equation}
When the central pressure is varied over a grid, the resulting sequence can equivalently be described as a sequence in \(\varepsilon_c\).

In some extensions of the MIT Bag Model, the finite strange-quark mass and perturbative QCD corrections are incorporated phenomenologically through an approximately linear relation of the form
\begin{equation}
p = a\left(\varepsilon-\varepsilon_0\right),
\label{eq:MIT_linear_general}
\end{equation}
where \(a\) plays the role of an effective squared sound speed and \(\varepsilon_0\) is the energy density at zero pressure. The massless MIT Bag Model corresponds to
\begin{equation}
a=\frac{1}{3},
\qquad
\varepsilon_0=4B.
\end{equation}
In this thesis, such generalized forms are useful for comparing the massless case with configurations in which finite strange-quark mass effects effectively soften the quark-matter EoS.

\section{Walecka-type relativistic mean-field model}
\label{app:walecka_model}

The Walecka model, also known as Quantum Hadrodynamics I (QHD-I), provides a relativistic field-theoretical description of dense nuclear matter. In this approach, baryons are treated as Dirac fields interacting through meson exchange. The main physical idea is that the attractive part of the nuclear interaction is modeled by a scalar meson field, usually denoted by \(\sigma\), while the repulsive part is modeled by a vector meson field, usually denoted by \(\omega_\mu\). Extensions of the model include the isovector-vector \(\rho_\mu\) meson, which controls the isospin dependence of the interaction and is therefore essential for neutron-rich matter.

In its simplest form, the Walecka Lagrangian density for nucleons interacting with scalar and vector mesons is
\begin{equation}
\mathcal{L}
=
\bar{\psi}
\left[
i\gamma^\mu\partial_\mu
-
M
+
g_\sigma\sigma
-
g_\omega\gamma^\mu\omega_\mu
\right]\psi
+
\frac{1}{2}
\left(
\partial_\mu\sigma\,\partial^\mu\sigma
-
m_\sigma^2\sigma^2
\right)
-
\frac{1}{4}
\Omega_{\mu\nu}\Omega^{\mu\nu}
+
\frac{1}{2}m_\omega^2\omega_\mu\omega^\mu,
\label{eq:walecka_lagrangian_simple}
\end{equation}
where \(\psi\) is the nucleon field, \(M\) is the bare nucleon mass, \(m_\sigma\) and \(m_\omega\) are meson masses, \(g_\sigma\) and \(g_\omega\) are coupling constants, and
\begin{equation}
\Omega_{\mu\nu}
=
\partial_\mu\omega_\nu-\partial_\nu\omega_\mu
\end{equation}
is the field tensor associated with the vector meson.

For neutron-star matter, one usually considers an extension containing protons, neutrons, leptons, and the \(\rho\)-meson. The baryonic part of the Lagrangian can be written schematically as
\begin{equation}
\mathcal{L}_B
=
\sum_B
\bar{\psi}_B
\left[
i\gamma^\mu\partial_\mu
-
M_B
+
g_{\sigma B}\sigma
-
g_{\omega B}\gamma^\mu\omega_\mu
-
g_{\rho B}\gamma^\mu I_{3B}\rho_\mu
\right]\psi_B,
\end{equation}
where \(B\) labels the baryon species, \(I_{3B}\) is the third component of isospin, and the couplings \(g_{\sigma B}\), \(g_{\omega B}\), and \(g_{\rho B}\) determine the strength of the scalar, vector, and isovector interactions.

In the relativistic mean-field approximation, the meson fields are replaced by their expectation values. For static, uniform matter, only the time components of the vector fields survive:
\begin{equation}
\sigma \rightarrow \sigma_0,
\qquad
\omega^\mu \rightarrow \delta^\mu_0 \omega_0,
\qquad
\rho^\mu \rightarrow \delta^\mu_0 \rho_{03}.
\end{equation}
The scalar field modifies the effective baryon mass,
\begin{equation}
M_B^\ast = M_B - g_{\sigma B}\sigma_0,
\label{eq:effective_mass}
\end{equation}
while the vector fields shift the baryon chemical potentials. For a baryon \(B\), the chemical potential is
\begin{equation}
\mu_B
=
\sqrt{k_{F,B}^2+{M_B^\ast}^2}
+
g_{\omega B}\omega_0
+
g_{\rho B}I_{3B}\rho_{03}.
\label{eq:chemical_potential_baryon}
\end{equation}
The baryon number density is
\begin{equation}
n_B = \frac{k_{F,B}^3}{3\pi^2},
\end{equation}
and the scalar density is
\begin{equation}
n_{s,B}
=
\frac{1}{\pi^2}
\int_0^{k_{F,B}}
\frac{M_B^\ast}{\sqrt{k^2+{M_B^\ast}^2}}\,k^2\,dk.
\end{equation}

The mean-field equations for the meson fields are obtained by minimizing the energy density, or equivalently by applying the Euler-Lagrange equations to the Lagrangian. In the presence of nonlinear scalar self-interactions, the scalar potential is often written as
\begin{equation}
U(\sigma)
=
\frac{1}{3}b\,M\left(g_\sigma\sigma\right)^3
+
\frac{1}{4}c\left(g_\sigma\sigma\right)^4,
\end{equation}
where \(b\) and \(c\) are dimensionless parameters. The field equations then take the schematic form
\begin{equation}
m_\sigma^2\sigma_0
+
\frac{dU}{d\sigma_0}
=
\sum_B g_{\sigma B} n_{s,B},
\label{eq:sigma_field}
\end{equation}
\begin{equation}
m_\omega^2\omega_0
=
\sum_B g_{\omega B} n_B,
\label{eq:omega_field}
\end{equation}
and
\begin{equation}
m_\rho^2\rho_{03}
=
\sum_B g_{\rho B} I_{3B} n_B.
\label{eq:rho_field}
\end{equation}
These equations must be solved self-consistently for each value of the total baryon density.

The total energy density of cold beta-equilibrated matter is given by
\begin{equation}
\varepsilon
=
\sum_B
\frac{1}{\pi^2}
\int_0^{k_{F,B}}
\sqrt{k^2+{M_B^\ast}^2}\,k^2\,dk
+
\frac{1}{2}m_\sigma^2\sigma_0^2
+
U(\sigma_0)
+
\frac{1}{2}m_\omega^2\omega_0^2
+
\frac{1}{2}m_\rho^2\rho_{03}^2
+
\sum_l \varepsilon_l,
\label{eq:rmf_energy_density}
\end{equation}
where \(l=e,\mu\) denotes the leptons. The pressure is
\begin{equation}
p
=
\sum_B
\frac{1}{3\pi^2}
\int_0^{k_{F,B}}
\frac{k^4}{\sqrt{k^2+{M_B^\ast}^2}}\,dk
-
\frac{1}{2}m_\sigma^2\sigma_0^2
-
U(\sigma_0)
+
\frac{1}{2}m_\omega^2\omega_0^2
+
\frac{1}{2}m_\rho^2\rho_{03}^2
+
\sum_l p_l.
\label{eq:rmf_pressure}
\end{equation}
Equations~\eqref{eq:rmf_energy_density} and \eqref{eq:rmf_pressure} define the hadronic EoS once the coupling constants and meson masses have been specified.

For neutron-star matter, weak interactions impose beta equilibrium. In nucleonic matter composed of neutrons, protons, electrons, and muons, the equilibrium conditions are
\begin{equation}
\mu_n = \mu_p+\mu_e,
\end{equation}
and
\begin{equation}
\mu_e=\mu_\mu,
\end{equation}
whenever muons are present. Charge neutrality requires
\begin{equation}
n_p=n_e+n_\mu.
\end{equation}
Solving these equations together with the meson field equations determines the particle fractions and the pressure as functions of the total energy density.

The GM1 parametrization used in this thesis belongs to this class of nonlinear relativistic mean-field models. It was introduced by Glendenning and Moszkowski and has been widely used in neutron-star calculations. In the present work, GM1 is adopted as a representative hadronic EoS because it provides a standard description of cold, charge-neutral, beta-equilibrated nucleonic matter and is sufficiently stiff to support massive neutron-star configurations. Its role in this thesis is not to perform a detailed EoS inference analysis, but to provide a realistic hadronic input with which the effects of deformation and modified gravity can be systematically compared.

\section{Connection with the stellar-structure equations}
\label{app:eos_connection_structure}

The MIT Bag Model and Walecka-type RMF models describe different microscopic compositions. The MIT Bag Model is used for deconfined quark matter and self-bound strange-star configurations, while the Walecka-type RMF framework underlies hadronic equations of state such as GM1, appropriate for conventional neutron-star matter. Despite this difference, both models ultimately provide the same type of input required by the stellar-structure equations: a relation between pressure and total energy density,
\begin{equation}
p=p(\varepsilon).
\end{equation}

Once an EoS is specified, the hydrostatic equilibrium equations determine the radial profiles \(m(r)\), \(p(r)\), and \(\varepsilon(r)\). In the standard spherical case, these profiles are obtained from the TOV system. In the deformed case, the same EoS is inserted into the \(\mathcal{D}\)-TOV equations. In modified gravity, the trace dependence introduces additional terms involving
\begin{equation}
T=\varepsilon-3p
\end{equation}
for the signature convention adopted in this thesis. Therefore, a consistent treatment requires not only \(p(\varepsilon)\), but also the ability to compute \(\varepsilon(p)\) and, in some formulations, derivatives such as \(d\varepsilon/dp\).

For the massless MIT Bag Model, this inversion is analytic:
\begin{equation}
\varepsilon(p)=3p+4B.
\end{equation}
For RMF models such as GM1, the relation is obtained numerically from the solution of the mean-field equations and is typically supplied as a tabulated EoS. In the stellar-structure codes used in this thesis, the tabulated relation is interpolated and inserted into the corresponding structure equations.

This appendix therefore clarifies the physical origin of the EoS inputs used in the numerical analysis. The MIT Bag Model allows us to study the behavior of self-bound quark stars, while the Walecka-type RMF framework provides the basis for modeling hadronic neutron stars. Together with the polytropic baseline discussed in the main text, these models allow us to compare how different matter descriptions respond to the effects of deformation and modified gravity.

\end{appendices}

\begin{appendices}
\chapter{Computational framework: the \textsc{SAURON} code}
\label{app:sauron_code}

The numerical results presented throughout this thesis were obtained with a computational
framework developed during this doctoral work, which we refer to as \textsc{SAURON}:
\textit{System for Astrophysical Understanding and Resolution of Neutron stars}. The code
was designed to solve the structure equations of compact stars in different physical
scenarios, including the standard TOV system, the \(\mathcal{D}\)-TOV formalism,
modified-gravity extensions, strange-star configurations described by the MIT Bag Model,
hybrid-star configurations, and radial-oscillation calculations.

The development of \textsc{SAURON} followed an incremental strategy. Each physical module
was implemented, tested, and validated as the research progressed. In this sense, the code
should not be understood as a single isolated routine, but as a computational environment
built to support the broader investigation of compact objects carried out in this thesis.
The numerical framework allowed different equations of state, gravitational models, and
deformation parameters to be compared within a unified structure.

The main modules implemented in \textsc{SAURON} include routines for the integration of
the standard Tolman--Oppenheimer--Volkoff equations, the effective \(\mathcal{D}\)-TOV
equations, the \(f(R,T)\) modified-gravity structure equations, and the deformed
\(f(R,T)\) stellar equations. Additional routines were developed to handle tabulated and
analytic equations of state, identify stellar surfaces through the condition \(p(R)=0\),
construct mass--radius sequences, locate maximum-mass configurations, and generate
normalized pressure and energy-density profiles.

For tabulated equations of state, the code interpolates the relation between pressure and
energy density and evaluates the required thermodynamic quantities during the integration.
For analytic equations of state, such as the massless MIT Bag Model and the polytropic
model, the corresponding relations \(\varepsilon(p)\) are computed directly. This modular
structure made it possible to apply the same numerical strategy to hadronic stars,
strange stars, hybrid configurations, and controlled polytropic models.

The integration of the stellar-structure equations was performed using Runge--Kutta
methods, with the integration starting from a small nonzero radius in order to avoid the
coordinate singularity at the origin. Regular central conditions were imposed by using the
central pressure and the corresponding central energy density obtained from the equation
of state. The stellar radius was determined by the vanishing of the pressure, \(p(R)=0\),
and the gravitational mass was extracted from the mass function at the stellar surface.

The code was also extended to compute radial oscillation frequencies for selected
equilibrium configurations. In this case, the equilibrium solution obtained from the
structure equations was used as the background on which the perturbation equations were
integrated. The eigenfrequencies were then obtained through a shooting procedure based on
the surface boundary condition for the radial perturbation variables.

Artificial-intelligence tools were used as auxiliary resources during the development of
some numerical routines, particularly for code organization, debugging, and implementation
support. However, the physical formulation of the problems, including the definition of
the governing equations, boundary conditions, equations of state, numerical tests, and
interpretation of the results, remained under the direct responsibility of the author.
All equations implemented in the code were derived, selected, and validated within the
theoretical framework developed in this thesis.

Therefore, \textsc{SAURON} should be understood as a research tool developed to support
the numerical investigation of deformed compact objects in general relativity and
modified gravity. Its role in this thesis is methodological: it provides the computational
basis for obtaining the equilibrium sequences, maximum masses, radial profiles, and
selected oscillation spectra discussed in the main chapters.

\end{appendices}

\begin{appendices}
\chapter{Friedmann equations in the Energy--Curvature Association model}
\label{app:ECA_friedmann}

In addition to the compact-star applications developed in the main chapters of this thesis, modified gravity was also explored during the doctoral work in a cosmological context. This appendix presents the step-by-step derivation of the Friedmann-like equations associated with a phenomenological modified-gravity ansatz referred to here as the Energy--Curvature Association (ECA) model. The purpose is to record this complementary formalism, which was developed in parallel with the neutron-star studies, and to show explicitly how the cosmological background equations follow from the proposed action.

The ECA model introduces a coupling between the trace of the energy--momentum tensor, \(T\), and the Ricci scalar, \(R\), through the ratio \(T/R\). Since this construction is still under development and is not used in the stellar-structure calculations presented in the main text, the discussion is placed in this appendix. Nevertheless, it illustrates how the modified-gravity ideas explored throughout the thesis can also be applied to homogeneous and isotropic cosmology. We also clarify below the role and dimensionality of the coupling parameter that controls the strength of the ECA contribution.

The ECA model is based on the functional dependence
\begin{equation}
	f(R,T) = R + \xi \frac{T}{R},
\end{equation}
where \(R\) is the Ricci scalar, \(T\equiv g^{\mu\nu}T_{\mu\nu}\) is the trace of the energy--momentum tensor, and \(\xi\) is a coupling constant. The term \(T/R\) introduces an explicit association between the matter sector and the curvature scalar. Therefore, this model belongs to the class of matter--geometry coupled theories and should be interpreted as a phenomenological extension of GR.

In geometrized units, \(G=c=1\), the quantities \(R\) and \(T\) have the same physical dimension, namely
\begin{equation}
	[R]=[T]=L^{-2}.
\end{equation}
Consequently, the ratio \(T/R\) is dimensionless. Since \(f(R,T)\) must have the same dimension as the Ricci scalar, the coupling parameter \(\xi\) must have dimensions of curvature,
\begin{equation}
	[\xi]=L^{-2}.
\end{equation}
Equivalently, \(\xi\) defines a characteristic curvature scale for the additional ECA contribution. The GR limit is recovered when \(\xi\to 0\). We also note that, because the model contains inverse powers of \(R\), the formulation assumes \(R\neq 0\) in the regime under consideration.

For the functional above, one has
\begin{equation}
	f_R = \frac{\partial f}{\partial R}
	=
	1 - \xi \frac{T}{R^2},
	\qquad
	f_T = \frac{\partial f}{\partial T}
	=
	\frac{\xi}{R}.
\end{equation}
The field equations in the \(f(R,T)\) formalism may be written as
\begin{equation}
	f_R R_{\mu \nu}
	-\frac{1}{2}f(R,T)g_{\mu \nu}
	+\left(g_{\mu \nu}\Box-\nabla_\mu\nabla_\nu\right)f_R
	=
	8\pi T_{\mu\nu}
	-
	f_T\left(T_{\mu\nu}+\Theta_{\mu\nu}\right),
\end{equation}
where
\begin{equation}
	\Theta_{\mu\nu}
	\equiv
	g^{\alpha\beta}
	\frac{\delta T_{\alpha\beta}}{\delta g^{\mu\nu}}.
\end{equation}
Here \(\nabla_\mu\) denotes the covariant derivative compatible with the metric \(g_{\mu\nu}\), namely the derivative associated with the Levi-Civita connection. The operator \(\Box\) is the corresponding covariant d'Alembertian, defined by
\begin{equation}
\Box \equiv g^{\alpha\beta}\nabla_\alpha\nabla_\beta .
\end{equation}
When acting on a scalar quantity \(\phi\), one has
\begin{equation}
\nabla_\mu \phi = \partial_\mu \phi,
\qquad
\nabla_\mu\nabla_\nu\phi
=
\partial_\mu\partial_\nu\phi
-
\Gamma^\lambda_{\mu\nu}\partial_\lambda\phi .
\end{equation}

Substituting the ECA functional in E.6 we obtain
\begin{equation}
	\left(1-\xi\frac{T}{R^2}\right)R_{\mu\nu}
	-\frac{1}{2}g_{\mu\nu}\left(R+\xi\frac{T}{R}\right)
	+\left(g_{\mu\nu}\Box-\nabla_\mu\nabla_\nu\right)
	\left(1-\xi\frac{T}{R^2}\right)
	=
	8\pi T_{\mu\nu}
	-\frac{\xi}{R}\left(T_{\mu\nu}+\Theta_{\mu\nu}\right).
\end{equation}

Defining
\begin{equation}
	\chi \equiv \frac{T}{R^2},
\end{equation}
and using
\begin{equation}
	\left(g_{\mu\nu}\Box-\nabla_\mu\nabla_\nu\right)(1)=0,
\end{equation}
we find
\begin{equation}
	(1-\xi\chi)R_{\mu\nu}
	-\frac{1}{2}Rg_{\mu\nu}
	-\xi\frac{T}{2R}g_{\mu\nu}
	-\xi\left(g_{\mu\nu}\Box-\nabla_\mu\nabla_\nu\right)\chi
	=
	8\pi T_{\mu\nu}
	-\frac{\xi}{R}\left(T_{\mu\nu}+\Theta_{\mu\nu}\right).
\end{equation}
Isolating the Einstein tensor gives
\begin{equation}
	G_{\mu\nu}
	=
	8\pi T_{\mu\nu}
	+
	\xi
	\left[
	\chi R_{\mu\nu}
	+
	\left(g_{\mu\nu}\Box-\nabla_\mu\nabla_\nu\right)\chi
	+
	\frac{T}{2R}g_{\mu\nu}
	-
	\frac{1}{R}\left(T_{\mu\nu}+\Theta_{\mu\nu}\right)
	\right].
\end{equation}
Introducing the operator
\begin{equation}
	A_{\mu\nu}
	\equiv
	R_{\mu\nu}
	+
	g_{\mu\nu}\Box
	-
	\nabla_\mu\nabla_\nu,
\end{equation}
the field equations can be written as
\begin{equation}
	G_{\mu\nu}
	=
	8\pi T_{\mu\nu}
	+
	\xi
	\left[
	A_{\mu\nu}\chi
	+
	\frac{T}{2R}g_{\mu\nu}
	-
	\frac{1}{R}\left(T_{\mu\nu}+\Theta_{\mu\nu}\right)
	\right].
\end{equation}

For the matter Lagrangian choice adopted here, one has
\begin{equation}
	\Theta_{\mu\nu}
	=
	-2T_{\mu\nu}
	+
	g_{\mu\nu}\mathcal{L}_m.
\end{equation}
Therefore,
\begin{equation}
	G_{\mu\nu}
	=
	8\pi T_{\mu\nu}
	+
	\xi
	\left[
	A_{\mu\nu}\chi
	+
	\frac{T}{2R}g_{\mu\nu}
	+
	\frac{T_{\mu\nu}}{R}
	-
	\frac{g_{\mu\nu}\mathcal{L}_m}{R}
	\right].
\end{equation}
The last three terms can be combined as
\begin{equation}
	\frac{T}{2R}g_{\mu\nu}
	+
	\frac{T_{\mu\nu}}{R}
	-
	\frac{g_{\mu\nu}\mathcal{L}_m}{R}
	=
	\frac{1}{R}
	\left[
	T_{\mu\nu}
	+
	\left(\frac{T}{2}-\mathcal{L}_m\right)g_{\mu\nu}
	\right].
\end{equation}
Thus, the ECA field equations take the compact form
\begin{equation}
	G_{\mu\nu}
	=
	8\pi T_{\mu\nu}
	+
	\xi
	\left\{
	A_{\mu\nu}\chi
	+
	\left[
	T_{\mu\nu}
	+
	\left(\frac{T}{2}-\mathcal{L}_m\right)g_{\mu\nu}
	\right]\frac{1}{R}
	\right\}.
	\label{eq:ECA_field_appendix}
\end{equation}

Taking the trace of Eq.~\eqref{eq:ECA_field_appendix}, one obtains
\begin{equation}
	-R
	=
	8\pi T
	+
	\xi
	\left[
	\left(R+3\Box\right)\chi
	+
	\frac{3T-4\mathcal{L}_m}{R}
	\right].
\end{equation}
This relation shows explicitly that the model introduces a scalar matter--curvature coupling through the quantities \(T\), \(R\), and their derivatives.

We now consider the spatially flat FLRW metric,
\begin{equation}
	ds^2
	=
	-dt^2
	+
	a(t)^2
	\left[
	dr^2
	+
	r^2
	\left(d\theta^2+\sin^2\theta\,d\varphi^2\right)
	\right],
	\label{eq:FLRW_ECA_appendix}
\end{equation}
together with the matter Lagrangian density
\begin{equation}
	\mathcal{L}_m=-\varepsilon,
\end{equation}
and the perfect-fluid energy--momentum tensor
\begin{equation}
	T_{\mu\nu}
	=
	(\varepsilon+p)u_\mu u_\nu
	+
	pg_{\mu\nu}.
\end{equation}
This choice of \(\mathcal{L}_m\) is commonly adopted for cosmological perfect fluids. In matter--geometry coupled theories, however, different choices of \(\mathcal{L}_m\) may lead to different forms of \(\Theta_{\mu\nu}\), and therefore to different effective field equations. In this appendix, we keep the choice \(\mathcal{L}_m=-\varepsilon\) fixed in order to derive the background equations used in the ECA analysis.

The Hubble function is defined as
\begin{equation}
	H\equiv\frac{\dot a}{a},
\end{equation}
and the relevant curvature quantities are
\begin{equation}
	R_{00}
	=
	-3\frac{\ddot a}{a}
	=
	-3\left(H^2+\dot H\right),
	\qquad
	G_{00}=3H^2,
\end{equation}
and
\begin{equation}
	R^1{}_{1}=3H^2+\dot H,
	\qquad
	G^1{}_{1}=-3H^2-2\dot H.
\end{equation}
The Ricci scalar is
\begin{equation}
	R=6\left(\dot H+2H^2\right).
\end{equation}
For the perfect fluid,
\begin{equation}
	T=-\varepsilon+3p,
	\qquad
	T_{00}=\varepsilon,
	\qquad
	T^1{}_{1}=p.
\end{equation}

For the temporal component, Eq.~\eqref{eq:ECA_field_appendix} gives
\begin{equation}
	G_{00}
	=
	8\pi T_{00}
	+
	\xi
	\left\{
	A_{00}\chi
	+
	\left[
	T_{00}
	+
	\left(\frac{T}{2}-\mathcal{L}_m\right)g_{00}
	\right]\frac{1}{R}
	\right\}.
\end{equation}
Using \(T=-\varepsilon+3p\), \(\mathcal{L}_m=-\varepsilon\), \(T_{00}=\varepsilon\), and \(g_{00}=-1\), the algebraic term becomes
\begin{equation}
	T_{00}
	+
	\left(\frac{T}{2}-\mathcal{L}_m\right)g_{00}
	=
	-\frac{3p-\varepsilon}{2}.
\end{equation}
Hence,
\begin{equation}
	3H^2
	=
	8\pi\varepsilon
	+
	\xi
	\left[
	A_{00}\chi
	-
	\frac{3p-\varepsilon}{2R}
	\right].
\end{equation}

We now evaluate
\begin{equation}
	A_{00}\chi
	=
	R_{00}\chi
	+
	g_{00}\Box\chi
	-
	\nabla_0\nabla_0\chi.
\end{equation}
For a homogeneous scalar \(\chi=\chi(t)\),
\begin{equation}
	\nabla_0\nabla_0\chi=\ddot\chi,
\end{equation}
and
\begin{equation}
	\Box\chi
	=
	\frac{1}{\sqrt{-g}}
	\partial_\mu
	\left(
	\sqrt{-g}\,g^{\mu\nu}\partial_\nu\chi
	\right)
	=
	-\ddot\chi-3H\dot\chi.
\end{equation}
Therefore,
\begin{equation}
	A_{00}\chi
	=
	-3\left(H^2+\dot H\right)\chi
	+
	3H\dot\chi.
\end{equation}
Substituting this result, we obtain the first ECA Friedmann-like equation,
\begin{equation}
	3H^2
	=
	8\pi\varepsilon
	+
	\xi
	\left[
	3H\dot\chi
	-
	3\left(H^2+\dot H\right)\chi
	-
	\tilde\chi
	\right],
	\label{eq:ECA_friedmann1_appendix}
\end{equation}
where
\begin{equation}
	\tilde\chi
	\equiv
	\frac{3p-\varepsilon}{2R}.
\end{equation}

To derive the second Friedmann-like equation, we consider a spatial component in mixed form:
\begin{equation}
	G^1{}_{1}
	=
	8\pi T^1{}_{1}
	+
	\xi
	\left\{
	A^1{}_{1}\chi
	+
	\left[
	T^1{}_{1}
	+
	\left(\frac{T}{2}-\mathcal{L}_m\right)\delta^1{}_{1}
	\right]\frac{1}{R}
	\right\}.
\end{equation}
Since \(\delta^1{}_{1}=1\), we have
\begin{equation}
	-2\dot H-3H^2
	=
	8\pi p
	+
	\xi
	\left\{
	A^1{}_{1}\chi
	+
	\frac{\varepsilon+5p}{2R}
	\right\}.
\end{equation}

We now evaluate
\begin{equation}
	A^1{}_{1}\chi
	=
	R^1{}_{1}\chi
	+
	\delta^1{}_{1}\Box\chi
	-
	\nabla^1\nabla_1\chi.
\end{equation}
For a homogeneous scalar,
\begin{equation}
	\nabla^1\nabla_1\chi=-H\dot\chi.
\end{equation}
Using \(\Box\chi=-\ddot\chi-3H\dot\chi\), one obtains
\begin{equation}
	A^1{}_{1}\chi
	=
	\left(3H^2+\dot H\right)\chi
	-
	\ddot\chi
	-
	2H\dot\chi.
\end{equation}
Substituting this result into the spatial field equation gives
\begin{equation}
	-2\dot H-3H^2
	=
	8\pi p
	+
	\xi
	\left[
	\left(3H^2+\dot H\right)\chi
	-
	\ddot\chi
	-
	2H\dot\chi
	+
	\frac{\varepsilon+5p}{2R}
	\right].
\end{equation}
Equivalently,
\begin{equation}
	2\dot H+3H^2
	=
	-8\pi p
	+
	\xi
	\left[
	\ddot\chi
	+
	2H\dot\chi
	-
	\left(\dot H+3H^2\right)\chi
	-
	\tilde{\tilde\chi}
	\right],
	\label{eq:ECA_friedmann2_appendix}
\end{equation}
where
\begin{equation}
	\tilde{\tilde\chi}
	\equiv
	\frac{\varepsilon+5p}{2R}.
\end{equation}

Equations~\eqref{eq:ECA_friedmann1_appendix} and \eqref{eq:ECA_friedmann2_appendix} reduce to the standard GR Friedmann equations in the limit \(\xi\to0\). For nonzero \(\xi\), the additional terms depend on
\begin{equation}
	\chi=\frac{T}{R^2}
	=
	\frac{-\varepsilon+3p}{R^2},
\end{equation}
and on its time derivatives. These contributions can be interpreted as an effective modification of the cosmic source sector at the background level.

\end{appendices}

\end{document}